\documentclass[a4paper,12pt]{article}
\pdfoutput=1 

\usepackage{jheppub_X} 

\usepackage{booktabs,colortbl}
\usepackage{cancel,xcolor,multirow}
\usepackage[normal]{caption}
\usepackage{subcaption}
\usepackage{cleveref}
\usepackage{hyperref}
\usepackage{comment}
\usepackage[normalem]{ulem}
\usepackage{tabu}
\usepackage{slashed}
\usepackage{setspace}
\usepackage{physics}
\usepackage{bm}
\usepackage{enumerate}
\usepackage{cancel}

\usepackage{tikz}
\usetikzlibrary{shapes,shapes.misc,decorations.pathmorphing,decorations.markings,arrows.meta}
\tikzset{cross/.style={cross out, draw=black, minimum size=2*(#1-\pgflinewidth), inner sep=0pt, outer sep=0pt},cross/.default={1pt}}

\usepackage{tcolorbox}
\definecolor{light-gray}{gray}{0.97}

\crefname{table}{Table}{Tables}
\crefname{equation}{Eq.}{Eqs.}
\crefname{appendix}{App.}{Apps.}
\crefname{section}{Sec.}{Secs.}
\crefname{figure}{Fig.}{Figs.}

\usepackage[T1]{fontenc}
\usepackage{listings}
\usepackage[utf8]{inputenc}
\usepackage{lineno}

\newcommand{\pathd}{\mathcal{D}}

\newcommand{\toFP}{\,\,\xrightarrow[\hspace{3pt}{h\,\to\, h_*}\hspace{3pt}]{}\,\,}

\newcommand{\be}{\begin{equation}}
\newcommand{\ee}{\end{equation}}
\newcommand{\bea}{\begin{eqnarray}}
\newcommand{\eea}{\end{eqnarray}}
\newcommand{\bs}{\begin{split}}
\newcommand{\es}{\end{split}}

\providecommand{\Refs}[1]{Refs.~\cite{#1}}

\newcommand{\Lag}{\mathcal{L}}
\newcommand{\dif}{\mathrm{d}}
\newcommand{\rr}{\mathbb{R}}

\def\hc{\text{h.c.}}
\def\eg{\textit{e.g.}}
\def\ie{\textit{i.e.}}

\newcommand{\s}{\hspace{0.8pt}}

\allowdisplaybreaks

 \def\stop{\ensuremath{\mathchoice%
      {\displaystyle\raise.0ex\hbox{$\displaystyle\tilde t$}}%
         {\textstyle\raise.0ex\hbox{$\textstyle\tilde t$}}%
       {\scriptstyle\raise.0ex\hbox{$\scriptstyle\tilde t$}}%
 {\scriptscriptstyle\raise.0ex\hbox{$\scriptscriptstyle\tilde t$}}}}
\def\stopone{\ensuremath{\mathchoice%
      {\displaystyle\raise.0ex\hbox{$\displaystyle\tilde t_1$}}%
         {\textstyle\raise-.2ex\hbox{$\textstyle\tilde t_1$}}%
       {\scriptstyle\raise.0ex\hbox{$\scriptstyle\tilde t_1$}}%
 {\scriptscriptstyle\raise.0ex\hbox{$\scriptscriptstyle\tilde t_1$}}}}

\def\nino{\ensuremath{\mathchoice%
      {\displaystyle\raise.4ex\hbox{$\displaystyle\tilde\chi^0$}}%
         {\textstyle\raise.4ex\hbox{$\textstyle\tilde\chi^0$}}%
       {\scriptstyle\raise.3ex\hbox{$\scriptstyle\tilde\chi^0$}}%
 {\scriptscriptstyle\raise.3ex\hbox{$\scriptscriptstyle\tilde\chi^0$}}}}
\def\ninoone{\ensuremath{\mathchoice%
      {\displaystyle\raise.4ex\hbox{$\displaystyle\tilde\chi^0_1$}}%
         {\textstyle\raise.1ex\hbox{$\textstyle\tilde\chi^0_1$}}%
       {\scriptstyle\raise.3ex\hbox{$\scriptstyle\tilde\chi^0_1$}}%
 {\scriptscriptstyle\raise.3ex\hbox{$\scriptscriptstyle\tilde\chi^0_1$}}}}

\lstdefinestyle{myCustomMatlabStyle}{
  language=Matlab,
  stepnumber=1,
  numbersep=10pt,
  tabsize=4,
  showspaces=false,
  showstringspaces=false
}

\title{\Large
Is SMEFT Enough? \\[5pt]
}

\author[a]{Timothy Cohen,}
\author[b]{Nathaniel Craig,}
\author[a]{Xiaochuan Lu,}
\author[c]{and Dave Sutherland\hspace{1pt}}

\affiliation[a]{Institute for Fundamental Science, University of Oregon, Eugene, Oregon 97403, USA}
\affiliation[b]{Department of Physics, University of California, Santa Barbara, CA 93106, USA}
\affiliation[c]{INFN Sezione di Trieste, via Bonomea 265, 34136 Trieste, Italy}

\emailAdd{tcohen@uoregon.edu}
\emailAdd{ncraig@ucsb.edu}
\emailAdd{xlu@uoregon.edu}
\emailAdd{dave.sutherland@sissa.it}

\abstract{
There are two canonical approaches to treating the Standard Model as an Effective Field Theory (EFT): Standard Model EFT (SMEFT), expressed in the electroweak symmetric phase utilizing the Higgs doublet, and Higgs EFT (HEFT), expressed in the broken phase utilizing the physical Higgs boson and an independent set of Goldstone bosons. 
HEFT encompasses SMEFT, so understanding whether SMEFT is sufficient motivates identifying UV theories that \emph{require} HEFT as their low energy limit. This distinction is complicated by field redefinitions that obscure the naive differences between the two EFTs.
By reformulating the question in a geometric language, we derive concrete criteria that can be used to distinguish SMEFT from HEFT independent of the chosen field basis.
We highlight two cases where perturbative new physics must be matched onto HEFT: ($i$) the new particles derive all of their mass from electroweak symmetry breaking, and ($ii$) there are additional sources of electroweak symmetry breaking.
Additionally, HEFT has a broader practical application: it can provide a more convergent parametrization when new physics lies near the weak scale.
The ubiquity of models requiring HEFT suggests that SMEFT is not enough.
}

\begin{document}
\maketitle
\flushbottom
\setcounter{page}{2}
\newpage

\begin{spacing}{1.1}
\parskip=0ex
\section{Introduction}
\label{sec:Intro}
Treating the Standard Model (SM) as an Effective Field Theory (EFT) is a principled way to organize the observable impact of new physics \cite{Weinberg:1980wa}.\footnote{For reviews of various aspects of EFTs, see \eg~\cite{Polchinski:1992ed, Georgi:1994qn, Manohar:1995xr,  Kaplan:1995uv, Rothstein:2003mp, Kaplan:2005es, Skiba:2010xn, Petrov:2016azi, Manohar:2018aog,  Neubert:2019mrz, Cohen:2019wxr, Penco:2020kvy}.}
Defining an EFT entails choosing a set of low energy degrees of freedom, and specifying a UV cutoff and a set of symmetries.
The EFT Lagrangian is then expressed as an expansion in terms of operators that respect the symmetries and is truncated to a given order as determined by the power counting.
One major benefit of this approach is that constraints on an EFT are ``model independent'' in that different ``UV completions'' can be matched onto the same set of low energy EFT parameters.
This can be contrasted with the ``model dependent'' approach, where one searches for the signatures of a specific new particle(s).
However, in choosing among EFT frameworks that are distinguished by manifesting different symmetries, one necessarily assumes certain properties of the UV completion.
The goal of this paper is to elucidate how aspects of short-distance physics determine the appropriate EFT framework for extensions of the SM.

There are two logically distinct EFT frameworks for capturing Beyond the Standard Model (BSM) effects.
The first approach is the ``Standard Model Effective Field Theory'' (SMEFT), involving the most general set of local operators invariant under an $SU(3)_C \times SU(2)_L \times U(1)_Y$ gauge symmetry, see \eg~\cite{Weinberg:1979sa, Buchmuller:1985jz, Leung:1984ni} along with recent reviews~\cite{deFlorian:2016spz, Brivio:2017vri}.
Here irrelevant operators are suppressed by a new physics scale $\Lambda$.
Insofar as all low energy states are modeled using fields that transform linearly under the assumed symmetries, the observed Higgs boson $h$ is a component of an electroweak doublet scalar $H$.
The second approach is what has become known as the ``Higgs Effective Field Theory'' (HEFT),\footnote{Alternately, the Higgs-Electroweak Chiral Lagrangian (EWCh$\mathcal{L}$).} in which the only manifest gauge symmetry is $SU(3)_C \times U(1)_\text{em}$, see \eg~\cite{Feruglio:1992wf, Bagger:1993zf, Koulovassilopoulos:1993pw, Burgess:1999ha, Grinstein:2007iv, Alonso:2012px, Espriu:2013fia, Buchalla:2013rka, Brivio:2013pma, Alonso:2015fsp, Alonso:2016oah,Buchalla:2017jlu, Alonso:2017tdy,deBlas:2018tjm, Falkowski:2019tft}.
The fundamental $SU(2)_L \times U(1)_Y$ electroweak symmetry is non-linearly realized utilizing a multiplet of Goldstone bosons.
But since a non-linearly realized gauge symmetry is equivalent to no symmetry at all, no relation is presumed between $h$ and the Goldstones.
In HEFT, irrelevant operators can be suppressed by the electroweak breaking scale $v$.

From symmetry considerations alone, HEFT is clearly the most general parametrization of low energy physics involving only the SM degrees of freedom:
\begin{align}
\text{HEFT } \supset \text{ SMEFT } \supset \text{ SM}\,.
\label{eq:HEFTsupsetSMEFT}
\end{align}
However, SMEFT is generically more straightforward to work with than HEFT -- as it befits from a more restrictive symmetry structure -- and has rapidly become the pre-eminent framework for interpreting LHC data in terms of an EFT.
To the extent that this choice imposes assumptions about the properties of physics in the UV, we are confronted with a pressing question: what UV physics is precluded by working with SMEFT rather than HEFT?
Equivalently, when is HEFT the only appropriate EFT for describing the effects of BSM physics at low energies?
The self-evident answer -- that HEFT is the unique choice when the interactions among the low-energy degrees of freedom only respect  $U(1)_\text{em}$~\cite{Burgess:1999ha} -- is too general to shed much light on the relevant properties of the microscopic theory.
In particular, it does not illuminate how UV theories respecting $SU(2)_L \times U(1)_Y$ give rise to an EFT respecting only $U(1)_\text{em}$.
A more UV-centric answer to this question would both delineate the space of SM extensions while also guiding the interpretation of data, especially in the event that experimental discrepancies with SM predictions are discovered.

Historically, concrete motivation for HEFT has drawn on a connection to non-decoupling BSM physics, especially that which could result from new strong dynamics.
The earliest work on HEFT~\cite{Feruglio:1992wf} emphasized this aspect of the EFT, and explained that power counting follows in close analogy with chiral perturbation theory.
Thus, HEFT is clearly appropriate when new non-decoupling strong dynamics spontaneously break electroweak symmetry, giving rise to a Higgs-like scalar~\cite{Bagger:1993zf, Koulovassilopoulos:1993pw, Grinstein:2007iv, Espriu:2013fia, Brivio:2013pma}.
But as emphasized in \Refs{Alonso:2012px, Buchalla:2013rka, Brivio:2013pma}, whether BSM strong dynamics favors HEFT over SMEFT is ultimately a question of decoupling.
Obviously, this implies that HEFT is required if the SM is not recovered as all BSM masses are taken to infinity.
A more subtle case occurs if a SMEFT can in principle be written down; then HEFT can be used to effectively resum a series expansion in terms of $g\s v/\Lambda$ for some coupling $g$, which is relevant if the new physics scale is close to the weak scale.

These previous studies made significant progress towards understanding when HEFT might be necessary.
However, they largely take a bottom up point of view (though some scenarios have been studied from the top down, \eg~\cite{Buchalla:2016bse}), and do not systematically characterize the underlying microscopic physics. They also generally leave unresolved the ambiguities raised by field redefinitions \cite{Criado:2018sdb}, which can appear to blur distinctions between EFTs formulated in different bases.
One of our main goals of this paper is to develop robust criteria that can be used to determine when it is possible to write down SMEFT, which we will then utilize to explore the implications for perturbative UV models.

In recent years, significant progress has been made towards identifying general criteria for when HEFT is required.
A seminal step in this direction was made by Alonso, Jenkins, and Manohar (AJM) in \Refs{Alonso:2015fsp, Alonso:2016oah}, in which they introduce a natural geometric interpretation of HEFT by treating the Higgs and Goldstone bosons as the coordinates on a Riemannian manifold.
They argue that HEFT is required if no $O(4)$ fixed (invariant) point exists on the moduli space.\footnote{This point, which maps to itself under the action of all $O(4)$ group elements, corresponds to the point at which electroweak symmetry is restored. The appearance of the $O(4)$ symmetry group here is due to the assumption of custodial invariance in the UV, which is assumed in this paper as well.  For the fully general HEFT, this condition would be to identify an $SU(2)_L\times U(1)_Y$ fixed point, see Sec.~6 of AJM~\cite{Alonso:2016oah} for the explicit construction. The corresponding geometric interpretation of SMEFT was recently formulated in \cite{Helset:2020yio}. 
}
This provides a natural criterion at the level of the two-derivative part of the action, but leaves open the possibility that the $O(4)$ symmetry is violated in the zero-derivative sector, \ie, the potential.
A complementary approach was introduced by Falkowski and Rattazzi (FR) in \cite{Falkowski:2019tft}, where it was argued that HEFT is required when the scalar potential as expressed in terms of the electroweak doublet $H$ is non-analytic at $H = 0$.\footnote{More precisely, HEFT is required when the non-analyticities cannot be removed by a field redefinition. This argument is anticipated, to a certain extent, in \cite{Feruglio:1992wf}, \emph{c.f.}~footnote 9.}
As~FR~\cite{Falkowski:2019tft} emphasizes, such non-analyticity is a hallmark of integrating out new states that acquire all of their mass from electroweak symmetry breaking, thereby violating decoupling.
Both approaches highlight the irrelevance of the linear versus non-linear parametrization of the Higgs field itself in distinguishing HEFT from SMEFT, insofar as the two are related by field redefinitions; the distinction ultimately depends on the properties of the Lagrangian in a given parametrization.

In this work, we develop a unified set of geometric criteria for distinguishing HEFT from SMEFT based on the properties of the scalar sector up to two derivatives, extending the framework of AJM~\cite{Alonso:2015fsp, Alonso:2016oah} and unifying it with the perspective of FR~\cite{Falkowski:2019tft}.
In extending the geometric approach of AJM, we introduce generalized invariants on the scalar manifold whose finiteness yields criteria that are sensitive to the structure of the scalar potential and robust against field redefinitions.\footnote{We explore the detailed connection between the geometric approach in AJM~\cite{Alonso:2015fsp, Alonso:2016oah} and the unitarity arguments in FR~\cite{Falkowski:2019tft} in a companion paper \cite{amplitudes}.}

The existence of an $O(4)$ fixed point on the EFT manifold and the smoothness of functions in its vicinity determine whether a SMEFT expansion is admissible around $v = 0$, but are not sufficient to determine whether a SMEFT expansion captures physics around the observed vacuum where $v \neq 0$. To this end, we extend the geometric approach (for the case of BSM scalar sectors at tree level) by identifying the submanifold of the UV scalar manifold on which the EFT resides.
This allows us to determine if the $O(4)$ invariant fixed point where electroweak symmetry is restored connects smoothly to the observed vacuum.
If such a submanifold exists and satisfies certain conditions, then we will argue that such a model can be matched onto a physically relevant realization of SMEFT, in the sense that the SMEFT expansion converges in the observed vacuum.
What is surprising about this picture is that it is sensitive to the properties of the entire submanifold, as opposed to only requiring properties near one of the special points. Although our submanifold analysis is formulated strictly for extended scalar sectors at tree level, it illustrates a number of qualitative features that hold more generally.

Our geometric criteria are illuminated by a number of concrete examples, which highlight the relevance of HEFT even when the microscopic physics is weakly coupled. We emphasize that these examples are primarily for illustrative purposes; some examples are likely ruled out by current Higgs data, while others remain viable. The phenomenological relevance of these examples will be explored in further detail in a pair of companion papers \cite{twohdm, higglons}.
The matching calculations in these examples are facilitated by a new formalism enabling the computation of the two-derivative contribution to the effective Lagrangian to all orders in the fields.
Insights garnered from these examples motivate us to take a further step beyond our unified geometric picture, conjecturally classifying microscopic theories that lead uniquely to HEFT.
Our conjectured physical interpretation of HEFT is that it uniquely describes the low energy physics when there is a source of electroweak symmetry breaking that persists when the Higgs vacuum expectation value (vev) is zero, or when a state whose mass is fully determined by the Higgs vev has been integrated out, building on observations in FR~\cite{Falkowski:2019tft}.

Although the majority of this paper is devoted to distinctions between HEFT and SMEFT in principle, we also explore distinctions between HEFT and SMEFT in practice, highlighting aspects of the rate of convergence of the two EFTs.
While the formal criterion for when HEFT is required can be considered as a condition near the point in field space where the Higgs vev vanishes, considerations of convergence imply that HEFT has broader applications even when the SMEFT expansion exists.

This paper is organized as follows. We begin in \cref{sec:Define}, by reviewing concrete parametrization of SMEFT and HEFT.  The rest of the paper is then devoted to more formal aspects of the distinction between SMEFT and HEFT. To the extent that the non-analyticity of the potential expressed in terms of the electroweak doublet $H$ is a possible hallmark of HEFT, in \cref{sec:PolarCoords} we discuss the interpretation of non-analyticities present in a given EFT parametrization, and investigate their properties under field redefinitions. Distinguishing between physical non-analyticities (which manifest invariant properties of the EFT) and unphysical non-analyticities (which do not) motivates considering geometric quantities related to the scalar manifold of the EFT. We introduce generalized curvature invariants built upon the scalar curvature $R$ of the EFT manifold, and analogous invariants built upon the scalar potential $V$. 

At this point, we apply the geometric description to establish criteria for the existence and convergence of a SMEFT parametrization. We begin in \cref{sec:WhatIsHEFT} with the existence of an $O(4)$ invariant point and the smoothness of functions defined in its vicinity, reviewing and extending the AJM curvature criteria for SMEFT. %
In order to understand the physical scenarios that violate the SMEFT criteria (requiring HEFT), in ~\cref{sec:eftsubmani} we develop the notion of the EFT submanifold for extended scalar sectors at tree level. This sheds light on both the types of microscopic theories that require HEFT, and also the circumstances under which a SMEFT description formulated around the $O(4)$ invariant point extends all the way to the observed vacuum. Ultimately both conditions -- the existence of a SMEFT expansion around the invariant point and its convergence at the observed vacuum -- must be satisfied in order for SMEFT to describe observations, and the violation of either requires HEFT. A variety of examples showing how to apply the geometric criteria when integrating out new heavy particles are presented in \cref{sec:Examples} and \cref{sec:BSMSymBreak}; the former section concerns particles acquiring all of their mass from electroweak symmetry breaking, while the latter concerns additional sources of symmetry breaking. We then explore practical aspects of the distinction between SMEFT and HEFT in \cref{sec:Truncate}, emphasizing the role of the radius of convergence in the two EFTs when integrating out states with mass near the weak scale. This highlights a final consideration for the validity of a SMEFT expansion, involving properties of the EFT submanifold away from either the invariant point or the observed vacuum.

After concluding in \cref{sec:Conclusions}, we provide a number of technical details in a series of appendices. These include details of the geometric formulation of HEFT in \cref{appsec:ScalarCurveature}; a proof in \cref{appsec:CurvatureCriterionK1} of the criteria presented in \cref{sec:WhatIsHEFT}; a technical discussion of singularities arising from integrating out massless states in \cref{appsec:SubmanifoldSingularities}; a new formalism for computing the two-derivative contribution to the effective Lagrangian to all orders in the fields using functional methods in \cref{appsec:TwoDerUniversal}; and caveats to our criteria stemming from truncating the EFT Lagrangian at two-derivative order in \cref{sec:swampland}.

\subsection{Guide for the Reader}
Since this is a rather lengthy paper interweaving qualitative points with technical details, we have provided the reader with a few different suggestions for where to focus their attention.
As we have already emphasized in the Introduction, the conclusions of this work are easy to state but difficult to prove.
As you will see in what follows, our goal is to convince you that HEFT is \emph{required} if one integrates out a BSM state that gets all of its mass from electroweak symmetry breaking or if there are BSM sources of electroweak symmetry breaking.
Furthermore, in practice one should use HEFT to describe the EFT that results from integrating out a state whose mass is near the electroweak scale.
Since this is an intuitive conclusion, the reader may wish to focus on only a subset of the results germane to their interests.
Note that the purpose of~\cref{sec:Define} is to set up conventions and notation, while~\cref{sec:PolarCoords} reviews textbook theory of analytic manifolds, and so one can explore these sections with a depth commensurate with their level of background.
We have four suggested paths through the paper. 
\begin{enumerate}
\item The most obvious option is for the ambitious reader to proceed from top to bottom.
In this case, we suggest keeping in mind the following big picture sketch of the flow.  
First, we study what could possibly go wrong at the putative $O(4)$ fixed point~\cref{sec:WhatIsHEFT}.
We then determine whether a SMEFT description at the fixed point can be extended to make predictions in our physical vacuum using the notion of the EFT submanifold in~\cref{sec:eftsubmani}.
After seeing these ideas applied to a number of concrete examples in~\cref{sec:Examples,sec:BSMSymBreak}, we finally discuss a practical point in \cref{sec:Truncate} regarding the rate of convergence of the SMEFT expansion, where features of the EFT manifold away from either the fixed point or the observed vacuum may favor HEFT.

\item For the reader who is interested in learning more about the analytic structure of EFTs, we recommend first reviewing the physics in~\cref{sec:PolarCoords}.  
Then understanding the curvature criterion presented in~\cref{sec:WhatIsHEFT} provides the foundation to understand the application to the examples in~\cref{sec:Examples}.
We note that these examples provide an additional benefit by exploring some subtleties that one can encounter when matching at tree or one-loop level.
They also rely on a new general formalism for matching using functional methods that is derived in \cref{appsec:TwoDerUniversal}.
\item For the reader who is interested in EFTs of extended scalar sectors, studying the EFT submanifold discussion in~\cref{sec:eftsubmani} will illuminate concrete examples of extended scalar sectors in~\cref{sec:Examples,sec:BSMSymBreak}.
In particular, this path is relevant to those who are interested in the structure of electroweak singlet, doublet, or triplet extensions of the SM.
\item For the experimentally focused reader, we suggest going from \cref{sec:Define} to the phenomenologically viable examples of microscopic theories that give HEFT in~\cref{sec:Examples}, and thereafter to the discussion of EFT convergence in \cref{sec:Truncate}.  
These are likely the most important sections for those who are thinking about the relevance of HEFT to searches at colliders.
\end{enumerate}
Buon viaggio!

\section{Defining SMEFT and HEFT}
\label{sec:Define}
We begin our comparison of SMEFT and HEFT by defining the two EFTs at a level appropriate for our analysis.
For simplicity, we assume the UV description respects an $O(4)$ custodial symmetry, which is of course violated by the SM itself due to the non-zero hypercharge gauge coupling and the mass splitting between fermions from the same doublets.
Relaxing the assumption of custodial symmetry would introduce a larger class of curvature invariants, expanding the analysis relative to the custodially symmetric case pursued here, but the approach would be analogous.
Note that we are typically interested in the impact of integrating out UV physics on the Higgs interactions -- we will not include the impact on operators involving the gauge bosons or SM matter fermions in what follows. Our classification may be extended to include interactions involving fermions and gauge bosons in terms of a suitably generalized field space supermanifold \cite{Finn:2020nvn}.

\clearpage
\subsection*{SMEFT}
We begin by defining SMEFT at the level of the bosonic fields.
The Higgs bosons and Goldstones in a custodially symmetric SMEFT can be conveniently parameterized by a field $\vec\phi$ that is a vector under the global $O(4)$ symmetry:
\begin{equation}
\vec \phi  = \mqty( \phi_1 \\ \phi_2 \\ \phi_3 \\ \phi_4 )  \qquad\text{with} \qquad \vec\phi \to O\s \vec\phi \,,
\label{eqn:CartesianCoor}
\end{equation}
where $O$ is a $4\times 4$ orthogonal transformation.
When parameterizing the theory in terms of a complex Higgs field that transforms as a doublet under $SU(2)_L$, we repackage the $\vec \phi$ into
\begin{equation}
H = \frac{1}{\sqrt{2}}\s \mqty( \phi_1 + i\s\phi_2 \\ \phi_4 + i\s\phi_3 ) \,.
\end{equation}

Subject to the assumption of custodial symmetry, the terms in the Lagrangian that only depend on this scalar take the general form\footnote{Note that we omit the other possible $SU(2)$ contraction involving two Higgses and a spacetime derivative because it violates the assumed custodial symmetry.   
}
\begin{equation}
\Lag_\text{SMEFT} = A\big( |H|^2 \big)\, | \partial H|^2 + \frac{1}{2}\s B\big(  |H|^2 \big)\,{\Big[ \partial \big(|H|^2\big) \Big]^2} - \tilde{V}\big( |H|^2 \big) + \mathcal{O}\big(\partial^4\big) \,,
\label{eqn:LSMEFT}
\end{equation}
where $A$, $B$, and $\tilde{V}$ are functions of $|H|^2$ that are real-analytic at the origin $|H|=0$.
The critical assumption of analyticity means that $A$, $B$ and $\tilde V$ (as well as similar functions multiplying higher derivative terms in the Lagrangian) can be expressed as convergent Taylor series about $|H|^2=0$, see~\cref{sec:PolarCoords} for a more detailed discussion. 
In a basis where the kinetic term is canonically normalized, $A(0) = 1$.

When investigating SMEFT from a geometric point of view, we interpret the scalar fields $\phi_1, \cdots, \phi_4$ as a Cartesian-like coordinate system on the scalar manifold.
The origin $\vec\phi=0$ is the fixed point of the $O(4)$ transformation, where electroweak symmetry is restored.

\clearpage

\subsection*{HEFT}
Next, we turn to the construction of HEFT.
The key assumption of HEFT is that electroweak symmetry is non-linearly realized, and so the most general Lagrangian can be constructed using the classic CCWZ prescription~\cite{Coleman:1969sm, Callan:1969sn}.
Assuming custodial symmetry, the Goldstone fields $\vec \pi$ chart the coset space $O(4)/O(3)$. 
No relationship is assumed between the real CP-even Higgs scalar $h$ and the Goldstone fields. 
A convenient parametrization of the Higgs scalar and Goldstones is
\begin{equation}
\hspace{60pt} h \qquad \text{and} \qquad
\vec{n} = \mqty( n_1=\pi_1/v \\[3pt] n_2=\pi_2/v \\[3pt] n_3=\pi_3/v \\[8pt] n_4=\sqrt{1-n_1^2-n_2^2-n_3^2} ) \,.
\label{eqn:PolarCoor}
\end{equation}
Here $\vec n\left(\vec\pi\right)\in S^3$ is a four-component unit vector subject to the constraint $\vec n \cdot \vec n = 1$. 
Our notation is the same as in Eq.~(2.14) in~AJM~\cite{Alonso:2016oah}, where various possible parametrization are discussed in detail.
These states have well-defined $O(4)$ transformations:
\begin{equation}
h \,\to\, h  \qquad\text{and} \qquad \vec n \,\to\, O\s\vec n \,,
\end{equation}
where as in the previous section, $O$ is a $4\times 4$ orthogonal matrix.
The constrained multiplet $\vec n$ transforms linearly under the $O(4)$ symmetry, while the physical degrees of freedom $\vec \pi$ furnish a non-linear realization.

In terms of these coordinates, the scalar part of the custodially symmetric HEFT Lagrangian is
\begin{equation}
\Lag_\text{HEFT} = \frac{1}{2}\s \big[ K( h ) \big]^2\s ( \partial h )^2 + \frac{1}{2}\s \big[ v\s F( h ) \big]^2\s ( \partial \vec n )^2 - V(h) + {\cal O}\big( \partial^4 \big) \,.
\label{eqn:LHEFT}
\end{equation}
where $K$, $F$, and $V$ are functions of $h$ that are real-analytic about the physical vacuum where $h=0$. 
However, they are not necessarily real-analytic outside the neighborhood of $h=0$.
Although $K(h)$ is formally redundant and could be removed by a field redefinition, it is typically generated in matching to perturbative UV completions, and so we retain it here for convenience.
$K(0)= 1$ when the field $h$ is canonically normalized.
We have used the freedom to shift $h$ by a constant to set $h=0$ at the minimum of the potential $V(h)$ that corresponds to the physical vacuum.
We have chosen $v$ to normalize $F(h)$ such that $F(0)=1$.

\subsection{Mapping Between SMEFT and HEFT}
\label{sec:MapBetween}
While our goal is to understand the circumstances under which a UV theory must be matched onto HEFT, any SMEFT can of course be rewritten as a HEFT.
Therefore, it is useful to briefly discuss how to map between the two ways of formulating the EFTs, assuming there is no obstruction to expressing the theory as SMEFT.
Using the notation introduced above, one can identify
\begin{align}
H = \frac{1}{\sqrt{2}}\s \mqty( \phi_1 + i\s\phi_2 \\ \phi_3 + i\s\phi_4 )
\qquad\Longleftrightarrow\qquad
\vec\phi = \mqty(\, \phi_1\, \\ \,\phi_2\, \\ \,\phi_3\, \\ \,\phi_4\, ) = (v_0+h)\s \vec{n} \,,
\label{eqn:Doublethn}
\end{align}
where $v_0 = - h_*$ is determined by the condition $F(h_*)=0$, which can be satisfied when there is an $O(4)$ invariant point on the scalar manifold such that the EFT can be expressed as SMEFT; see~AJM~\cite{Alonso:2016oah} for the derivation and discussion of this point. 
Note that $v_0$ need not be the same as the parameter $v$ appearing in \cref{eqn:PolarCoor}, the relation between $\vec n$ and $\vec \pi$; note that $v$ determines the gauge boson masses. 
Assuming both SMEFT and HEFT are valid, we use $H$ when working with SMEFT, and $\big(h, \vec{n}\s\big)$ when working with HEFT.

At zero-derivative order, there is only one $O(4)$-symmetric building block:
\begin{equation}
|H|^2 = \frac{1}{2}\s \vec\phi \cdot \vec\phi\ = \frac{1}{2}\s ( v_0 + h )^2 \,.
\label{eqn:HTohnPotential}
\end{equation}
There are two additional $O(4)$-symmetric building blocks at two-derivative order:\footnote{Throughout this paper, we use partials instead of covariant derivatives since we are focused on the scalar sector alone, which only manifests the global part of the $O(4)$ symmetry.}
\begin{subequations}
\begin{align}
|\partial H|^{2} & = \frac{1}{2}\s\big(\s\partial \vec \phi \,\big)^2 = \frac{1}{2}\s(\partial h)^2 + \frac{1}{2}\s(v_0 + h)^2\s\big(\partial \vec n\big)^2 \,, \label{eqn:HTohnTwoDerivativeA} \\[5pt]
\big(\s\partial |H|^2\s\big)^2 &= \big(\vec \phi  \cdot \partial \vec \phi\, \big)^2 = (v_0+h)^2\s (\partial h)^2 \,.  \label{eqn:HTohnTwoDerivativeB}
\end{align}
\label{eqn:HTohnTwoDerivative}%
\end{subequations}
The natural generalization to higher derivatives is straightforward.

Assuming the SMEFT parametrization exists, the relationship between HEFT and SMEFT is clear from \cref{eqn:HTohnPotential,eqn:HTohnTwoDerivative}.
Comparing \cref{eqn:HTohnTwoDerivativeA} against \cref{eqn:HTohnTwoDerivativeB}, it is interesting to note that not all operators with derivatives depend on the Goldstone boson vector $\vec{n}$; extracting a factor of $\vec{n}$ requires acting with a derivative on a factor of $H$ that has open gauge indices.

One can always use the relations in \cref{eqn:HTohnPotential,eqn:HTohnTwoDerivative} to write a given SMEFT Lagrangian into a HEFT Lagrangian. 
Therefore, a SMEFT is always a HEFT as expressed in \cref{eq:HEFTsupsetSMEFT}. 
On the other hand, a HEFT Lagrangian is not always a SMEFT. It is true that one can always rewrite a given HEFT Lagrangian into a SMEFT-like doublet form, by using the mapping in the opposite direction:
\begin{align}
\hspace{-6pt}\Lag_\text{HEFT} &= \frac{1}{2}\s \big[ K(h) \big]^2\s (\partial h)^2 + \frac{1}{2}\s\big[v\s F(h) \big]^2\s (\partial \vec n )^2 - V(h) + {\cal O}\big(\partial^4 \big) \notag\\[9pt]
 &= \frac{v^2\s F^2}{2\s |H|^2}\s |\partial H|^2 + \frac{1}{2}\s \big(\partial |H|^2 \big)^2\s \frac{1}{2\s|H|^2}\bigg(K^2 - \frac{v^2\s F^2}{2\s|H|^2} \bigg)  - \tilde{V}\big(|H|^2\big) + {\cal O}\big(\partial^4 \big) \,.
\label{eqn:SMEFTfromHEFT}
\end{align}
However, we can see that this Lagrangian is generically non-analytic at the origin $|H|^2=0$ and hence is not SMEFT, as emphasized in~AJM~\cite{Alonso:2016oah} and~FR~\cite{Falkowski:2019tft}. 
More generally, a given HEFT Lagrangian is a SMEFT only when the functions $K(h), F(h), V(h)$ satisfy certain conditions. 
In this case, there are either no non-analyticities, or more specifically only ``unphysical non-analyticities'' are allowed (to be defined in a geometric sense in \cref{subsec:NonAnalyticities} below) since these can be removed by a field redefinition (as mentioned in~AJM~\cite{Alonso:2016oah}). 
In this paper, we prove concrete conditions on $K(h), F(h), V(h)$ that imply a SMEFT description exists. 
Along the way, we also explore the implications of physical non-analyticities from the UV point of view.

Given these two coordinate systems, we now turn to defining a geometric description in terms of a manifold in field space, which will allow us to explore the interplay of analyticity and field redefinitions.

\section{Geometry, Analyticity, and Field Redefinitions}
\label{sec:PolarCoords}
As explained in \cref{sec:Define}, one can always try to write a HEFT Lagrangian into a SMEFT-like doublet form, but the resultant Lagrangian may exhibit non-analyticities. 
The interpretation is complicated by the fact that some non-analyticities can be removed by field redefinitions, implying that they are not physical. 
The issue can be made clear by carefully studying the analytic structure of functions as expressed in different coordinate systems.
Specifically, we will discuss the implications for analyticity when one attempts to map between polar-like (HEFT) and Cartesian-like (SMEFT) coordinates.
For a concrete example that demonstrates some of the pitfalls one can encounter when performing field redefinitions of $h$ as opposed to $H$, see~\cref{subsubsec:Loophole}

Our main goal in this section is to review the mathematical framework for using charts to define analytic manifolds.
This will set the stage for our later explorations of moving between HEFT and SMEFT descriptions of an EFT.
In particular, we will argue that some non-analyticities can be ``unphysical'' artifacts of having made a particular field basis choice.
This will then motivate our introduction of curvature invariants, since they are unchanged by field redefinitions. 
We will conclude this section by defining various generalized curvature invariants whose (in)finiteness will play a central role in distinguishing SMEFT from HEFT.

\clearpage

\subsection{Analyticity of Field Manifolds and Functions}
Consider a generic EFT Lagrangian built out of $n$ real scalar fields $\phi^i$, $i=1, 2, \ldots, n$,
\begin{align}
\Lag = \frac{1}{2}\s g_{ij}\s\big(\vec\phi\,\big)\s \partial_\mu \phi^i\s \partial^\mu \phi^j - V\big(\vec\phi\,\big) \,. 
\label{eqn:LagAnalyticity}
\end{align}
The fields $\phi^i: U \to \rr^n$ can be viewed as a coordinate chart on the field manifold $M$, \ie, an invertible map from an open subset $U \subset M$ of the manifold of field configurations to (an open subset of) $\rr^n$. 
Of course, some arbitrary choices must be made when defining these coordinates that will have no impact on the physics. 
This is the freedom to make field redefinitions. 
For example, identifying $\vec\phi = \vec{0}$ with the vacuum state, it can be shown (see \eg~\cite{Coleman:1969sm}) that the $S$-matrix of the theory is unaffected when making field redefinitions of the form
\begin{equation}
\phi^i = \varphi^j\s F^{ij}\big(\vec\varphi\,\big) \,, \label{eqn:FieldRedef}
\end{equation}
where $F^{ij}$ are some real analytic functions of the new fields $\varphi^i$, \ie, they have a convergent Taylor expansion, in the neighborhood of $\vec\varphi = \vec{0}$, with $F^{ij}\big(\vec{0}\,\big)=\delta^{ij}$. 
Therefore, performing a field redefinition can be viewed as defining a new coordinate chart on the field manifold. 
Restricting to real analytic field redefinitions of the form in \cref{eqn:FieldRedef} means that we require our field manifold to be a \textit{real analytic manifold}~\cite{Coleman:1969sm}: a manifold where any two charts have the property that they are invertible real analytic functions of each other in the region where they overlap.

A useful fact about real analytic manifolds is that the real analyticity property of functions at any given point is coordinate chart independent. 
This can be understood intuitively using the fact that different charts are invertible real analytic functions of one another. 
Therefore, if a function is real analytic at a point in one chart, it must also be real analytic when expressed in terms of any other valid chart, due to the analyticity of analytic function composition~\cite[Prop. 2.2.8]{krantz2002primer}. 
Conversely, if a function is non-analytic at a certain point in a given chart, it remains non-analytic at that point when using any other valid chart on the manifold.

Here, we are concerned with the analyticity of the metric and potential at particular points on the field space manifold. Being able to find a convergent Taylor expansion of the metric and potential at particular points (the $O(4)$ invariant fixed point and our observed vacuum respectively) amounts to being able to write down a local operator expansion for the Lagrangian in terms of SMEFT and HEFT fields respectively.

It bears emphasizing that the field redefinitions defined by \cref{eqn:FieldRedef} are only a subset of the possible field redefinitions that leave the $S$-matrix invariant. 
Most notably, this omits field redefinitions that involve derivatives. 
Derivative field redefinitions are generally admissible, and as we observe in \cref{sec:swampland}, can even be induced in an EFT when performing non-derivative field redefinitions in the UV description. 
In what follows, we will restrict ourselves to non-derivative field redefinitions, which is consistent with our focus on the geometric quantities that are defined by the two-derivative action. 
To the extent that derivative field redefinitions induce higher-derivative terms, capturing their effects geometrically requires a corresponding generalization of curvature invariants beyond two-derivative order (and the scope of this work).

Our restricted scope leaves open the possibility of encountering non-analyticities in an EFT that can be removed exclusively using derivative field redefinitions. 
However, as a practical matter, when matching between the full UV theory and an EFT it is possible to choose a basis in the UV theory such that unphysical non-analyticities can always be removed by zero-derivative field redefinitions.

\subsection{Polar Coordinates Obscures Analyticity of the Origin}
\label{subsec:NonAnalyticities}

We wish to understand the analyticity properties of the metric and potential on the field space manifold. 
On a real analytic manifold, the arbiter of their analyticity would be any valid chart; we would test to see if we can Taylor expand the metric and potential in said coordinates. 
The problem with a Lagrangian written exclusively in HEFT coordinates is that this chart alone does not cover the whole manifold. 
Like any polar coordinate system, which does not include the origin, the HEFT chart cannot be used to investigate physics about a putative $O(4)$ fixed point. 
In other words, the analyticity of the HEFT Lagrangian at the fixed point is not manifest.

Therefore, our goal is to build a real analytic manifold that can possibly include the origin, \ie, we want to build an atlas. 
Each new chart must overlap with an existing one, and their respective coordinates must be real analytic invertible functions of each other in the region of overlap. 
However, we are presented with mutually incompatible choices for how to do this, which may yield different conclusions regarding the analyticity of the Lagrangian at the fixed point.

As an analogy, consider the manifold $\rr^2$. 
As a proxy for the fields of HEFT, we define a polar chart P on the manifold that maps all points \emph{except the origin} onto the usual $(r,\theta)$ coordinates. 
As a proxy for the Lagrangian of HEFT, we define a Riemannian metric by the line element
\begin{equation}
  \dd s^2 = \dd r^2 + r^2\s \dd \theta^2 \, .
  \label{eq:goodmetric}
\end{equation}
Next, consider two new Cartesian-like charts, C1 and C2, which map all points respectively to $(x_1,y_1)$ and $(x_2,y_2)$. 
\emph{Away} from the origin, these new coordinates are both invertible and analytic in terms of the old polar coordinates:
\begin{subequations}
\begin{align}
x_1 &= r \cos\theta  \\[5pt]
y_1 &= r \sin\theta \,,
\end{align}
\label{eq:CartToPolar}%
\end{subequations}
and
\begin{subequations}
\begin{align}
x_2 &= \left(r+r^2\right) \cos\theta  \\[8pt]
y_2 &= \left(r+r^2\right) \sin\theta \,.
\end{align}
\end{subequations}
In this sense, each chart C1 and C2 is individually compatible with P. However, expressing the two sets of Cartesian coordinates in terms of each other,
\begin{subequations}
\begin{align}
x_2 &= x_1 \left(1+\sqrt{x_1^2+y_1^2}\right)  \label{eqn:x2x1} \\[8pt]
y_2 &= y_1 \left(1+\sqrt{x_1^2+y_1^2}\right) \,, \label{eqn:y2y1}
\end{align}
\end{subequations}
shows that the new coordinates are not real analytic functions of each other \emph{at} the origin. 
This implies that the two charts C1 and C2 are mutually incompatible, and they define distinct real analytic manifolds.

As the coordinates of C1 and C2 are not analytic functions of each other at the origin, they will not necessarily agree when investigating the analyticity of a third set of functions about the origin, such as the components of a metric. 
Explicitly
\begin{align}
  \dd s^2 &= \dd r^2 + r^2\s \dd \theta^2 \nonumber\\[5pt]
  &= \dd x_1^2 + \dd y_1^2  \nonumber  \\[7pt]
                                       &=\frac{1}{x_2^2+ y_2^2} \left[  \frac{\big( x_2\s \dd x_2 + y_2\s \dd y_2 \big)^2}{1+4\s \sqrt{x_2^2 + y_2^2}} + \frac{\big( x_2\s \dd y_2 - y_2\s \dd x_2 \big)^2}{\bigg( 1 + \sqrt{1+4\s \sqrt{x_2^2 + y_2^2}}\s \bigg)^2}   \right] \, .
\end{align}
Clearly, the metric components are analytic at the origin in C1, but not in C2. 
The existence of C1 shows that we can incorporate the metric \cref{eq:goodmetric} into a real analytic manifold that includes the origin, and for our purposes we can interpret the mapping \cref{eq:CartToPolar} as the field redefinition $h \to H$ that allows us to write an amenable HEFT as SMEFT. 
By contrast, the apparent non-analyticities of the metric in chart C2 are spurious and unphysical such that they can be ``defined away,'' and provide an analog of making a poor choice when searching for SMEFT-like coordinates.

For our purposes here, it is critical that we can distinguish this situation from having a metric with a ``physical'' non-analyticity at the origin.
In other words, situations arise where there can be no chart that simultaneously: (a) includes the origin, (b) is compatible with the existing polar chart, and (c) has a manifestly analytic metric. 
This is the analog of a HEFT Lagrangian that cannot be SMEFT.

For example, taking the same polar chart P, consider the more general metric,
\begin{equation}
  \dd s^2 = \dd r^2 + T(r) \, \dd \theta^2 \, .
\end{equation}
for some real analytic function $T$ of $r$.
We can find a physical singularity in the metric using tools that are familiar from General Relativity. 
Suppose there exists a chart including the origin with a corresponding metric that is analytic. 
The Ricci scalar computed in this chart, comprising the first and second derivatives of the metric, will also be analytic, and therefore finite and smoothly varying, at the origin. 
However, as a coordinate invariant quantity, it must agree with the Ricci scalar computed in the polar chart in the neighborhood of the origin,
\begin{equation}
  R(r) =  \frac{\left( T^\prime \right)^2}{2\s T^2} - \frac{T^{\prime \prime}}{T} \, .
\end{equation}
Such agreement would be impossible if $R(r)$ diverged in the polar chart approaching the origin, as happens, for example, when $T(r)=r$ implying that $R(r) = 1/(2\s r^2)$. 
We can contrast this against the metric \cref{eq:goodmetric}, for which $R(r)=0$ everywhere, which is sufficient to guarantee that there exists a way to incorporate the origin through the introduction of an appropriate Cartesian chart, namely C1.

\subsection{Curvature Invariants for HEFT}

In the case of writing a HEFT Lagrangian as a SMEFT, we want to find a field redefinition that renders both the metric and potential real analytic in valid coordinates at the fixed point.
If the metric is real analytic, then the curvature invariants built from the Ricci scalar curvature $R$
\begin{equation}
\nabla^{2n}\s R = \nabla^{\mu_1}\nabla_{\mu_1} \cdots \nabla^{\mu_n}\nabla_{\mu_n}\s R \,, \label{eqn:RInvariants}
\end{equation}
 will be finite for all $n\in\mathbb{N}$.\footnote{We take the natural numbers $\mathbb{N}=\{0,1,2,\ldots\}$ to include $0$.} Here $\nabla_\mu$ is the covariant derivative operator derived as usual using the metric connection. 
Similarly, if an arbitrary scalar function $V$ is real analytic, then the invariants\footnote{
Note that there is a potential ambiguity in the meaning of the analyticity of a function $V(r)$ at the point $r=0$. 
There are two scenarios, either we are talking about a single-argument function $V(r)$, with $r\in\left(-\infty, +\infty\right)$, charting a one dimensional manifold, or $V(r)$ is defined on a higher dimensional manifold, with $r\in\left(0, +\infty\right)$ being the radial direction in a polar coordinate system. 
The real analyticity of $V(r)$ at $r=0$ can differ in these two scenarios. 
For example, $V(r)=r$ is clearly real analytic at the point $r=0$ in the first scenario. 
On the other hand, if it is considered as a function on a two dimensional flat manifold with the usual polar coordinate chart $\left(r, \theta\right)$ and metric $\dd s^2=\dd r^2+r^2\dd \theta^2$, then it is not analytic at $r=0$. 
One can for example check that the curvature invariant $\nabla^2 V = \left( \frac{\dd^2}{\dd r^2}+\frac{1}{r}\frac{\dd}{\dd r} \right) V = \frac{1}{r}$ diverges at the origin in this case.
}
\begin{equation}
\nabla^{2n}\s V = \nabla^{\mu_1}\nabla_{\mu_1} \cdots \nabla^{\mu_n}\nabla_{\mu_n}\s V \,, \label{eqn:VInvariants}
\end{equation}
will be finite for all $n\in\mathbb{N}$. In the following section, we will show that \cref{eqn:RInvariants,eqn:VInvariants} are a sufficient set: if these curvature invariants all have defined limits approaching the fixed point in the HEFT parametrization, we can find an (analytic) SMEFT Lagrangian. 
We will also find in many practical examples that most of the diagnostic power lies in the finiteness of the Ricci scalar curvature, $R$. Explicitly, given the HEFT Lagrangian defined in \cref{eqn:LHEFT},
\begin{equation}
R =  - \frac{{2\s{N_\varphi }}}{{{K^2}\s F}}\Bigg[ {\big( {\partial_h^2F} \big) - \big( {{\partial_h}K} \big)\bigg( {\frac{1}{K}{\partial_h}F} \bigg)} \Bigg] + \frac{{{N_\varphi }\s( {{N_\varphi } - 1} )}}{{v^2\s{F^2}}}\Bigg[ {1 - {\bigg( {\frac{{{v}}}{K}\s{\partial_h}F} \bigg)^2}} \Bigg] \,,
\label{eq:RmainText}
\end{equation}
where $N_\varphi = 3$ counts the number of Goldstone bosons. 
Various geometric quantities for HEFT, including the Riemann tensor, Ricci tensor, and Ricci scalar, are presented in \cref{appsec:ScalarCurveature}.

\section{Does SMEFT Exist at the Fixed Point?}
\label{sec:WhatIsHEFT}
%
In this section, we take the first step towards answering ``Is SMEFT Enough?''~by developing a concrete set of tests that allow us to determine if it is possible to define a SMEFT description of the EFT at the $O(4)$ invariant fixed point, \ie, where electroweak symmetry is unbroken.
This is a necessary condition as we work towards our ultimate goal, which is to determine when it is possible to use the SMEFT description to probe BSM physics in the physical vacuum, see \cref{sec:eftsubmani}.
As we have emphasized extensively so far, the answer to this question can be easily obscured by performing field redefinitions.
This motivates the ``Curvature Criteria'' in \cref{subsec:CurvatureCriterion}, which are expressed in terms of the geometric invariant quantities introduced in \cref{eqn:RInvariants,eqn:VInvariants}.
Although the behavior of the EFT at the fixed point is just the first step towards understanding when HEFT is required, as we will show below in \cref{sec:Examples}, there are many interesting examples that can already be elucidated by studying the EFT here.

We can summarize the logic followed in this section as
\begin{enumerate}
\item We express the most general set of SMEFT Higgs operators up to the two-derivative level in terms of HEFT degrees of freedom.  When written in this basis, the potential is an even convergent power series in $v_0+h$, and the form factors multiplying the $h$ ($\vec{n}$) kinetic terms will be an even (odd) convergent power series in $v_0+h$.
\item Next, we canonically normalize the $h$ kinetic term.  We prove that this eliminates our ability to perform subsequent field redefinitions, thereby fixing the basis. In this specific basis, the potential will still be an even convergent power series in $v_0+h$, and the form factor multiplying the $\vec{n}$ kinetic term will be an odd convergent power series.  This yields our ``basis dependent criteria.''
\item We then generalize this basis specific statement through the use of curvature invariants, since these are the natural objects for expressing physical statements that are independent of coordinate transformations.  This yields our ``curvature criteria.''
\end{enumerate}

At this point, we will have firmly established the properties a HEFT must satisfy such that it can be expressed as a SMEFT at the fixed point.\footnote{As we will emphasize in \cref{appsec:CurvatureCriterionK1}, our criteria only hold for $O(N)$ groups with $N > 2$, \ie, when the Higgs transforms as a non-trivial representation of a non-Abelian group.}
Unsurprisingly, the strict criteria require that an infinite number of curvature invariants must be evaluated, since we are making an all-orders statement about the properties of an EFT.
It is typically not possible to use this in practice, which motivates us to conjecture physically motivated ``LO Criteria'' that only require checking a finite number of conditions.
We conjecture that for HEFT Lagrangians that come from perturbative physical examples, its parameters are tied together in such a way that the following simplified criteria would work to distinguish HEFT from SMEFT at the fixed point.  
\vspace{3mm}
\begin{tcolorbox}[colback=light-gray]
\begin{center}
\begin{minipage}{5.5in}
\textbf{Leading Order (LO) Criteria:} A physical HEFT can be converted to a SMEFT at the fixed point if and only if the following three conditions hold:
\begin{enumerate}
  \item The function $F(h)$ has a zero at some real value of $h=h_*$. This $h_*$ is then a candidate for an $O(4)$ invariant point.
  \item The functions $K(h)$, $F(h)$, and $V(h)$ all have convergent single-argument Taylor expansions in $h$ at $h=h_*$.
  \item The scalar curvature $R(h)$ is finite at $h_*$.
\end{enumerate}
\end{minipage}
\end{center}
\end{tcolorbox}
\vspace{3mm}
This will set us up for an exploration of concrete UV scenarios in~\cref{sec:Examples}, where these leading order criteria are applied to expose that HEFT is required at the fixed point when one is integrating out a state that receives all of its mass from electroweak symmetry breaking.
For completeness, we discuss some limitations of this approach in~\cref{sec:swampland}.

Having set the stage, we now turn to the detailed derivation of the rigorous criteria, first in the context of a basis specific statement in \cref{subsec:BasisDependentCriterion}, followed by a basis independent generalization utilizing curvature invariants in \cref{subsec:CurvatureCriterion}.

\subsection{Basis Dependent Criteria}
\label{subsec:BasisDependentCriterion}

We begin with the fact that the scalar sector of SMEFT (up to two derivatives and imposing custodial symmetry) can be expressed as three functions of $|H|^2$: the zero-derivative potential $\tilde V$, and form factors $A$ and $B$ multiplying the two derivative terms, see \cref{eqn:LSMEFT}.
These can be written into the HEFT form using \cref{eqn:HTohnPotential,eqn:HTohnTwoDerivative}, which yields
\begin{align}
{{\cal L}_{{\text{SMEFT}}}} &= A\big(|H|^2\big)\, {| {\partial H}|^2} + \frac{1}{2}\s B\big(|H|^2\big)\, {\big( {\partial {{|H|}^2}} \big)^2} - \tilde{V}(|H|^2) + {\cal O}\big( {{\partial^4}} \big) \notag\\[5pt]
 &= \frac{1}{2}\s\big[ {A + ( v_0 + h )^2 \s B} \big]\s{( {\partial h} )^2} + \frac{1}{2}\s ( {v_0 + h} )^2\s A{( {\partial \vec n} )^2} - V(h) + {\cal O}\big( {{\partial^4}} \big) \,.
\end{align}
Comparing this with the generic HEFT form up to two derivatives given in \cref{eqn:LHEFT}, we identify
\begin{subequations}\label{eqn:Kv0FgenericSMEFT}
\begin{align}
K &= \sqrt {A + {( {v_0 + h} )^2}B} \\[5pt]
v\s F &= ( {v_0 + h} )\s\sqrt A \,.
\end{align}
\end{subequations}
As $A$, $B$, and $\tilde{V}$ in SMEFT are real analytic functions of $|H|^2=(v_0+h)^2/2$, with $A(|H|^2=0)=1$ and $\tilde{V}'\big(|H|^2=v_0^2/2\big)=0$ (such that the potential is minimized at the vev), we identify the following three features of the class of HEFT Lagrangians that can be rewritten as SMEFT (we will refer to these as the ``naive criteria'' in what follows):
\begin{enumerate}
  \item The function $F(h)$ has a zero at the real valued point $h=h_*\equiv -v_0$.
      Since the HEFT Lagrangian is expressed in terms of $h$, the potential $V(h)$ should already satisfy the minimization condition $V'(h=0)=0$.\footnote{If this is not the case, we would shift $h$ to redefine its origin, removing any vev. Throughout this paper, we take the condition $V'(h=0)=0$ as an implicit condition in all field bases.}
      Then clearly the potential cannot be used to determine $v_0=\langle H\rangle$.
      In HEFT, locating the zero of $F(h)$ determines $v_0$, \ie, $F(h_*) = 0$.
  \item At this special point $h=h_*$, the functions $K(h)$, $F(h)$, and $V(h)$ are real-analytic functions of $h$.
  \item When Taylor expanded about $h=h_*$, $K$ and $V$ have only even powers of $\left(h-h_*\right)=\left(v_0+h\right)$, and $F$ has only odd powers.
      Furthermore, the fact that $A\left(|H|^2=0\right)=1$ implies that their leading terms are $K\left(h_*\right)=v\s F'\left(h_*\right)=1$.
\end{enumerate}

If one expresses SMEFT in terms of the HEFT fields, the resultant HEFT Lagrangian satisfies the above three conditions.
Conversely, these three conditions are also \emph{sufficient} to guarantee that a HEFT Lagrangian can be written as an analytic SMEFT Lagrangian (about the fixed point).
To see this explicitly, we can use \cref{eqn:HTohnPotential,eqn:HTohnTwoDerivative} to derive~\cref{eqn:SMEFTfromHEFT} in which the quantities are manifestly analytic.
However, the three conditions are not \emph{necessary} to successfully write a HEFT as a SMEFT, as condition 3 is field redefinition dependent.

\subsubsection{Loophole Using Field Redefinitions}
\label{subsubsec:Loophole}
Unfortunately, the ability to perform field redefinitions implies that the above naive test fails to capture all possible HEFT Lagrangians that can be expressed as SMEFT. We give an explicit example here, which is reminiscent of the issues mapping between polar and Cartesian coordinates discussed in~\cref{subsec:NonAnalyticities} above.

Let
\begin{align}
\Lag &= \frac{1}{2}{\bigg( {1 + \frac{h}{{2\s v_0}}} \bigg)^2}{( {\partial h} )^2} + \frac{1}{2}{( {v_0 + h} )^2}{\bigg( {\frac{3}{4} + \frac{h}{{4\s v_0}}} \bigg)^2}{( {\partial \vec n} )^2} - V \notag\\[10pt]
 &= \frac{1}{4}\s \left( {1 + \frac{{\sqrt {2\s{{|H|}^2}} }}{v_0} + \frac{{{{|H|}^2}}}{{2\s{v_0^2}}}} \right){| {\partial H} |^2} + \frac{1}{{4\s{v_0^2}}}\left( {\frac{v_0}{{\sqrt {2\s{{|H|}^2}} }} + \frac{3}{4}} \right)\s\frac{1}{2}\s{\big( {\partial {{|H|}^2}} \big)^2} - \tilde{V} \,, \label{eqn:LRedefExample}
\end{align}
where the implicit potential $V = V(h)$ satisfies $V'(h=0)=0$ and we have chosen explicit forms for $K(h)$ and $F(h)$:
\begin{subequations}
\begin{align}
K &= 1 + \frac{h}{{2\s v_0}} = \frac{1}{2} + \frac{1}{{2\s v_0}}( {v_0 + h} )  \\[5pt]
{v}\s F &= ( {v_0 + h} )\bigg( {\frac{3}{4} + \frac{h}{{4\s v_0}}} \bigg) = ( {v_0 + h} )\bigg[ {\frac{1}{2} + \frac{1}{{4\s v_0}}( {v_0 + h} )} \bigg] \,.
\end{align}
\end{subequations}
These do not satisfy the naive criteria: conditions 1 and 2 are satisfied, but condition 3 is violated.
From the second line of \cref{eqn:LRedefExample}, we also see that a direct mapping via \cref{eqn:HTohnPotential} and~\cref{eqn:HTohnTwoDerivative} yields functions $A\left(|H|^2\right)$ and $B\left(|H|^2\right)$ that are not real analytic about $|H|^2=0$.
However, this can be fixed using a field redefinition.

Send $h \to h_1\equiv h+\frac{1}{4\s v_0}h^2$.\footnote{Note that this field redefinition preserves the minimization condition for the potential.}
This gives
\begin{subequations}
\begin{align}
\partial_\mu h_1 &= \bigg( 1 + \frac{h}{2\s v_0} \bigg)\s \partial_\mu h \\[8pt]
  ( v_1 + h_1 )^2 &= ( v_0 + h )^2 \bigg( \frac{3}{4} + \frac{h}{4\s v_0} \bigg)^2 \,,
\end{align}
\end{subequations}
noting that $v_1=3\s v_0/4$.
The HEFT Lagrangian in terms of the field $h_1$ is then
\begin{align}
\Lag &= \frac{1}{2}{\bigg( {1 + \frac{h}{{2\s v_0}}} \bigg)^2}{( {\partial h} )^2} + \frac{1}{2}{( {v_0 + h} )^2}{\bigg( {\frac{3}{4} + \frac{h}{{4\s v_0}}} \bigg)^2}{( {\partial \vec n} )^2} \notag\\[5pt]
 &= \frac{1}{2}\s{( {\partial {h_1}} )^2} + \frac{1}{2}\s{( {{v_1} + {h_1}} )^2}{( {\partial \vec n} )^2} = {| {\partial {H_1}} |^2} \,.
\end{align}
We see that after the field redefinition, the HEFT Lagrangian now satisfies the naive criteria. The direct mapping via \cref{eqn:HTohnPotential,eqn:HTohnTwoDerivative} simply yields the $H_1$ kinetic term, which is obviously a valid SMEFT.

\subsubsection{Canonicalizing the $\bm{h}$ Kinetic Term to Fix the Basis}
\label{subsubsec:FixBasis}
The above example makes clear how our naive criteria do not account for the equivalence relation among different HEFT Lagrangians resulting from field redefinitions.
To derive a necessary and sufficient set of criteria we must work in a specific fixed basis; our goal in this section is to argue that we can fully specify a field basis by requiring that the kinetic term for the field $h$ is canonical, \ie, that $K(h)=1$ and that there is no kinetic mixing $(\partial^\mu h)(\partial_\mu n^i)$.\footnote{We reemphasize that we are restricting ourselves to the set of field redefinitions that do not involve derivatives.}

First, we show that any HEFT can be rewritten in this basis. We start with the general HEFT Lagrangian of \cref{eqn:LHEFT}, but with the arguments of its functions shifted without loss of generality
\begin{equation}
\Lag_\text{HEFT} = \frac{1}{2}\s{\big[ {K( v_0 + h )} \big]^2}{( {\partial h} )^2} + \frac{1}{2}{\big[ {v\s F( v_0 + h )} \big]^2}{( {\partial \vec n} )^2} - V( v_0 + h ) + {\cal O}\big( {{\partial^4}} \big) \, .
\label{eqn:startNaive}
\end{equation}
The potential is minimized when $h=0$.
Let
\begin{equation}
  v_1 + h_1 = Q(v_0 + h) = \int_0^{v_0+h} \dd t \, K(t) \, ,
  \label{eqn:canonhRedef}
\end{equation}
such that
\begin{equation}
  \dd h_1 = K(h)\, \dd h \, .
\end{equation}
As $K$ is always strictly positive and real analytic, the function $Q$ is real analytic and invertible, and therefore defines a valid field redefinition. It yields the canonicalized Lagrangian
\begin{equation}
  \Lag_\text{HEFT} = \frac{1}{2}\s{( {\partial h_1} )^2} + \frac{1}{2}\s{\big[ {v\s F\big( Q^{-1}(v_1+h_1) \big)} \big]^2}{( {\partial \vec n} )^2} - V\big( Q^{-1}(v_1+h_1) \big) + {\cal O}\big( {{\partial^4}} \big) \, ,
  \label{eqn:endCanonical}
\end{equation}
We define $v_1 = Q(v_0)$, such that the above potential is minimized when $h_1=0$.

Second, we show that canonicalizing the $h$ kinetic term fully fixes the field basis, exhausting our field redefinition freedom. We begin from the canonical two-derivative Lagrangian 
\begin{align}
\Lag_\text{HEFT} &= \frac{1}{2}\s{( {\partial h} )^2} + \frac{1}{2}\s{[ {v\s F\left( h \right)} ]^2}{( {\partial \vec n} )^2} \notag\\[8pt]
 &= \frac{1}{2}\s{( {\partial h} )^2} + \frac{1}{2}\s{[ {v\s F( h )} ]^2}\bigg( {{\delta^{ij}} + \frac{{{n^i}\s{n^j}}}{{1 - {n^2}}}} \bigg)\big( {{\partial^\mu }{n^i}} \big)\big( {{\partial_\mu }{n^j}} \big) \, , 
\end{align}
where in the second line, we expressed the $\vec{n}$ in terms of its three independent components $n^i$. We attempt the general field redefinition
\begin{subequations}\label{eqn:hnRedef}
\begin{align}
h &= h \left(h_1\right) \\[4pt]
\vec n &= \vec n \left(h_1, \vec n_1\right) \, .
\end{align}
\end{subequations}
Note that $h$ has no $\vec n_1$ dependence, since $h$ is an $O(4)$ scalar and $\vec n_1$ is a unit $O(4)$ vector that cannot be used to generate a non-trivial $O(4)$ scalar apart from $\vec n_1 \cdot \vec n_1 =1$. The redefined Lagrangian reads 
\begin{align}
\Lag_\text{HEFT} 
 &= \frac{1}{2}\Bigg[ {{{\bigg( {\frac{{\dd h}}{{\dd h_1}}} \bigg)}^2} + {{[ {v\s F\left( h \right)} ]}^2}\bigg( {{\delta^{ij}} + \frac{{{n^i}\s{n^j}}}{{1 - {n^2}}}} \bigg)\s\frac{{\partial {n^i}}}{{\partial {h_1}}}\s\frac{{\partial {n^j}}}{{\partial {h_1}}}} \Bigg]{( {\partial {h_1}} )^2} \notag\\[5pt]
 &\hspace{15pt} + {[ {v\s F( h )} ]^2}\bigg( {{\delta^{ij}} + \frac{{{n^i}\s{n^j}}}{{1 - {n^2}}}} \bigg)\frac{{\partial {n^i}}}{{\partial {h_1}}}\s\frac{{\partial {n^j}}}{{\partial n_1^k}}\s\big( {{\partial^\mu }{h_1}} \big)\s\big( {{\partial_\mu }n_1^k} \big) \notag\\[5pt]
 &\hspace{15pt} + \frac{1}{2}{[ {v\s F\left( h \right)} ]^2}\bigg( {{\delta^{ij}} + \frac{{{n^i}{n^j}}}{{1 - {n^2}}}} \bigg)\s\frac{{\partial {n^i}}}{{\partial n_1^k}}\s\frac{{\partial {n^j}}}{\partial n_1^l}\s\big( {{\partial^\mu }n_1^k} \big)\s\big( {{\partial_\mu }n_1^l} \big) \,,
 \label{eqn:Lagh1n1}
\end{align}
For $h$ to still have a canonical kinetic term, our field redefinition must result in a vanishing mixing term $\big(\partial^\mu h_1\big)\s\big(\partial_\mu n_1^k\big)$, which implies
\begin{equation}
\bigg( {{\delta^{ij}} + \frac{{{n^i}{n^j}}}{{1 - {n^2}}}} \bigg)\s\frac{{\partial {n^j}}}{{\partial n_1^k}}\s \frac{{\partial {n^i}}}{{\partial {h_1}}} = 0 \,. \label{eqn:ZeroMixing}
\end{equation}
Note that the matrix $\delta^{ij} + \frac{n^i n^j}{1 - n^2}$ is invertible, since
\begin{equation}
\bigg( \delta^{ij} + \frac{n^i n^j}{1 - n^2} \bigg)\s \big( \delta^{jk} - n^j n^k \big) = \delta^{ik} \,.
\end{equation}
In addition, the matrix $\partial n^j/\partial n_1^k$ is also invertible due to the assumption that this field redefinition is well defined.
Therefore, \cref{eqn:ZeroMixing} defines a vanishing vector
\begin{equation}
 \frac{\partial n^i}{\partial h_1} = 0 \,,
\end{equation}
since its coefficient matrix is invertible.
This implies that $\vec n$ has to be independent of $h_1$, namely $\vec n = \vec n\s(h_1, \vec n_1) = \vec n \s(\vec n_1)$. This is an important result and its consequence is twofold. One, when the $O(4)$ vector $\vec n$ is purely a function of another vector $\vec n_1$, the $O(4)$ symmetry guarantees that
\begin{equation}
\big(\partial \vec n\big)^2 = \big(\partial \vec n_1\big)^2 \,.
\end{equation}
Therefore, the last line of \cref{eqn:Lagh1n1} is trivial and the form factor $v\s F$ is unchanged. Two, for the $(\partial h_1)^2$ term in \cref{eqn:Lagh1n1}, the constraint $K=1$ implies that the $h$ redefinition is trivial:
\vspace{-5pt}
\begin{align}
K=1  \qquad\Longrightarrow\qquad  \frac{\dd h}{\dd h_1}=1  \qquad\Longrightarrow\qquad  h=h_1 \,.\\[-30pt]
\notag
\end{align}
Note that there is no constant shift allowed, due to maintaining the potential minimization condition.

To summarize, canonicalizing the $h$ kinetic term uniquely determines $h$ and $\vec{n}$, and hence fixes the field basis. In this basis, the Lagrangian takes the form
\begin{equation}
\Lag_\text{HEFT} = \frac{1}{2}\s{( {\partial h} )^2} + \frac{1}{2}\s{[ {v\s F\left( h \right)} ]^2}\s{( {\partial \vec n} )^2} - V(h) + \mathcal{O}\big(\partial^4\big) \,.
\label{eqn:LHEFTK1}
\end{equation}

\subsubsection{Fixed Basis Criteria}
\label{subsubsec:K1Criterion}
Now we are positioned to close the loophole we found for the naive criteria proposed in the beginning of \cref{subsec:BasisDependentCriterion}.
We have already established that if a HEFT Lagrangian can be written as SMEFT, it will satisfy the naive criteria in \emph{some} field basis.
We can imagine starting from this basis that satisfies the naive criteria, and then perform the field redefinition in \cref{eqn:canonhRedef} to canonicalize the $h$ kinetic term.

In our starting basis that satisfies the naive criteria, $K$ and $V$ in \cref{eqn:startNaive} are even functions, and $F$ is odd. Moreover, $K(0)=1$ and $v\s F^\prime(0)=1$.
This implies that $Q$ in \cref{eqn:canonhRedef} is odd, and satisfies $Q^\prime(0)=1$.
Therefore, in \cref{eqn:endCanonical},
\begin{itemize}
  \setlength\itemsep{1pt}
  \item the form factor of the $(\partial \vec n)^2$ term, $F \big( Q^{-1} ( v_1 + h_1 ) \big)$, is odd and has unit first derivative about the point $h_1=-v_1$;
  \item the potential $V \big( Q^{-1} ( v_1 + h_1 ) \big)$ is even about $h_1=-v_1$.
\end{itemize}
In other words, we derived the following criteria in the basis \cref{eqn:LHEFTK1}, defined by having a canonical $h$ kinetic term:


\begin{center}
\begin{tcolorbox}[colback=light-gray]
\begin{minipage}{5.5in}
\textbf{Fixed Basis Criteria:} Given a generic HEFT Lagrangian, one must use a field redefinition to canonicalize the kinetic term for $h$, while maintaining the minimization condition for the potential $V'(h=0)=0$.
This theory can be expressed as a SMEFT at the fixed point \emph{if and only if} the resulting HEFT Lagrangian satisfies the following three conditions:
\vspace{10pt}
\begin{enumerate}
  \item The function $F(h)$ has a zero at some real value of $h=h_*$. This $h_*$ is then a candidate for an $O(4)$ invariant point. We will define $v_0\equiv-h_*$.
  \item At $h=h_*$, the functions $F(h)$ and $V(h)$ both have convergent Taylor expansions, namely they are a nonnegative power series in $h-h_*=v_0+h$.
  \item For these Taylor expansions about $h_*$, $V$ is an even-power series, and $F$ is an odd power series with the leading term $v\s F'\left(h_*\right)=1$.
\end{enumerate}
\end{minipage}
\end{tcolorbox}
\end{center}

\subsection{Basis Independent Curvature Criteria}
\label{subsec:CurvatureCriterion}
In the previous section, we have derived criteria that can be applied to the HEFT Lagrangian to determine when it is possible to express it as SMEFT.
However, this result has the drawback that it must be evaluated in the basis where the $h$ kinetic term is canonical.
As we will see in \cref{sec:Examples} below, where we work out various detailed matching examples, integrating out a UV sector typically results in a non-trivial $K(h)$ function, \ie, the kinetic term for $h$ is non-canonical.
In principle, one could apply the field redefinition in \cref{eqn:canonhRedef} to set $K(h) = 1$.
However, in practice this procedure typically involves inverting the function $Q$, which comes with its own technical challenges (for example one might encounter a transcendental equation).
Therefore, it is of conceptual and practical use to develop basis independent criteria.
Unsurprisingly, a natural framework for doing so is through the use of curvature invariants that are evaluated on the Higgs manifold, which are manifestly independent of coordinate changes.

We first assume that the functions $K(h)$, $F(h)$ and $V(h)$ all have convergent single-argument Taylor expansions about $h=h_*$.\footnote{Clearly, we are assuming that this statement is basis independent, which places a mild restriction on the set of possible field redefinitions we are allowed to make.}
Then in \cref{appsec:CurvatureCriterionK1}, we start with our fixed basis form of the HEFT Lagrangian given in \cref{eqn:LHEFTK1} and show the following two sets of conditions are equivalent as applied to \cref{eqn:LHEFTK1}:
\begin{equation}\renewcommand\arraystretch{1.6}
\left. \begin{array}{l}
{F^{( {2k} )}}( {{h_*}} ) = 0, \hspace{15pt} \forall k \in \mathbb{N} \\
v\s F'( {{h_*}} ) = 1\\
{V^{( {2k + 1} )}}( {{h_*}} ) = 0, \,\, \forall k \in \mathbb{N}
\end{array} \right\}
\quad\Longleftrightarrow\quad
\left\{ \begin{array}{l}
F( {{h_*}} ) = 0\\
{\left. {{\nabla^{{\mu_1}}}{\nabla_{{\mu_1}}} \cdots {\nabla^{{\mu_n}}}{\nabla_{{\mu_n}}}R} \right|_{{h_*}}} < \infty , \,\, \forall n \in \mathbb{N} \\
{\left. {{\nabla^{{\mu_1}}}{\nabla_{{\mu_1}}} \cdots {\nabla^{{\mu_n}}}{\nabla_{{\mu_n}}}V} \right|_{{h_*}}} < \infty , \,\, \forall n \in \mathbb{N}
\end{array} \right. \, , \label{eqn:Equivalence}
\end{equation}
where $R$ is given in~\cref{eq:RmainText} and is derived in \cref{appsec:ScalarCurveature}, and $\nabla_{\mu}$ is the covariant derivative defined on the Higgs manifold.
This justifies our basis independent curvature based criteria:\footnote{We emphasize again that the potential minimization condition $V'(h=0)=0$ is taken as the definition of the origin of the field $h$, which is assumed in all field bases throughout this paper.  Additionally, we leave generalizing these criteria to account for the ability to perform field redefinitions that include derivatives for future work.}
\clearpage
\begin{center}
\vspace{3mm}
\begin{tcolorbox}[colback=light-gray]
\begin{minipage}{5.5in}
\textbf{Curvature Criteria:} A HEFT can be expressed as a SMEFT at the fixed point \emph{if and only if} the following three conditions hold:
\vspace{10pt}
\begin{enumerate}
  \item The function $F(h)$ has a zero for some real value $h=h_*$.
  This is a candidate $O(4)$ invariant fixed point.
  \item The metric is analytic at $h_*$.
  This requires ($i$) the functions $K(h)$ and $F(h)$ have convergent single-argument Taylor expansions about $h=h_*$, and ($ii$) curvature invariants $\left( \nabla^{\mu_1} \nabla_{\mu_1} \cdots \nabla^{\mu_n} \nabla_{\mu_n} R \right)$ built from the scalar curvature $R$ are finite at $h_*$ for all integers $n \geq 0$.
  When these conditions hold, the special point $h_*$ exists on the scalar manifold and is the location of the $O(4)$ invariant fixed point.
  \item The potential $V$ is analytic at $h_*$.
  This requires ($i$) the function $V(h)$ has a convergent single-argument Taylor expansions about $h=h_*$, and ($ii$) curvature invariants $\left( \nabla^{\mu_1} \nabla_{\mu_1} \cdots \nabla^{\mu_n} \nabla_{\mu_n} V \right)$ built from $V$ are finite about $h=h_*$ for all integers $n \geq 0$.
\end{enumerate}
\end{minipage}
\end{tcolorbox}
\vspace{3mm}
\end{center}
We have provided a straightforward mathematical derivation of these curvature criteria in \cref{appsec:CurvatureCriterionK1}.
It is additionally useful to develop some physical intuition.
As pointed out by AJM in Refs.~\cite{Alonso:2015fsp, Alonso:2016oah}, a necessary and sufficient condition for a HEFT to be expressible as a SMEFT is the existence of an $O(4)$ invariant point $h_*$ on the scalar manifold.
The necessity is obvious, since the $H=0$ in any SMEFT theory is an $O(4)$ invariant point.
The sufficiency is guaranteed by the CWZ linearization lemma~\cite{Coleman:1969sm}, which states that when an $O(4)$ invariant point exists, one can construct a set of coordinates that transform in the (linear) fundamental representation of $O(4)$; these are the defining coordinates of SMEFT.
However, the linearization lemma assumes that the manifold under discussion is analytic, which is why we need the requirements in condition 2.

Finally, we note that it might be the case that for any ``physical'' HEFT Lagrangian, \ie, one that can be obtained from integrating out a UV state, condition 2 always guarantees condition 3.
This will be true for the concrete examples that we present in the next section.

\section{Can Our Physical Vacuum Be Described by SMEFT?}
\label{sec:eftsubmani}

Thus far, we have established {\it when} HEFT is required to describe physics about the $O(4)$ invariant fixed point.
However, this leaves open the question of {\it why} HEFT is required, \ie, what properties of an $SU(2)_L \times U(1)_Y$-symmetric UV theory result in a low-energy description that preserves only $U(1)_\text{em}$. 
Additionally, we need to extend our analysis of the EFT about the fixed point to ensure that the SMEFT we write down there can be expanded about the physical vacuum, since this is required to make phenomenologically relevant predictions.
We will address this question by leveraging the same geometric framework that underlies our criteria to distinguish SMEFT from HEFT about the fixed point.
In particular, when the UV theory is renormalizable, it is automatically truncated at two-derivative order. 
This implies that all of the technology discussed in \cref{sec:PolarCoords} applies, and the UV dynamics may be entirely encoded by a metric and potential function defined on a field space manifold.
We will show how the emergence of an EFT that only exhibits a $U(1)_\text{em}$ symmetry, even though the UV theory manifests the full $SU(2)_L\times U(1)_Y$ symmetry, can be understood using the language of the EFT submanifold.

As we will show in this section, integrating out heavy scalar states (at tree-level) leaves behind an EFT which lives on a slice of the full manifold. 
In other words, the EFT can be identified with a submanifold of the UV manifold, thereby providing a natural framework for understanding how properties of the UV theory can give rise to non-analyticities in the EFT.
Then by recognizing that there are different possible ``EFT branches,'' we will argue that a well defined SMEFT must live on a smooth branch that connects the fixed point to the physical vacuum.
Exploring the potential obstructions that can emerge along such branches will provide us with additional insight into situations when HEFT is required.
This submanifold point of view also provides intuitive visual representations of the various scenarios that can occur.

Our goal here is twofold: to define the EFT submanifold, and to understand how non-analyticities arise on it. 
Although the analysis in this section is restricted to tree-level matching among scalar field theories, the lessons are more broadly applicable.

\subsection{The EFT Submanifold}
First, we will explain what is meant by the ``EFT submanifold.''
Consider a generic UV action $S_\text{UV}[\phi,\Phi]$, which describes the dynamics of the SM scalar fields $\phi$ and the BSM fields we plan to integrate out $\Phi$.
From the functional point of view, integrating out UV dynamics essentially boils down to solving for the effective action $S_\text{Eff}[\phi]$ from the partition function directly:
\begin{equation}
  \int \pathd \phi\, \exp \left( i\s S_\text{Eff}[\phi] + i\s \int \phi\s J \right) = \int \pathd \phi\, \pathd \Phi\, \exp\left( i\s S_\text{UV}[\phi,\Phi] + i\s \int \phi \s J  \right)  \,.
  \label{eqn:matchpathint}
\end{equation}
We can develop some intuition by considering the Euclidean path integral with $\phi$ treated as a fixed background field.
Then $S_\text{Eff}[\phi] \sim \ln \left( \int \pathd \Phi\, \exp(-S_\text{UV}) \right)$ computes the free energy of the $\Phi$ modes, which will typically respond smoothly to changes in the external $\phi$ configuration.
The exception is for $\phi$ values where the UV system exhibits some critical behavior, the harbingers of which are massless excitations that are responsible for generating non-analyticities in the effective Lagrangian.
As was emphasized by FR~\cite{Falkowski:2019tft}, such effective Lagrangians must be matched onto HEFT.
We will show how this physical story manifests when using the language of the EFT submanifold.

Our goal is to match onto an EFT action up to two derivative order. In the case of an extended scalar sector, the configuration space of the UV theory generates a manifold with coordinates $(\phi,\Phi)$, upon which the zero- and two-derivative terms in the UV Lagrangian define a potential and metric respectively.
As we will now show, matching the UV theory onto an EFT at tree-level is equivalent to simply solving for the EFT submanifold on which the induced zero- and two-derivative EFT Lagrangian terms live.
Approximating the path integral \cref{eqn:matchpathint} using the method of steepest descent gives
\begin{equation}
  S_\text{Eff}[\phi] = S_\text{UV}\big[\phi,\Phi_\mathbf{c}[\phi]\big] \, ,
\end{equation}
where the classical field $\Phi_\mathbf{c}[\phi]$ extremizes the action for a given $\phi$:
\begin{equation}
  \frac{\delta S_\text{UV}}{\delta \Phi} \big[\phi,\Phi_\mathbf{c}[\phi]\big]= 0 \, .
  \label{eqn:steepestdescent}
\end{equation}
Next, we assume that the UV Lagrangian admits a quasi-local derivative expansion
\begin{align}
S_\text{UV}[\phi,\Phi] = \sum_{k=0}^\infty S^{(2k)}_\text{UV}[\phi,\Phi] \, ,
\end{align}
where terms that contain $2\s k$ derivatives are contained in $S^{(2k)}_\text{UV}[\phi,\Phi]$, and are themselves local and analytic functions of the fields $\phi$ and $\Phi$.
We wish to solve~\cref{eqn:steepestdescent} for $\Phi_\mathbf{c}[\phi] = \sum_{k=0}^\infty \Phi_\mathbf{c}^{(2k)}[\phi]$ as a derivative expansion.
Then the zero derivative term $\Phi_\mathbf{c}^{(0)}[\phi]$ is a functional of the light field coordinates $\phi$, which satisfies 
\begin{equation}
  \frac{\partial V}{\partial \Phi}\Big(\phi,\Phi_\mathbf{c}^{(0)}\Big) = 0 \qquad\text{ where } \qquad S^{(0)}_\text{UV} = -\int \dd^4 x\, V \, ,
  \label{eqn:phic0extrempot}
\end{equation}
as a consequence of \cref{eqn:steepestdescent}.
The higher derivative terms $\Phi^{(2k)}_\mathbf{c}$ can in principle be solved for iteratively by expanding \cref{eqn:steepestdescent} order-by-order in derivatives.
However, we only need $\Phi^{(0)}_\mathbf{c}$ to calculate contributions to the effective Lagrangian when truncated at the two-derivative order:
\begin{align}
  S_\text{Eff}^{(0)}[\phi] + S_\text{Eff}^{(2)}[\phi] &= S_\text{UV}^{(0)}\Big[\phi,\Phi_\mathbf{c}^{(0)}+\Phi_\mathbf{c}^{(2)}\Big] + S_\text{UV}^{(2)}\Big[\phi,\Phi_\mathbf{c}^{(0)}\Big] \,\nonumber \\[5pt]
                                                                          &= S_\text{UV}^{(0)}\Big[\phi,\Phi_\mathbf{c}^{(0)}\Big] + \Phi_\mathbf{c}^{(2)}\s \frac{\delta S_\text{UV}^{(0)}}{\delta \Phi}\Big[\phi,\Phi_\mathbf{c}^{(0)}\Big] + S_\text{UV}^{(2)}\Big[\phi,\Phi_\mathbf{c}^{(0)}\Big]  \,\nonumber \\[5pt]
                                                                          &= S_\text{UV}^{(0)}\Big[\phi,\Phi_\mathbf{c}^{(0)}\Big]  + S_\text{UV}^{(2)}\Big[\phi,\Phi_\mathbf{c}^{(0)}\Big]  \,,
\end{align}
where the third line follows as a consequence of \cref{eqn:phic0extrempot}.

Therefore, when matching perturbatively at tree-level, the effective potential and metric are simply those induced by taking the $\big(\phi,\Phi_\mathbf{c}^{(0)}(\phi)\big)$ slice of the UV manifold.
This submanifold is charted by $\phi$ and is determined in practice by profiling over the $\Phi$ direction to find a point which extremizes the UV potential.
This makes the geometric interpretation of the low energy theory completely manifest.

\subsection*{Generalization for Non-trivial Representations}
These ideas straightforwardly generalize to our case of interest involving non-trivial representations.
If the EFT can be expressed as SMEFT, the UV manifold must have at least one $O(4)$ fixed point, such that the EFT manifold can possibly inherit it. 
Then the linearization lemma of \cite{Coleman:1969sm} implies that there exist a set of UV coordinates about the fixed point which arrange themselves into irreps of $SO(4)$; at least one of these irreps must be a 4-plet. 
If there is only one 4-plet, it can be identified with our $\phi^i$ field ($i=1,\dots,4$) and the remaining irreps form the $n$ real coordinates $\Phi^a$ ($a=1,\dots,n$). 
The $O(4)$ fixed point when on the manifold defined using these coordinates is at $\left(\phi^i=0,\Phi^a=0\right)$.\footnote{If $\{\Phi^a\}$ contains $O(4)$ singlets, then there are infinitely many $O(4)$ fixed points along their axes.} 
We then attempt to integrate out the coordinates $\Phi^a$, assuming the generic form for the UV Lagrangian:\footnote{Should there exist more than one 4-plet, we have the freedom to specify a basis in which to carry out the matching, such that one of the 4-plets is identified with our $\phi^i$ field while the remaining 4-plets and any other irreps become the coordinates $\Phi^a$. In  \cref{sec:BSMSymBreak} and \cref{sec:swampland}, we will explore aspects of choosing a basis in such situations; note that changes of basis in the UV theory can correspond to derivative field redefinitions in the EFT, and can result in different conclusions on whether HEFT is required.}
\begin{equation}
  \mathcal{L}_\text{UV} = \frac{1}{2}
  \begin{pmatrix} \partial_\mu \phi^i & \partial_\mu \Phi^a \end{pmatrix}
  \begin{pmatrix}\, g_{ij}(\phi,\Phi) & g_{ib}(\phi,\Phi) \\ g_{aj}(\phi,\Phi) & g_{ab}(\phi,\Phi) \,\end{pmatrix}
  \begin{pmatrix} \partial^\mu \phi^j \\ \partial^\mu \Phi^b \end{pmatrix} - V(\phi^i,\Phi^a) \,,
  \label{eqn:genUVlag}
\end{equation}
where the potential and positive-definite metric are assumed to be analytic in the coordinates $(\phi^i,\Phi^a)$.

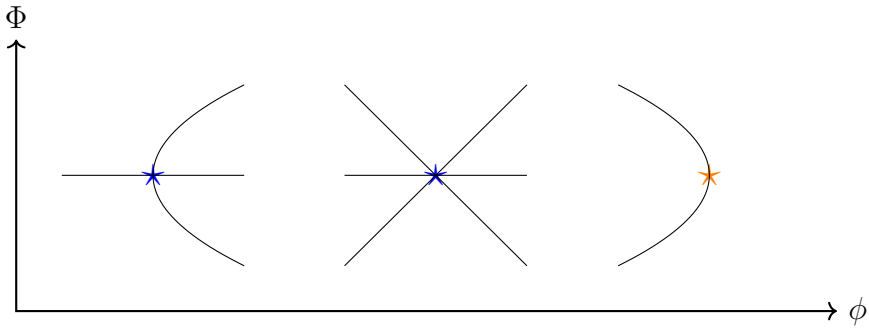
\begin{figure}
  \centering
  \begin{tikzpicture}[scale=1.2]
  \draw[thick,<->] (0,3) node[above] {$\Phi$} -- (0,0) -- (9,0) node[right] {$\phi$};
  \foreach \x/\y/\clr in {1.5/1.5/blue,4.6/1.5/blue,7.6/1.5/orange} {
  \node[\clr] at (\x,\y) {\Large$\star$};}
  \draw[domain=-1:1, smooth, variable=\t] plot ({7.6-\t*\t}, {1.5+\t});
  \draw[domain=-1:1, smooth, variable=\t] plot ({4.6+\t}, {1.5});
  \draw[domain=-1:1, smooth, variable=\t] plot ({4.6+\t}, {1.5+\t});
  \draw[domain=-1:1, smooth, variable=\t] plot ({4.6+\t}, {1.5-\t});
  \draw[domain=-1:1, smooth, variable=\t] plot ({1.5+\t*\t}, {1.5+\t});
  \draw[domain=-1:1, smooth, variable=\t] plot ({1.5+\t}, {1.5});
\end{tikzpicture}  
\caption{In a schematic UV theory of two fields ($\phi,\Phi$), possible behaviors of regions of the EFT submanifold (solutions $(\phi,\Phi_\textbf{c})$ of $\frac{\partial V}{\partial \Phi}=0$) about points `$\star$' where $\frac{\partial^2 V}{\partial \Phi^2}=0$.
  For the two lefthandmost (blue) points, $\frac{\partial^2 V}{\partial \phi\s \partial \Phi} = 0$, and the gradient \cref{eqn:maniGradient} is ill-defined. Generically, different branches of solution will meet. $\Phi_\textbf{c}$ may or may not be analytic in $\phi$ on each branch at this point.
  For the righthandmost (orange) point, $\frac{\partial^2 V}{\partial \phi \s\partial \Phi} \neq 0$, and the gradient \cref{eqn:maniGradient} diverges. $\Phi_\textbf{c}$ ceases to be an analytic function of $\phi$.
See \cref{appsec:SubmanifoldSingularities} for a fuller discussion.
\label{fig:singularEFTSubmanifold}}
\end{figure}

In analogy with~\cref{eqn:phic0extrempot}, the $n$ equations of motion (EOMs) are
\begin{equation}
  \frac{\partial V}{\partial \Phi^a}\Big(\phi^i,\Phi_\mathbf{c}^a\Big) = 0 \,.
   \label{eq:eomvariety}
\end{equation}
This system of equations is solved by a set of points $(\phi^i,\Phi_\mathbf{c}^a)$ that arrange themselves to lie along different ``branches,'' each a contiguous four-dimensional surface.
These branches are either disconnected from each other or can join at singular points of \cref{eq:eomvariety}, see~\cref{fig:singularEFTSubmanifold}.
Each of these branches defines an EFT submanifold expressed in terms of the $\phi^i$ coordinates.

\subsection{Implications of the Physical Branch \label{sec:scalarsmeft}}
A phenomenologically viable EFT (not necessarily a SMEFT) corresponds to the ``physical branch'' that includes the observed vacuum, which is a local minimum of the potential $V$. 
First, we can check if there are any intersection point(s) between the physical branch and the $\phi^i=0$ plane, \ie, the plane which includes all of the possible $O(4)$ fixed points.
If there is no $O(4)$ fixed point on the physical EFT submanifold then it is necessarily a HEFT. 
This phenomenon occurs when at $\phi^i=0$, BSM $O(4)$ non-singlet fields have a vev $\Phi_\mathbf{c}^a \neq 0$, which breaks the $O(4)$ symmetry.
Therefore, HEFT results because there is \emph{extra electroweak symmetry breaking}.

Having covered the obvious case, we can assume that our physical branch includes an $O(4)$ fixed point, such that it could potentially be written as SMEFT. 
Next, we need to check whether the EFT Lagrangian is analytic in the coordinates $\phi$.
The analytic implicit function theorem~\cite{krantz2012implicit} tells us that the solutions $\Phi_\textbf{c}$ of \cref{eq:eomvariety}, whose first derivatives are given by
\begin{equation}
\frac{\dd \Phi^b_\mathbf{c}}{\dd \phi^i} (\phi) = - \left( \frac{\partial^2 V}{\partial \Phi\s \partial \Phi} (\phi,\Phi_\mathbf{c}) \right)^{-1}_{ba} \frac{\partial^2 V}{\partial \phi^i\s \partial \Phi^a} (\phi,\Phi_\mathbf{c}) \,,
\label{eqn:maniGradient}
\end{equation}
will be analytic in $\phi$ if $\frac{\partial^2 V}{\partial \Phi^a\s \partial \Phi^b} (\phi,\Phi_\mathbf{c})$ is invertible. 
Upon substitution into \cref{eqn:genUVlag}, an analytic $\Phi_\textbf{c}$ yields an analytic EFT Lagrangian. 
By contrast, when the matrix $\frac{\partial^2 V}{\partial \Phi^a\s \partial \Phi^b} (\phi,\Phi_\mathbf{c})$ is singular, we expect non-analyticities in $\Phi_\textbf{c}$, and by extension the EFT Lagrangian. 
Their exact form is considered in more detail in \cref{appsec:SubmanifoldSingularities}, and we show schematic examples of non-analytic behavior in \cref{fig:singularEFTSubmanifold}.

When the matrix $\frac{\partial^2 V}{\partial\Phi^a\s \partial\Phi^b}$ is singular at the $O(4)$ fixed point, this corresponds to a BSM state that acquires all of its mass from electroweak symmetry breaking. 
We generically expect any EFT Lagrangian to be non-analytic at the fixed point in the sense defined in \cref{sec:WhatIsHEFT}, and for HEFT to be required, see~FR~\cite{Falkowski:2019tft}.

Furthermore, we may now consider the analytic properties of the EFT as we move along the physical branch between the fixed point and our observed vacuum.
Let there be an $O(4)$ fixed point on the physical branch, and let the sub-mass matrix $\frac{\partial^2 V}{\partial\Phi^a \partial\Phi^b}$ be invertible at this putative $O(4)$ fixed point.
This only guarantees that we have a SMEFT expansion in the \emph{neighborhood} of the $O(4)$ fixed point. 
When applying this expansion to physical observables, there is no guarantee that the predictions made about our vacuum will be convergent. 
For example, if the sub-mass matrix $\frac{\partial^2 V}{\partial\Phi^a \partial\Phi^b}$ includes a tachyonic direction at the $O(4)$ fixed point (even though it is invertible), then the convergence of the SMEFT expansion cannot be preserved all the way from the fixed point to the observed vacuum.  
The issue is that on any such path, there exists a point at which $\frac{\partial^2 V}{\partial\Phi^a \partial\Phi^b}$ becomes non-invertible, which implies the existence of a singularity.

Furthermore, even when the (real) analyticity of SMEFT is preserved all the way along a path connecting the fixed point with the our observed vacuum, the convergence of physical predictions is still not guaranteed. 
We will discuss a concrete example of this phenomenon in \cref{sec:Truncate}.

\section{When HEFT is Required: Light BSM States}
\label{sec:Examples}
In this section, we show a number of concrete examples where the low energy theory must be expressed as HEFT by starting with a UV description and explicitly integrating out the BSM particles.
In particular, our example models include new singlets, since these present some of the most challenging scenarios for direct exploration.
Our focus in this section is on the limit where these new physics states get all of their mass from the Higgs vev, since this is one situation where we expect HEFT to be relevant.

These examples are simple enough that we can perform the matching calculation and obtain the potential term $\Lag_\text{Eff}^{(0)}$ and the two-derivative term $\Lag_\text{Eff}^{(2)}$ of the effective Lagrangian to all-orders in the light fields, namely that we can get the full form factor functions $K(h)$, $F(h)$, and $V(h)$ in \cref{eqn:LHEFT}; truncating $\Lag_\text{Eff}$ to finite order yields an EFT description.
This enables us to check the claims of our LO Criteria in \cref{sec:WhatIsHEFT} for each of these examples.
Then in \cref{sec:BSMSymBreak}, we will discuss the implications of UV models with new sources of electroweak symmetry breaking beyond the SM Higgs doublet.

\subsection{Integrating Out a Singlet Scalar at Tree Level}
\label{sec:SingletModel}
In this section, we introduce one of the simplest scenarios where a BSM state -- a new scalar singlet $S$ -- gets all of its mass from the Higgs vev through a coupling $\kappa\s |H|^2\s S^2$. Recent matching calculations for this scenario in the context of HEFT and SMEFT include \cite{Henning:2014wua, deBlas:2014mba, Gorbahn:2015gxa, Chiang:2015ura, Brehmer:2015rna, Buchalla:2016bse,Jiang:2016czg, Corbett:2017ieo, Dawson:2017vgm}.
Specifically, the UV Lagrangian is
\begin{equation}
  \Lag_\text{UV} = {\left| {\partial H} \right|^2} + \frac{1}{2}\s (\partial S)^2 - V \,,
\label{eqn:LUVSingletZ2}
\end{equation}
with the potential
\begin{equation}
  V = - \mu_H^2 \s |H|^2 + \lambda_H\s |H|^4 + \frac{1}{2} \big(\s m^2 + \kappa\s |H|^2\s \big)\s S^2 + \frac{1}{4}\s \lambda_S\s  S^4 \,.
\label{eqn:VUVSingletZ2}
\end{equation}
We require $\lambda_H, \lambda_S >0$ and $4 \s \lambda_H\s \lambda_S > \kappa^2$ to enforce that the potential is bounded from below.
We have imposed a $\mathbb{Z}_2$ symmetry $S\to -S$ to make the model as simple as possible. 
Naively, this $\mathbb{Z}_2$ symmetry would forbid any non-trivial tree-level effects unless it is spontaneously broken by the vev of the singlet $v_S$. 
Once written in terms of the dynamic field $S=v_S+s$, the Lagrangian is
\begin{equation}
\Lag_\text{UV} = {\left| {\partial H} \right|^2} + \frac{1}{2}\s (\partial s)^2 - V \,,
\label{eqn:LUVSingletZ2Broken}
\end{equation}
with the potential
\begin{align}
V &=  - \Big(\mu_H^2 - \frac{1}{2}\s \kappa\s v_S^2\Big)\s |H|^2 +  \lambda_H \s |H|^4 + \kappa\s v_S \s |H|^2 \s s \notag\\[5pt]
&\quad + \frac{1}{2}\s \Big( 2\s \lambda_S \s v_S^2 + \kappa\s |H|^2 \Big) \s s^2 + \lambda_S \s v_S \s s^3 + \frac{1}{4} \s \lambda_S\s s^4 \,.
\label{eqn:VUVSingletZ2Broken}
\end{align}
Then ${\cal L} \supset -\kappa\s v_S\s |H|^2\s s$, the interaction term linear in $s$, generates non-zero Wilson coefficients at tree-level. 
The minimization condition for the vev is 
\begin{align}
v_S\,\big(m^2+\lambda_S\s v_S^2\big) = 0\,.
\label{eq:vSmin}
\end{align}
Clearly there are three solutions, and we choose to work in the parameter space where $m^2 \le 0$, implying that $v_S \neq 0$ corresponds to the physical vacuum; solving \cref{eq:vSmin} implies $m^2=-\lambda_S\s v_S^2$.
Then the field-dependent mass of the singlet $s$ is
\begin{equation}
m_s^2(H) = 2\s\lambda_S\s v_S^2 + \kappa\s |H|^2 = -2\s m^2 + \kappa\s |H|^2 \,.
\label{eq:msSq}
\end{equation}

Next, we will integrate out the BSM singlet $s$ at the tree-level and obtain the effective Lagrangian for $H$.
Our expectation is that we can match onto SMEFT when $m^2\ne0$.
However, if $m^2=0$, the singlet $s$ acquires all of its mass from the Higgs vev $v_0$.
This can generate non-analyticities in the effective Lagrangian for $H$, such that one cannot match this theory onto SMEFT.
In the following, we will check if this is the case by performing the matching calculation, and also check if our LO Criteria in \cref{sec:WhatIsHEFT} hold.

\subsubsection*{Effective Lagrangian}
To integrate out the heavy singlet $s$ at the tree-level, we solve its zero-derivative equation of motion
\begin{equation}
0 = \frac{{\partial V}}{{\partial s }} = (v_S + s)\big(\s 2\s \lambda_S\s v_S\s s  + \lambda_S\s s^2 + \kappa\s |H|^2\s \big)   \,.
\label{eqn:EOMSingletZ2}
\end{equation}
Noting that $s_\mathbf{c}[H=0]=0$, the physically relevant solution is
\begin{equation}
s_\mathbf{c}[H] = - v_S + \sqrt{v_S^2 - \frac{\kappa}{\lambda_S}\s |H|^2} = - v_S + \sqrt {\frac{- m^2 - \kappa\s |H|^2}{\lambda_S}} \,,
\label{eqn:EOMSolutionZ2}
\end{equation}
where we have adopted the convention $v_S\ge 0$. 
Substituting $s_\mathbf{c}[H]$ back into~\cref{eqn:LUVSingletZ2Broken}, we derive the effective Lagrangian:
\begin{align}
\Lag_\text{Eff} &= |\partial H|^2 - \frac{\kappa^2}{8\s \lambda_S \big(\s m^2 + \kappa \s |H|^2\s \big)}\s\big(\s \partial |H|^2\s \big)^2 \notag\\[7pt]
&\hspace{15pt} + \mu_H^2\s |H|^2 - \lambda_H\s |H|^4 + \frac{1}{4\s \lambda_S}\big(\s m^2 + \kappa\s |H|^2\s\big)^2 \,.
\label{eqn:LEFTSingletZ2}
\end{align}

For this derivation, we were explicit about the fact that the $\mathbb{Z}_2$ symmetry was spontaneously broken. 
However, this required introducing the explicit vev $v_S$ and expanding $S = v_S + s$, so that the potential in~\cref{eqn:VUVSingletZ2Broken} includes more terms than in that for the unbroken phase~\cref{eqn:VUVSingletZ2}, thereby complicating the algebra.
In fact, we could have performed the same derivation directly using~\cref{eqn:VUVSingletZ2} by taking variation of \cref{eqn:VUVSingletZ2} with respect to the field $S$:
\begin{equation}
0= \frac{\partial V}{\partial S} = S\,\big(\s\lambda_S\s S^2 + m^2 + \kappa\s |H|^2\s\big) \,.
\end{equation}
Then as long as one is careful to work with the physical solution to this EOM, then the resulting effective Lagrangian will be identical.
We will follow this more concise strategy when performing matching calculations in~\cref{sec:BSMSymBreak} below.

\subsubsection*{Form Factors and Curvature Invariants}
To apply our criteria to~\cref{eqn:LEFTSingletZ2}, we use Eqs.~\eqref{eqn:Doublethn}-\eqref{eqn:HTohnTwoDerivative} to express $H$ in terms of $h$ and $\vec n$:
\begin{align}
\Lag_\text{Eff} &= \frac{1}{2}\s \Bigg( 1 - \frac{\kappa^2\s (v_0 + h )^2}{2\s \lambda_S\s \big[ 2\s m^2 +  \kappa\s (v_0 + h)^2\big]} \Bigg)\s (\partial h)^2 + \frac{1}{2}\s ( v_0 + h )^2\s (\partial \vec n)^2 \notag\\[5pt]
&\hspace{15pt} + \frac{1}{2}\s\mu_H^2\s ( v_0 + h)^2 - \frac{1}{4}\s\lambda_H\s(v_0 + h)^4 + \frac{1}{16\s \lambda_S}\Big[ 2\s m^2 + \kappa\s ( v_0 + h)^2\Big]^2 \,,
\end{align}
from which we identify the form factors $K$ and $F$, along with the potential $V$:
\begin{subequations}\label{eqn:KFVSingletz2}
\begin{align}
K(h) &= \sqrt {1 - \frac{ { {\kappa^2}{( {v_0 + h} )^2}}}{{{{2\s \lambda_S \big[ {2\s{m^2} +  \kappa\s {( {v_0 + h} )^2}} \big]}}}}}\,,
\qquad  {v}\s F( h ) = v_0 + h  \,, \\[5pt]
V(h) &= -\frac{1}{2}\s\mu_H^2\s{( {v_0 + h} )^2} + \frac{1}{4}\s{\lambda_H}{( {v_0 + h} )^4} - \frac{1}{16\s \lambda_S} \Big[ 2\s m^2 + \kappa\s {( {v_0 + h} )^2} \Big]^2 \,.
\end{align}
\end{subequations}
Note that $v_0 = v$ in this model.
First, we see that $F(h_*) = 0$ can be satisfied for
\begin{equation}
h_*=-v \,;
\end{equation}
this is the candidate $O(4)$ invariant point.
In addition, the functions $K(h)$, $F(h)$, and $V(h)$ all have convergent Taylor expansions as a function of $h$ about $h_*$.
Both conditions 1 and 2 of our LO Criteria are therefore satisfied, whether $m^2=0$ or not.

Next, we check the third condition, which requires computing the scalar curvature $R$.
Noting that
\begin{equation}
{\partial_h}F = \frac{1}{v}  \quad,\quad  \partial_h^2F = 0  \quad,\quad  {\partial_h}K = -\frac{\kappa^2\s m^2}{\lambda_S\s K}\frac{{( {v + h} )}}{ \big[ {2\s{m^2} + \kappa\s {( v + h )^2}} \big]^2}  \,,
\end{equation}
we plug into \cref{eqn:Rgeneral} to derive the scalar curvature
\begin{align}
R &= \frac{{2\s{N_\varphi }}}{{( {v + h} )\s{K^3}}}( {{\partial_h}K} ) + \frac{{{N_\varphi }\s( {{N_\varphi } - 1} )}}{{{{( {v + h} )}^2}}}\bigg( {1 - \frac{1}{{{K^2}}}} \bigg) \notag\\[7pt]
 &= N_\varphi\s \kappa^2\s \frac{(N_\varphi -1 )\s \kappa \s(v+h)^2\s (\kappa - 2\s \lambda_S) - 4\s (N_\varphi + 1)\s m^2\s \lambda_S }{\big[ \kappa\s (v+h)^2\s (\kappa - 2\s \lambda_S) - 4\s m^2\s \lambda_S   \big]^2} \,,
 \label{eqn:RSingletZ2}
\end{align}
where we have used $\partial_h F = 1/v$ and $\partial^2_h F = 0$ in deriving the first line.

\subsection*{$\bm{m^2\neq0}$: Matching Onto SMEFT is Allowed}
\label{subsec:SingletSMEFT}
We see from \cref{eqn:RSingletZ2} that when $m^2\ne0$, the scalar curvature is finite at the candidate $O(4)$ fixed point: 
\begin{equation}
R\left( {h =  - v} \right) = -\frac{\kappa^2}{ 4\s m^2\s \lambda_S }{N_\varphi }\s( {{N_\varphi } + 1} ) \,.
\end{equation}
Therefore, our LO Criteria imply that the theory can be matched onto SMEFT.
This can be verified by expanding the effective Lagrangian for $H$ we obtained in \cref{eqn:LEFTSingletZ2} in powers of $1/m^2$.
Performing this expansion, we find
\begin{align}
\Lag_\text{Eff} &= |\partial H|^2 + \mu_H^2\s|H|^2 - \lambda_H\s|H|^4 \notag\\[3pt]
&\hspace{15pt} + \frac{1}{4\s \lambda_S} \big(\s m^2 + \kappa\s |H|^2\s \big)^2 - \frac{1}{8}\s \frac{\kappa^2}{\lambda_S\s m^2} \big(\s \partial |H|^2\s \big)^2 + {\cal O}\left( \text{dim-8} \right) \,,
\end{align}
which is a SMEFT expansion.
In addition, we also see that the decoupling limit $m^2\to\infty$ is well behaved, \ie, the curvature $R(h)\to0$, see \cref{eqn:RSingletZ2}.

\subsection*{$\bm{m^2=0}$: Matching Onto HEFT is Required}
When $m^2=0$, the singlet $s$ gets all of its mass from the vev of $H$, see~\cref{eq:msSq}.
Therefore, we expect the effective Lagrangian to be non-analytic at $H=0$, and as such it cannot be mapped onto SMEFT.
This is reflected by the divergence of the scalar curvature at $h_*=-v$:
\begin{align}
R|_{m^2 = 0} =&\,
\frac{N_\varphi\s (N_\varphi-1)\s \kappa}{(\kappa -2\s \lambda_S) (v+h)^2} \toFP \infty\,.
\end{align}
Our LO Criteria imply that the effective Lagrangian with $m^2=0$ can only be matched onto HEFT.

\subsection{Integrating Out a Singlet Scalar at Loop Level}
\label{sec:IntOutDoublet}
Now that we have seen how our LO Criteria work in the context of a concrete example at tree-level, we will turn to the same $\mathbb{Z}_2$ singlet model in the regime where $S$ does not get a vev, so that the leading Wilson coefficients are generated at one-loop order.
We note that matching coefficients up to one-loop and dimension-6 for the more general parameter space of the singlet scalar model have been previously computed~\cite{Ellis:2017jns, Jiang:2018pbd, Haisch:2020ahr}.
The novel result derived here will be the all-orders form factors $F(h)$ and $K(h)$.

We set the singlet quartic coupling to zero for simplicity, such that the Lagrangian is
\begin{equation}
{\cal L}_\text{UV} = {\left| {\partial H} \right|^2} + \mu_H^2\s {|H|^2} - {\lambda_H\s}{|H|^4} + \frac{1}{2}\s S\s\big(\s { - {\partial^2} - {m^2} - \kappa\s {{|H|}^2}}\s \big)\s S \,,
\label{eqn:LUV2HDM}
\end{equation}
Our goal is to integrate out $S$ to obtain the effective Lagrangian for $H$, and then apply our LO Criteria to determine under what conditions one can match onto SMEFT.
The field-dependent mass of $S$ is
\begin{equation}
m_S^2[H] = m^2 + \kappa\s |H|^2 \,,
\end{equation}
and similarly, we expect to be able to match onto SMEFT when $m^2\ne0$.
If $m^2=0$, the mass of $S$ is purely from electroweak symmetry breaking, and we expect to be forced to match onto HEFT.
In what follows, we will see that this expectation is consistent with our LO Criteria.

\subsubsection*{Effective Lagrangian}
Starting with the Lagrangian in \cref{eqn:LUV2HDM}, it is straightforward to derive the tree-level equation of motion for $S$:
\begin{equation}
\big(\s { - {\partial^2} - m^2 - {\kappa}\s{{|{{H}}|}^2}} \s\big)\s{S} = 0 \quad\implies\quad {S_\mathbf{c}} = 0 \,.
\end{equation}
Since $S_\mathbf{c}$ vanishes at tree level, the new physics contribution to the effective Lagrangian for $H$ begins at one loop.
As is well known, the effective Lagrangian can be computed from a functional determinant:
\begin{subequations}\label{eqn:LEFT2HDM}
\begin{align}
\Lag_\text{Eff, tree}(H) &= |\partial H|^2 + \mu_H^2\s |H|^2 - \lambda_H\s |H|^4 \,, \\[5pt]
\int {\dd^4x\, \Lag_\text{Eff, 1-loop}(H)} &= \frac{i}{2}\s\log {\textstyle\det_S}\big(\s {{\partial^2} + m^2 + {\kappa}\s{{|H|}^2}}\s \big) \,.\label{eqn:LEFT2HDM1Loop}
\end{align}
\end{subequations}
No techniques exist to evaluate the functional determinant in~\cref{eqn:LEFT2HDM1Loop} to all orders.
However, we can make progress by organizing the effective Lagrangian as a derivative expansion:
\begin{equation}
\mathcal{L}_\text{Eff} = \mathcal{L}_\text{Eff}^{(0)} + \mathcal{L}_\text{Eff}^{(2)} + \mathcal{O}\big(\partial^4\big) \,, \label{eqn:LEFTDerExpansion}
\end{equation}
where $\mathcal{L}_\text{EFT}^{(k)}$ contains all the terms with $k$ derivatives, we are dropping the ``1-loop'' subscript for brevity, and terms with odd-powers of derivatives do not contribute since we only consider operators with bosonic fields.

In \cref{appsec:TwoDerUniversal}, we work out a \emph{new formalism} that allows one to calculate the two derivative contribution to the effective Lagrangian \emph{to all orders in the fields}, by evaluating the functional determinant directly from the path integral.
Here, we are interested in the potential $V = - \mathcal{L}_\text{EFT}^{(0)}$, and the form factors $K(h)$ and $F(h)$ that multiply the two-derivative terms $\mathcal{L}_\text{EFT}^{(2)}$.
Then we can apply~\cref{eqn:TraceFinal} with the identification $M^2=m^2$ and $U=\kappa\s|H|^2$, such that the terms of $\mathcal{O}([U,\partial_\mu U]) = 0$ identically since $|H|^2$ is a singlet and has a vanishing commutator.\footnote{In the next section, we will present a model where this commutator does not vanish, leading to more involved expressions for the form factors.}
The resulting effective Lagrangian is thus
\begin{subequations}\label{eqn:LEFT022HDM}
\begin{align}
\Lag_\text{Eff}^{\left( 0 \right)} &=  \mu_H^2\s {|H|^2} - \lambda_H\s{|H|^4} + \frac{1}{{{64\s\pi^2}}} {\big(\s {m^2 + {\kappa}\s{{|H|}^2}}\s\big)^2}\bigg( {\ln \frac{{{\mu^2}}}{{m^2 + {\kappa}\s{{|H|}^2}}} + \frac{3}{2}} \bigg) \\[10pt]
\Lag_\text{Eff}^{\left( 2 \right)} &= {| {\partial H} |^2} + \frac{1}{{{ 384\s\pi^2}}} \frac{{\kappa^2}}{{m^2 + {\kappa}\s{{|H|}^2}}}{\big(\s {\partial {{|H|}^2}} \s\big)^2} \,.
\end{align}
\end{subequations}
In this basis, it is already clear by inspection that when $m^2 \neq 0$, this Lagrangian can be expanded in $1/m^2$ and matched onto SMEFT, while when $m^2 = 0$ one encounters non-analytic behavior about $H = 0$.
We will see this intuition play out more rigorously in what follows.

\subsubsection*{Form Factors and Curvature Invariants}
Next, we use Eqs.~\eqref{eqn:Doublethn}-\eqref{eqn:HTohnTwoDerivative} to write \cref{eqn:LEFT022HDM} in terms of $h$ and $\vec n$:
\begin{subequations}
\begin{align}
\Lag_\text{Eff}^{\left( 0 \right)} &=  \frac{1}{2}\s\mu_H^2{( {v_0 + h} )^2} - \frac{1}{4}\s \lambda_H {( {v_0 + h} )^4} \notag\\[3pt]
&\hspace{20pt} + \frac{1}{{{64\s\pi^2}}} \s{\bigg[ {m^2 + \frac{{{\kappa}}}{2}{( {v_0 + h} )^2}} \bigg]^2}\Bigg[ {\ln \frac{{{\mu^2}}}{{m^2 + \frac{{{\kappa}}}{2}{( {v_0 + h} )^2}}} + \frac{3}{2}} \Bigg] \,, \label{eqn:LEFT02HDM} \\[10pt]
\Lag_\text{Eff}^{\left( 2 \right)}&= \frac{1}{2}{( {\partial h} )^2}\Bigg[ {1 + \frac{1}{{{96\s\pi^2}}} \frac{{\kappa^2\s{( {v_0 + h} )^2}}}{{2\s m^2 + {\kappa}{( {v_0 + h} )^2}}}} \Bigg] + \frac{1}{2}\s( {v_0 + h} )^2\s{( {\partial \vec n} )^2} \,, \label{eqn:LEFT22HDM}
\end{align}
\end{subequations}
where the potential $V = - \Lag_\text{Eff}^{\left( 0 \right)}$.
Noting that $v_0=v$, we then identify the form factors by comparing against~\cref{eqn:LHEFT}:
\begin{subequations}\label{eqn:KFV2HDM}
\begin{align}
K(h) &= \sqrt {1 + \frac{1}{{{96\s\pi^2}}} \frac{{\kappa^2\s{( {v + h} )^2}}}{{2\s m^2 + {\kappa}\s{( {v + h} )^2}}}}  \\[10pt]
F(h) &= 1 + \frac{h}{v}\,.\label{eq:F2HDM}
\end{align}
\end{subequations}
We use these to evaluate the scalar curvature $R$ via \cref{eqn:Rgeneral}:
\begin{align}
R &= \frac{{2\s {N_\varphi }}}{{( {v + h} )\s {K^3}}}( {{\partial_h}K} ) + \frac{{{N_\varphi }\s( {{N_\varphi } - 1} )}}{{{( {v + h} )^2}}}\bigg( {1 - \frac{1}{{{K^2}}}} \bigg) \notag\\[5pt]
 &= \frac{1}{{{96\s\pi^2}}} \Bigg(\frac{{2\s{N_\varphi }}}{{{K^4}}}\frac{{2\s m^2\s \kappa^2}}{{ {\big[\s {2\s m^2 + {\kappa}{( {v + h} )^2}}\s \big]^2}}}  + \frac{{{N_\varphi }\s( {{N_\varphi } - 1} )}}{{{K^2}}}\frac{{\kappa^2}}{{\big[\s {2\s m^2 + {\kappa}{( {v + h} )^2}} \s\big]}}\Bigg) \,.
\label{eqn:R2HDM}
\end{align}
Next, we can use these functions as input to our LO Criteria, to explore the ramifications for matching.

\subsection*{$\bm{m^2\neq0}$: Matching Onto SMEFT is Allowed}
From \cref{eq:F2HDM} above, $F(h_*) = 0$ has a solution for $h_*=-v$, which we identified with the candidate $O(4)$ invariant point on the manifold.
Hence, condition 1 in our LO Criteria is satisfied for any choice of $m^2=0$.
Next, we will check the other two conditions assuming $m^2\ne0$.
In this case, the potential in \cref{eqn:LEFT02HDM} and the form factors in \cref{eqn:KFV2HDM} are all real analytic single-argument functions at $h_*=-v$:
\begin{align}
V(h_*) = \frac{1}{{{64\s\pi^2}}}{m^4}\bigg[ {\ln \frac{{{\mu^2}}}{{m^2 }} + \frac{3}{2}} \bigg]\,,\qquad K(h_*) = 1\,,
\end{align}
and therefore condition 2 in our LO Criteria is satisfied.
In addition, the curvature scalar $R$ is also finite at $h_*=-v$:
\begin{equation}
R( h_* ) = \frac{1}{{{192\s\pi^2}}} \frac{{\kappa^2}}{{ m^2}}{N_\varphi}\s( {{N_\varphi} + 1} ) \,,
\end{equation}
which satisfies condition 3.
Therefore, our LO Criteria imply that the theory can be matched onto SMEFT when $m^2\ne0$.
Additionally, \cref{eqn:R2HDM} implies that the curvature $R(h)\to0$ as $m^2\to\infty$, implying that SMEFT will reduce to the SM in the decoupling limit as it must.

These claims can also be explicitly verified using our result in \cref{eqn:LEFT2HDM}. 
When $m^2\ne0$, we can expand the functional determinant in powers of $1/m^2$.
For example, we can apply the universal formula Eq.~(2.54) in~\cite{Henning:2014wua} truncating up to mass-dimension-6,
\begin{align}
{\Lag^\text{1-loop}_{{\text{EFT}}}}(H) =&\,\, \frac{1}{{{64\s\pi^2}}} \Bigg[ m^2\left( {\ln \frac{{{\mu^2}}}{{m^2}} + 1} \right)2{\kappa}{{|H|}^2} + \left( {\ln \frac{{{\mu^2}}}{{m^2}}} \right)\kappa^2{{|H|}^4} \notag\\[8pt]
&\hspace{50pt}+ \frac{{\kappa^3}}{{m^2}}\bigg( {\frac{1}{6}{{\left( {\partial {{|H|}^2}} \right)}^2} - \frac{1}{3}{{|H|}^6}} \bigg) \Bigg] \,.
\label{eqn:LEFT2HDMSMEFT}
\end{align}
which is obviously a well defined SMEFT expansion.
This same result can be derived by expanding \cref{eqn:LEFT022HDM} in $1/m^2$.

\subsection*{$\bm{m^2= 0}$: Matching Onto HEFT is Required}
Now we investigate the situation when $m^2=0$.
Using \cref{eqn:KFV2HDM}, we see that
\begin{align}
{K^2}|_{m^2 = 0} = 1+\frac{\kappa}{96\s\pi^2} \,.
\end{align}
while $F(h)|_{m^2 = 0} = 1 + h/v$ is unchanged.
These are clearly both analytic functions of $h$ about $h_*$.
However, condition 2 in the LO Criteria is violated because the potential is now
\begin{align}
V|_{m^2 = 0} &=  -\frac{1}{2}\s\mu_H^2{( {v + h} )^2} + \frac{1}{4}\s \lambda_H {( {v + h} )^4} \notag\\[3pt]
&\quad - \frac{1}{{{256\s\pi^2}}} \kappa^2 ( {v + h} )^4 \Bigg[ {\ln \frac{{{\mu^2}}}{{\frac{1}{2}\s{\kappa}{( {v + h} )^2}}} + \frac{3}{2}} \Bigg] \,,
\end{align}
which is not real-analytic in $h$ at $h_*=-v$, due to the presence of the logarithm.
Moreover, we can check that condition 3 is also violated by explicitly evaluating the scalar curvature in \cref{eqn:R2HDM} for $m^2 = 0$:
\begin{align}
R|_{m^2 = 0} &= \frac{1}{{{96\s\pi^2}}} \frac{{{N_\varphi }\s( {{N_\varphi } - 1} )}}{{{K^2}}}\frac{{{\kappa}}}{{{( {v + h} )^2}}} \notag\\[7pt]
&= \frac{{{N_\varphi }\s( {{N_\varphi } - 1} )}}{{{( {v + h} )^2}}}\frac{{{\kappa}}}{{{96\s\pi^2} + {\kappa}}} \toFP \infty \,.
\end{align}
This example again illustrates that when a state gets all of its mass from electroweak symmetry breaking, one is required to match onto HEFT.
We see that our LO Criteria hold for an EFT whose non-zero Wilson coefficients first appear at loop level.

\subsection{Integrating Out a Vector-like Fermion at Loop Level}
\label{sec:VectorLikeFermion}
Now that we have worked through the implications of integrating out new scalar particles, we will turn to an example with BSM fermions for completeness.
One new feature  is that the calculation of the form factors is significantly more involved.
In particular, this will be our first example with a non-trivial $F(h)$ form factor, see \cref{eqn:KFVFermionF} below.
Again our intuition will be in concert with our LO Criteria.
When the vector-like mass $M\ne 0$, we can match onto SMEFT, and if $M=0$, these fermions acquire all of their mass from electroweak symmetry breaking, and we must match onto HEFT. Our results are a natural generalization of the effective action arising from integrating out chiral fermions  \cite{DHoker:1984izu}; recent EFT calculations involving vector-like fermions include \cite{Huo:2015exa, Chen:2017hak, Ellis:2020ivx, Angelescu:2020yzf}.

Recall that in setting up the formalism implemented here, we made the simplifying assumption of custodial symmetry, see \eg~\cref{eqn:LHEFT}.
Therefore, we must introduce UV physics that respects $SU(2)_L \times SU(2)_R$, and in particular this implies that we cannot simply add a fourth generation.
Arguably the simplest UV Lagrangian involving BSM fermions that respect custodial symmetry is
\begin{align}
\Lag_\text{UV} &= \bar \psi \s\big( {i\s\slashed \partial  - M - \Xi } \big)\s\psi \notag\\[5pt]
&= {{\bar \psi }_L}\s\big( {i\s\slashed \partial - M} \big)\s{\psi_L} + {{\bar \psi }_R}\s\big( {i\s\slashed \partial  - M} \big)\s{\psi_R} - \big( {y\s{{\bar \psi }_L}\s\Sigma {\psi_R} + \text{h.c.}} \big)\,,
\label{eqn:LUVFermion}
\end{align}
where $\psi_L$ and $\psi_R$ transform as doublets under $SU(2)_L$ and $SU(2)_R$ respectively.
We can write this in a manifestly custodial invariant form using
\begin{equation}
\psi \equiv \mqty(\psi_L \\ \psi_R)  \,,\qquad\Sigma \equiv \mqty(\tilde{H} & H) \,,\qquad  \Xi \equiv \mqty( 0 & y\s\Sigma \\ y^*\Sigma^\dag & 0)\,,
\end{equation}
with $\tilde H \equiv i\s\sigma^2 H^*$, so that under a custodial symmetry transformation,
\begin{align}
\Sigma \,\to\, {U_L}\s\Sigma\s U_R^\dag  \,,\qquad  \psi_L = \mqty(\psi_{La} \\ \psi_{Lb}) \,\to\, U_L\s \psi_L  \,,\qquad  \psi_R = \mqty(\psi_{Ra} \\ \psi_{Rb}) \,\to\, U_R\s \psi_R \,,
\end{align}
where $\psi_{La}, \psi_{Lb}, \psi_{Ra}, \psi_{Rb}$ are all Dirac fermions.
Next, we will compute the contribution to the effective Lagrangian for $H$ that results from integrating out $\psi$.

\subsubsection*{Effective Lagrangian}
To begin, we note that there are no tree-level contributions to matching from integrating out $\psi$.
This is obvious, since we are investigating the impact of heavy fermions on the scalar Lagrangian.
One can also see this by deriving the tree-level equation of motion for $\psi$ from the Lagrangian in \cref{eqn:LUV2HDM}:
\begin{equation}
\big( {i\s\slashed \partial  - M - \Xi} \big)\s\psi = 0 \quad\implies\quad {\psi_\mathbf{c}} = 0 \,.
\end{equation}
Since $\psi_\mathbf{c}$ vanishes at tree level, the new physics contribution to the effective Lagrangian for $H$ begins at one loop.

We derive the one-loop effective Lagrangian using
\begin{align}
\int {\dd^4 x\, \Lag_\text{Eff}} &\supset  - i\ln \det \big( {i\s\slashed \partial  - M - \Xi} \big) \notag\\[3pt]
 &= - \frac{i}{2}\s\Big[ \Tr \ln \big( i\s\slashed \partial - M - \Xi \big) + \Tr \ln \big( - i\s\slashed \partial - M - \Xi \big) \Big] \notag\\[5pt]
 &= - \frac{i}{2} \Tr \ln \Big[ {\partial^2} + M^2 - \big( i\s\slashed \partial \Xi \big) + 2\s M\s\Xi + \Xi^2 \Big] \notag\\[5pt]
 &= - \frac{i}{2} \Tr \ln \big( \partial^2 + M^2 + U \big)\, ,
\end{align}
where in the last line we have implicitly defined
\begin{equation}
U \equiv  - i\s\slashed \partial\s \Xi + 2\s M\s\Xi + {\Xi^2} = {U^{(0)}} + {U^{(1)}}  \,,
\end{equation}
where we organize $U$ as a derivative expansion
\begin{equation}
U^{(0)} = \Xi^2 + 2\s M\s \Xi  \,,\qquad \qquad  U^{(1)} = - i\s \slashed \partial\s \Xi \,.
\end{equation}

We now have everything we need to apply the new technology developed in~\cref{appsec:TwoDerUniversal}.
In particular, we apply \cref{eqn:TraceFinal} to obtain the first two terms in the derivative expansion of the effective Lagrangian\footnote{The superscript notation on the $T$'s tracks the explicit derivatives.  There can additionally be implicit derivatives within $U$, as will occur in what follows.}
\begin{equation}
\int {\dd^4 x\, \Lag_\text{Eff}} \supset  - \frac{1}{2}\s T(U) \supset  - \frac{1}{2}\s{T^{(0)}}( U ) - \frac{1}{2}\s{T^{(2)}}\big({U^{(0)}}\big) \,,
\end{equation}
\clearpage
\noindent with
\begin{subequations}
\begin{align}
T^{(0)}\big(U\big) &= \int\dd^4 x\, \frac{1}{16\s\pi^2} \tr \bigg[ \frac{1}{2}\s \big( M^2 + U \big)^2 \s\bigg( \ln \frac{\mu^2}{M^2+U} + \frac{3}{2} \bigg) \bigg] \,, \\[10pt]
T^{(2)}\big(U^{(0)}\big) &= \int \dd^4 x\,  \frac{1}{16\s\pi^2} \int_0^\infty \dd t\, \frac{1}{4}\s t^2\s \tr \Big[ \partial_\mu U^{(0)} \left( t+M^2+U^{(0)} \right)^{-2} \nonumber \\[5pt]
                 & \hspace{155pt} \times \partial^\mu U^{(0)} \big( t+M^2+U^{(0)} \big)^{-2} \Big] \, ,
\end{align}
\end{subequations}
In comparison with \cref{sec:IntOutDoublet} where we integrated out a scalar at one-loop, the fermionic case is more involved.
In particular, complications arise because of the non-trivial commutators $\left[U^{(0)}, U^{(1)}\right]\ne0$ and $\left[U^{(0)}, \partial_\mu U^{(0)}\right]\ne0$.
Therefore, we must be careful when expanding $T^{(0)}\big(U^{(0)}+U^{(1)}\big)$, which requires including up to two factors of $U^{(1)}$, and the most general form of \cref{eqn:TraceFinal} is needed to evaluate $T^{(2)}\big(U^{(0)}\big)$.
We also note that some simplifications occur since
\begin{subequations}
\begin{align}
{\Xi^2} &= |y|^2{|H|^2} \,, \qquad\qquad {( {\partial\s \Xi} )^2} = |y|^2{| {\partial H} |^2} \,,
\end{align}
\end{subequations}
are both proportional to the identity.
Additionally, odd powers of $\Xi$ and $U^{(1)}$ have vanishing traces.
Carrying out the expansion, we find
\begin{subequations}\label{eqn:T0T2Fermion}
\begin{align}
{T^{(0)}}( U ) &\supset \int \dd^4 x\, \dfrac{1}{{{16\s\pi^2}}} \Bigg\{ 4\s{( {M + \xi } )^4}\bigg[ {\ln \dfrac{{{\mu^2}}}{{{( {M + \xi } )^2}}} + \dfrac{3}{2}} \bigg] \notag\\[3pt]
          &\hspace{90pt} + 4\s{( {M - \xi } )^4}\bigg[ {\ln \dfrac{{{\mu^2}}}{{{{( {M - \xi } )}^2}}} + \dfrac{3}{2}} \bigg] \notag\\[5pt]
          &\hspace{90pt} - \bigg[ {1 - \dfrac{{{M^2} + {\xi^2}}}{{4\s M\s \xi }}\ln \dfrac{{{( {M + \xi } )^2}}}{{{( {M - \xi } )^2}}} + \dfrac{1}{2}\s\ln \dfrac{{{\mu^4}}}{{{( {{M^2} - {\xi^2}} )^2}}}} \bigg]\s 8\s ( {\partial\s \Xi} )^2 \notag\\[5pt]
          &\hspace{90pt} + \bigg[ {1 - \dfrac{{{M^2} + {\xi^2}}}{{4\s M\s\xi }}\ln \dfrac{{{( {M + \xi } )^2}}}{{{( {M - \xi } )^2}}}} \bigg]\dfrac{2}{{{\xi^2}}}{( {\partial\s {\Xi^2}} )^2} \Bigg\} \,, \\[20pt]
{T^{(2)}}\big( {{U^{(0)}}} \big) &= \int \dd^4 x\, \dfrac{1}{{{16\s\pi^2}}} \Bigg\{ \bigg[ {{M^2} + {\xi^2} - \dfrac{{{( {{M^2} - {\xi^2}} )^2}}}{{4\s M\s\xi }}\ln \dfrac{{{( {M + \xi } )^2}}}{{{( {M - \xi } )^2}}}} \bigg]\s\dfrac{2}{{{\xi^2}}}\s{( {\partial\s \Xi} )^2} \notag\\[5pt]
          &\hspace{85pt} + \bigg[ { - {M^2} + \dfrac{5}{3}\s{\xi^2} + \dfrac{{{( {{M^2} - {\xi^2}} )^2}}}{{4\s M\s\xi }}\ln \dfrac{{{( {M + \xi } )^2}}}{{{{( {M - \xi } )}^2}}}} \bigg]\s \dfrac{1}{{2\s{\xi^4}}}{( {\partial\s {\Xi^2}} )^2} \Bigg\} \,, \notag\\
\end{align}
\end{subequations}
with
\begin{equation}
\xi \equiv \sqrt{\Xi^2} = \sqrt{|y|^2\s |H|^2} = \sqrt{\frac{1}{2}\s |y|^2\s (v_0+h)^2} \,.
\end{equation}
The resulting effective Lagrangian up to second order in the derivative expansion is
\begin{subequations}\label{eqn:LEFT02Fermion}
\begin{align}
\Lag_\text{Eff}^{(0)} &= -\dfrac{{1}}{{{8\s\pi^2}}} \bigg[ ( {M + \xi } )^4 \bigg( {\ln \dfrac{{{\mu^2}}}{{{( {M + \xi } )^2}}} + \dfrac{3}{2}} \bigg) + ( {M - \xi } )^4 \bigg( {\ln \dfrac{{{\mu^2}}}{{{( {M - \xi } )^2}}} + \dfrac{3}{2}} \bigg) \bigg] \,, \notag\\[3pt]
\\
\Lag_\text{Eff}^{(2)} &= \dfrac{1}{{{16\s\pi^2}}} \Bigg\{ - \Bigg[ M^2 - 3\s\xi^2 - 2\s\xi^2\ln \dfrac{{{\mu^4}}}{{{( {{M^2} - \xi^2} )^2}}} \notag\\
          &\hspace{80pt} - \dfrac{{{M^4} - 6\s{M^2}\s\xi^2 - 3\s\xi^4}}{{4\s M\s\xi }}\ln \dfrac{{{( {M + \xi } )^2}}}{{{( {M - \xi } )^2}}} \Bigg] \dfrac{{{| {\partial H} |^2}}}{{{{|H|}^2}}} \notag\\
          & + \Bigg[ {M^2} - \dfrac{{17}}{3}\s\xi^2 - \dfrac{{{M^4} - 6\s{M^2}\s\xi^2 - 3\s\xi^4}}{{4\s M\s\xi }}\ln \dfrac{{{( {M + \xi } )^2}}}{{{( {M - \xi } )^2}}} \Bigg]\s\dfrac{1}{{4{{|H|}^4}}}{\big(\s {\partial {{|H|}^2}}\s \big)^2} \Bigg\} \,.
\end{align}
\end{subequations}
Next, we will identify the form factors and explore the implications for matching.

\subsubsection*{Form Factors and Curvature Invariants}
We use Eqs.~\eqref{eqn:Doublethn}-\eqref{eqn:HTohnTwoDerivative} to write \cref{eqn:LEFT02Fermion} in terms of $h$ and $\vec n$, and compare this to~\cref{eqn:LHEFT} to read off the form factors
\begin{subequations}\label{eqn:KFVFermion}
\begin{align}
K(h) &= \sqrt {1 + \frac{|y|^2}{8\s\pi^2}\Bigg[ \ln \frac{\mu^4}{\big( M^2 - \xi^2 \big)^2} - \frac{4}{3} \Bigg]} \,, \\[10pt]
v\s F(h) &= (v_0 + h) \s\Bigg\{ 1 + \frac{|y|^2}{8\s\pi^2} \Bigg[ \ln \frac{\mu^4}{\big( M^2 - \xi^2 \big)^2} + \frac{3}{2}  - \frac{M^2}{2\s\xi^2} \bigg[ 1 - \frac{M}{4\s\xi} \ln \bigg( \frac{M+\xi}{M-\xi} \bigg)^2 \bigg] \notag\\[5pt]
 &\hspace{70pt} - \bigg( 1 + \frac{\xi^2}{2M^2} \bigg)\s \frac{3\s M}{4\s\xi} \ln \bigg( \frac{M+\xi}{M-\xi} \bigg)^2 \Bigg] \Bigg\}^{1/2} \,. \label{eqn:KFVFermionF}
\end{align}
\end{subequations}
We see that the form factor $F(h)$ has a zero at $h_*=-v_0$, which is the candidate $O(4)$ invariant point, so condition 1 in our LO Criteria holds.

Now we have the required ingredients to evaluate the scalar curvature $R$ using \cref{eqn:Rgeneral}, which is a straightforward exercise.
However, the non-trivial form of $F$ in \cref{eqn:KFVFermion}, results in a lengthy and unenlightening expression for $R$.
Hence, we will not provide the full expression for $R$, but will instead evaluate it at $h = h_*$ for $M\neq 0$, and taking $M\rightarrow 0$ in what follows to check the compatibility with our LO Criteria.

\subsubsection*{When Matching Onto SMEFT is Allowed}
First, we will explore the case with $M\neq 0$.
As noted above, using \cref{eqn:KFVFermionF} it is clear that $F(h_*)=0$ has a solution at $h_* = -v_0$.
It is also straightforward to check that all the form factors in \cref{eqn:KFVFermion} are real-analytic single-argument functions at the candidate $O(4)$ fixed point.\footnote{Note the Taylor expansion
\begin{equation}
\frac{M}{4\s \xi} \ln \frac{(M+\xi)^2}{(M-\xi)^2}  = \sum\limits_{k = 0}^\infty{ \frac{1}{2\s k+1} \bigg(\frac{\xi}{M}\bigg)^{2k} } = 1 + \frac{1}{3} \frac{\xi^2}{M^2} + \frac{1}{5}\frac{\xi^4}{M^4} + \cdots \,.
\end{equation}
}
Then we can compute the scalar curvature and evaluate it at $h_*$ by plugging \cref{eqn:KFVFermion} into \cref{eqn:Rgeneral}:
\begin{equation}
R(h=-v_0) = N_\varphi \left( N_\varphi + 1 \right) \frac{{{{\left( {|y|^2} \right)}^2}}}{{{16\s\pi^2}}}\frac{16}{{5{M^2}}}{\left[ {1 + \frac{{4|y|^2}}{{{16\s\pi^2}}}\left( {\ln \frac{{{\mu^2}}}{{{M^2}}} - \frac{2}{3}} \right)} \right]^{ - 2}} \,,
\label{eqn:RFermion}
\end{equation}
which is finite.
Therefore, conditions 2 and 3 in our LO Criteria hold, which implies that this theory can be matched onto SMEFT.
This can be explicitly verified by expanding \cref{eqn:LEFT02Fermion} up to mass-dimension-8, giving
\begin{subequations}\label{eqn:SMEFTFermion}
\begin{align}
\Lag_\text{EFT}^{(0)} &\supset- \dfrac{{ 1}}{{{4\s\pi^2}}} \Bigg[ {M^4}\s\bigg( {\ln \dfrac{{{\mu^2}}}{{{M^2}}} + \dfrac{3}{2}} \bigg) + 6\s{M^2}\s|y|^2\s {|H|^2}\bigg( {\ln \dfrac{{{\mu^2}}}{{{M^2}}} + \dfrac{1}{3}} \bigg) \notag\\[3pt]
&\hspace{40pt} + {{|y|^4} }\s{|H|^4}\s\bigg( {\ln \dfrac{{{\mu^2}}}{{{M^2}}} - \dfrac{8}{3}} \bigg) + \dfrac{{{{|y|^6}}}}{{15\s{M^2}}}{|H|^6} + \dfrac{{{{{|y|^8} }}}}{140\s M^4}\s{|H|^8} \Bigg] \,, \\[10pt]
\Lag_{{\text{EFT}}}^{\left( 2 \right)} &\supset \dfrac{1}{{{8\s\pi^2}}} \Bigg\{ \left[ {2\s\ln \dfrac{{{\mu^2}}}{{{M^2}}} - \dfrac{4}{3} - \dfrac{{2\s |y|^2}}{{5\s {M^2}}}{{|H|}^2} - \dfrac{{2\s{{ {|y|^4} }}}}{{35\s{M^4}}}\s{{|H|}^4}} \right]\s|y|^2{| {\partial H} |^2} \notag\\[3pt]
&\hspace{40pt} + \bigg( {\dfrac{3}{{5\s{M^2}}} + \dfrac{{9\s|y|^2}}{{35\s{M^4}}}{{|H|}^2}} \bigg){ {|y|^4} }{\big( {\partial {{|H|}^2}} \big)^2} \Bigg\} \,,
\end{align}
\end{subequations}
which is a well defined SMEFT expansion, up to the order specified.

\subsubsection*{When Matching Onto HEFT is Required}
When $M=0$, the form factors and potential become
\begin{subequations}
\begin{align}
{\left. {K( h )} \right|_{M = 0}} &= \sqrt {1 + \frac{{|y|^2}}{{{4\s\pi^2}}}\Bigg[ {\ln \frac{{2{\mu^2}}}{{|y|^2{{( {v_0 + h} )}^2}}} - \frac{2}{3}} \Bigg]} \,, \\[8pt]
{\left. {v\s F( h )} \right|_{M = 0}} &= ( {v_0 + h} )\sqrt {1 + \frac{{|y|^2}}{{{4\s\pi^2}}}\ln \frac{{2{\mu^2}}}{{|y|^2\s{{( {v_0 + h} )}^2}}}} \,, \\[8pt]
\left. V(h) \right|_{M=0} &= \frac{|y|^2}{16\s\pi^2}(v_0+h)^4 \Bigg[ \ln \frac{2\mu^2}{|y|^2\s(v_0+h)^2} + \dfrac{3}{2} \Bigg] \,.
\end{align}
\end{subequations}
Now due to the appearance of $\ln\left(v_0+h\right)$, they are all non-analytic at $h=h_*$, violating condition 2 in our LO Criteria.
We can also compute the Ricci curvature with $M = 0$:
\begin{align}
R|_{M = 0} &= \frac{|y|^2}{\pi^2} \frac{2}{(v_0+h)^2\s K^4}\bigg[1- \frac{5\s |y|^2}{48\s\pi^2}\Bigg(1+\frac{|y|^2}{4\s\pi^2} \log\frac{2\s \mu^2}{|y|^2\s (v_0+h)^2}\bigg)^{-1}\Bigg]\,,
\end{align}
where we have taken $N_\phi = 3$ to simplify the expression.
It is straightforward to check that
\begin{align}
R|_{M = 0}\toFP \infty\,.
\end{align}
Therefore, our criterion asserts that this theory must be matched onto HEFT.

Finally, to build intuition, we write the effective Lagrangian in~\cref{eqn:LEFT02Fermion} for the case when $M=0$:
\begin{subequations}
\begin{align}
{\left. {\Lag_{{\text{Eff}}}^{\left( 0 \right)}} \right|_{M = 0}} &= -\frac{|y|^4}{{{4\s\pi^2}}}\bigg[ |H|^4\bigg( {\ln \frac{{{\mu^2}}}{{|y|^2\s{{|H|}^2}}} + \frac{3}{2}} \bigg) \bigg] \,, \\[10pt]
{\left. {\Lag_{{\text{Eff}}}^{\left( 2 \right)}} \right|_{M = 0}} &= \frac{|y|^2}{{{4\s\pi^2}}}\Bigg[ {{\ln\bigg(  \frac{{{\mu^2}}}{{|y|^2{{|H|}^2}}}} \bigg){{| {\partial H} |}^2} - \frac{{1}}{{{{6\s|H|}^2}}}{{\big(\s {\partial {{|H|}^2}}\s \big)}^2}} \Bigg] \,.
\end{align}
  \label{eq:LEffFermionMeq0}%
\end{subequations}
These expressions are non-analytic about the origin $H=0$.
Keeping in mind that this can not be used to make any rigorous claims since it suffers from the ambiguities associated with field redefinitions, we see the relation between the results of our criterion that utilizes curvature invariants and the non-analyticity of this effective Lagrangian expressed using $H$.
We conclude that \cref{eq:LEffFermionMeq0} must be matched onto HEFT.

This completes our exploration of examples from integrating out BSM states that get all of their mass from electroweak symmetry breaking.
We have seen that models with new scalars and/or fermions which generate Wilson coefficients at tree and/or loop level can yield effective Lagrangians with conical singularities at the putative electroweak symmetric point in field space.
In the next section, we will explore the other perturbative scenario where HEFT is required, namely when there are BSM sources of electroweak symmetry breaking.

\section{When HEFT is Required: BSM Symmetry Breaking}
\label{sec:BSMSymBreak}

At this point, we have explored a number of concrete examples to demonstrate that HEFT is required if one integrates out states that get all of their mass from electroweak symmetry breaking.
The LO criteria for SMEFT were violated by singularities at the putative $O(4)$ invariant point on the manifold, \ie, the Ricci curvature was ill defined at $h=h_\star$.
In this section, we turn to the other perturbative situation where HEFT is required -- when one is integrating out states that are associated with new sources of symmetry breaking.
In these scenarios, we will show that there is an obstruction to reaching the $O(4)$ invariant point. This manifests the expectation that extra electroweak symmetry breaking gives rise to an EFT manifold that does not contain an $O(4)$ invariant point.

Of course, the two situations have a great deal in common; in both cases, the resulting HEFT exhibits unitarity violation by $4\s \pi\s v$ whose resolution is associated with the appearance of new states below this scale. However, the two cases remain distinguished by the geometry: in the case of massless particles, there are singularities associated with the $O(4)$ invariant point on the EFT manifold, while in the case of extra symmetry breaking, the $O(4)$ invariant point is absent entirely. As such, we treat the two cases separately.

To our knowledge, no matching calculations appear in the literature that include the impact of extra symmetry breaking on the zero- and two-derivative matching terms to all orders in the fields.
All calculations will be performed by solving the EOMs for the fields expressed in the unbroken phase; by being careful to enforce that we are then taking the particular solution of interest when performing the matching calculation, we fully capture the symmetry breaking effects in analogy with the derivation in~\cref{sec:SingletModel} above.

Due to the additional complexity inherent to working with this class of models, we will first work through the details using a toy example.
In particular, this will provide a platform to introduce the notion of the ``unitary basis'' in the UV theory, which yields a simpler EFT description when integrating out BSM states.
We will then turn to the simplest extended scalar sectors of phenomenological relevance, when we add a second Higgs doublet (\cref{sec:the_two_higgs_doublet_model}) or a Higgs triplet (\cref{appsec:Triplet}).
Since the triplet model breaks custodial symmetry, we will not analyze it in full detail using the tools developed in this paper.
However, there are still many interesting aspects of this model that we will be able to highlight.

\subsection{Abelian Toy Model}
\label{subsec:2AHM}

In this section, we will explore a simple Abelian model that has two sources of symmetry breaking.
This will provide us with a platform to discuss the implications of the EFT submanifold discussion in \cref{sec:eftsubmani}.
While a $U(1) \simeq SO(2)$ symmetry is significantly simpler than the non-Abelian $O(4)$ symmetry of the custodially symmetric SM, we will see that our intuition plays out as expected -- the low energy description when integrating out a state that contributes to symmetry breaking must be expressed as HEFT.

\subsubsection{UV Model}
To write down the model, we introduce two Abelian Higgs fields $H_A$ and $H_B$, which transform under a $U(1)$ symmetry with charge assignments:\\[-40pt]
\begin{center}
\begin{equation}
\renewcommand{\arraystretch}{1.8}
\setlength{\arrayrulewidth}{.3mm}
\setlength{\tabcolsep}{2 em}
\begin{tabular}{c|c}
\text{Field} & $Q$ \\
\hline
$H_A$ & $+2$\\
$H_B$ & $+1$
\end{tabular}
\label{eq:Charges}
\end{equation}
\end{center}

The most general renormalizable potential allowed by the charge assignment in \cref{eq:Charges} is
\begin{equation}
  \mathcal{L} = {\left| {\partial {H_A}} \right|^2} + {\left| {\partial {H_B}} \right|^2} - V \,,
\label{eq:2AHMLag}
\end{equation}
where 
\begin{align}
  V &= m_A^2\s | H_A |^2 + m_B^2\s | H_B |^2 + \lambda_A\s | H_A |^4 + \lambda_B\s | H_B |^4 \notag\\[7pt]
    &\hspace{20pt} + 2\s \kappa\s  | H_A |^2  | H_B |^2 + \Big[ \mu\s H_A\s \big(H_B^*\big)^2 + \hc \Big] \,,
    \label{eqn:HABpotential}
\end{align}
and we enforce that $\lambda_A,\lambda_B>0$ and $\lambda_A\s \lambda_B > \kappa^2$ to ensure that the potential is bounded from below. 

\subsubsection{Matching in the Unitary Basis}
Our goal is to integrate out $H_A$, leaving behind an EFT for $H_B$. 
The calculation is simplified by making a UV field redefinition for $H_A$ that takes us to the ``unitary basis.''  
The idea is to parameterize $H_A$ as an arbitrary rescaling and rephasing of $H_B^2$. 
Generically, the unitary basis is obtained by parameterizing the heavy field as a rescaling and rephasing of a polynomial of the light field that transforms in the representation of the heavy fields. 
For the example at hand, we let
\begin{subequations}
\begin{align}
  H_B &= \frac{1}{\sqrt{2}}\s r\s e^{i\s \pi} \, , \\[5pt]
  H_A &= \sqrt{2}\s\frac{f}{r^2} e^{i\s \beta}\s H_B^2 =  \frac{1}{\sqrt{2}}\s f\s e^{i\s \beta + 2\s i\s \pi } \, ,
\label{eq:abelfbetaparam}%
\end{align}
\end{subequations}
where the degrees of freedom are now the real fields $r,f,\pi$ and $\beta$.

Writing the UV Lagrangian~\cref{eq:2AHMLag} in terms of the unitary basis, we find
\begin{equation}
  \mathcal{L} =  \frac{1}{2}\s (\partial f)^2 + \frac{1}{2}\s f^2\s \big(\partial (\beta+ 2\s \pi)\big)^2 + \frac{1}{2} (\partial r)^2 + \frac{1}{2}\s r^2 (\partial \pi)^2 - V \,,  \label{eqn:abeluvfbetaparam}
\end{equation}
where the potential~\cref{eqn:HABpotential} is 
\begin{equation}
  V =  \frac{1}{2}\s m_A^2\s f^2 + \frac{1}{2}\s m_B^2\s r^2 + \frac{1}{4}\s \lambda_A\s f^4 + \frac{1}{4}\s \lambda_B\s r^4 + \frac{1}{2}\s \kappa\s r^2\s f^2 + \frac{1}{\sqrt{2}}\s f\s r^2\s \Re \big( \mu\s e^{i\s \beta} \big) \,. \label{eqn:abeluvVparam}
\end{equation}
Next, we integrate out $f$ and $\beta$. 
To obtain the EFT Lagrangian up to two-derivative order, we only need to solve the EOM for $f$ and $\beta$ to zero derivative order, as explained in \cref{sec:eftsubmani}.
Hence, we simply need the variation of the potential:
\begin{subequations}\label{eqn:fbetaEOM}
\begin{align}
  0= \frac{\partial V}{\partial f} &= \big(m_A^2+\kappa\s r^2\big)\s f + \lambda_A\s f^3 + \frac{1}{\sqrt{2}}\s r^2\s \Re\big(\mu\s e^{i\s \beta}\big) \,, \\[8pt]
  0= \frac{\partial V}{\partial \beta} &= - \frac{1}{\sqrt{2}}\s r^2\s f\s \Im\big(\mu\s e^{i\s \beta}\big) \,.
\end{align}
\end{subequations}
The $\beta$ EOM is solved by the constants (in the equation immediately below, $\pi$ denotes Archimedes' constant, not the phase of $H_B$)
\begin{equation}
  \beta = -\arg \mu\qquad \text{or} \qquad  \beta = -\arg \mu + \pi \,,
  \label{eqn:abelbetasol}
\end{equation}
and without loss of generality we choose the latter. 
Noting that
\begin{equation}
  \frac{\partial^2 V}{\partial \beta^2} = - \frac{1}{\sqrt{2}}\s r^2\s f\s \Re\big(\mu\s e^{i\s \beta}\big) \,,
  \label{eqn:massbeta}
\end{equation}
we see that $\beta$ fluctuations have positive mass when $f>0$, and the global minimum (which we identify with the ``observed vacuum'' in this toy model) is also in the region where $f>0$.\footnote{Choosing the former solution for $\beta$ in \cref{eqn:abelbetasol} will yield the same EFT solutions, but with the opposite sign for $f$. Note that this corresponds to our freedom to redefine $H_A \to -H_A$ in the original potential \cref{eqn:HABpotential}.}
Subbing the $\beta$ solution into the $f$ EOM yields
\begin{equation}
  \frac{\partial V}{\partial f} = \big(m_A^2+\kappa\s r^2\big)\s f + \lambda_A\s f^3  - \frac{1}{\sqrt{2}}\s r^2\s |\mu| = 0 \,,
  \label{eqn:depcubabelian}
\end{equation}
which is a depressed cubic equation for $f$ that has either one or three real $f$ solutions for any given $r$.

At this stage, the benefits of writing $H_A$ in the unitary basis~\cref{eq:abelfbetaparam} are clear; these features will have an analogies for the doublet and triplet models studied next. 
Specifically, the matching of the radial and angular modes nicely factorises in this basis. 
The classical solution for $\beta$ is simply a constant, and the general problem of tree-level matching reduces to solving the $f$ EOM, which is at most a cubic for any renormalizable UV model. 
Moreover, fluctuations of $\beta$ about its classical solution are always UV mass eigenstates.

Note that, physically, a constant $\beta$ solution is necessary to define the action of the $U(1)$ symmetry on the EFT submanifold.
It implies the relative phases of $H_A$ and $H_B^2$ are locked along the EFT submanifold, such that all points on this submanifold map onto each other under the action of the $U(1)$ symmetry defined in the UV.\footnote{While convenient, the unitary basis is not necessary to make this phase alignment manifest.  It is also apparent in the original parametrization due to the vanishing variation of the potential in \cref{eqn:HABpotential}.} 
This means that we can, in principle, find coordinates for the EFT manifold on which the $U(1)$ is non-linearly realized.
The resulting EFT can be written in terms of a complex field on which the $U(1)$ is linearly realized only if the EFT submanifold includes the $U(1)$ invariant point, as we discuss presently.

The solution $f(r)$ of the cubic equation~\cref{eqn:depcubabelian} that passes through the global minimum is an analytic even function for the full range of $r$. 
Its expansion about $r=0$ depends on the sign of $m_A^2$, in that
\begin{equation}
  f(r) = \begin{cases}
    \dfrac{|\mu|^2\s r^2}{m_A^2\s \sqrt{2}} + \mathcal{O}\big(r^4\big) & \text{ if } m_A^2>0  \\[20pt]
    \sqrt{\dfrac{-m_A^2}{\lambda_A}} + \mathcal{O}\big(r^2\big)  & \text{ if } m_A^2<0  \,.
  \end{cases}
  \label{eq:frCases}
\end{equation}
Substitution $f(r)$ and $\beta = -\arg \mu + \pi$ into the UV Lagrangian~\cref{eqn:abeluvfbetaparam} yields the tree-level effective Lagrangian up to two derivative order, that is analytic and even in $r$:
\begin{equation}
  \mathcal{L}_\text{Eff} = \frac{1}{2}\s \Bigg[ 1 + \left(\frac{\dd f}{\dd r} \right)^2 \Bigg] \s(\partial r)^2 + \frac{1}{2}\s \big[r^2 + 4\s f^2(r)\big] \s(\partial \pi)^2 - V \,.
   \label{eqn:HEFT2ABH}
\end{equation}

When $m_A^2>0$, the effective Lagrangian~\cref{eqn:HEFT2ABH} can be matched onto SMEFT. 
It has a zero in the form factor 
\begin{align}
v^2\s F(r)^2 = r^2 + 4\s f^2\, ,
\end{align}
 at the point $r=f(r=0)=0$; it has inherited the $U(1)$ invariant point from the UV manifold. 
 Given that it is manifestly even in $r$, the Lagrangian may be rewritten in SMEFT coordinates $H_B$ by the substitutions $r = |H_B|$, $\pi = \arg(H_B)$, in concert with \cref{sec:WhatIsHEFT}.

If, on the other hand, $m_A^2 < 0$, then we should use the other solution in~\cref{eq:frCases}.  
Then as $f(0) \neq 0$, the form factor of the angular Goldstone mode $\pi$ does not have a zero. 
The resulting EFT lacks a $U(1)$ invariant point, violating the Criteria, and thus it is necessary to represent it using HEFT.

Additionally, the HEFT-like behavior when $m_A^2 < 0$ is in accordance with the arguments of \cref{sec:eftsubmani}, as
\begin{equation}
  \frac{\partial^2 V}{\partial f^2} = m_A^2 +\kappa\s r^2 + 3\s \lambda_A\s f^2 \,,
  \label{eqn:massf}
\end{equation}
will be positive at the global minimum but negative at the UV fixed point where $(r,f)=(0,0)$.
This implies that the UV invariant point does not belong to the same connected region as the global minimum, where the mass matrix of the UV fluctuations $f,\beta$ is positive definite. 
We conclude that no analytic EFT submanifold connects the two points.

\subsubsection{Matching in the Mass Basis}
Given the novelty of the unitary basis, we will briefly discuss how the same conclusions follow from a more standard approach to matching using the basis where the fields are mass eigenstates about the global minimum.
There are two CP even mass eigenstates $h$, $H$, and two CP odd eigenstates $\pi$ and $\beta$, where $\pi$ is massless. 
Our goal is to integrate out $H$ and $\beta$.

\begin{figure}[b!]
  \centering
  \includegraphics[width=0.3\textwidth]{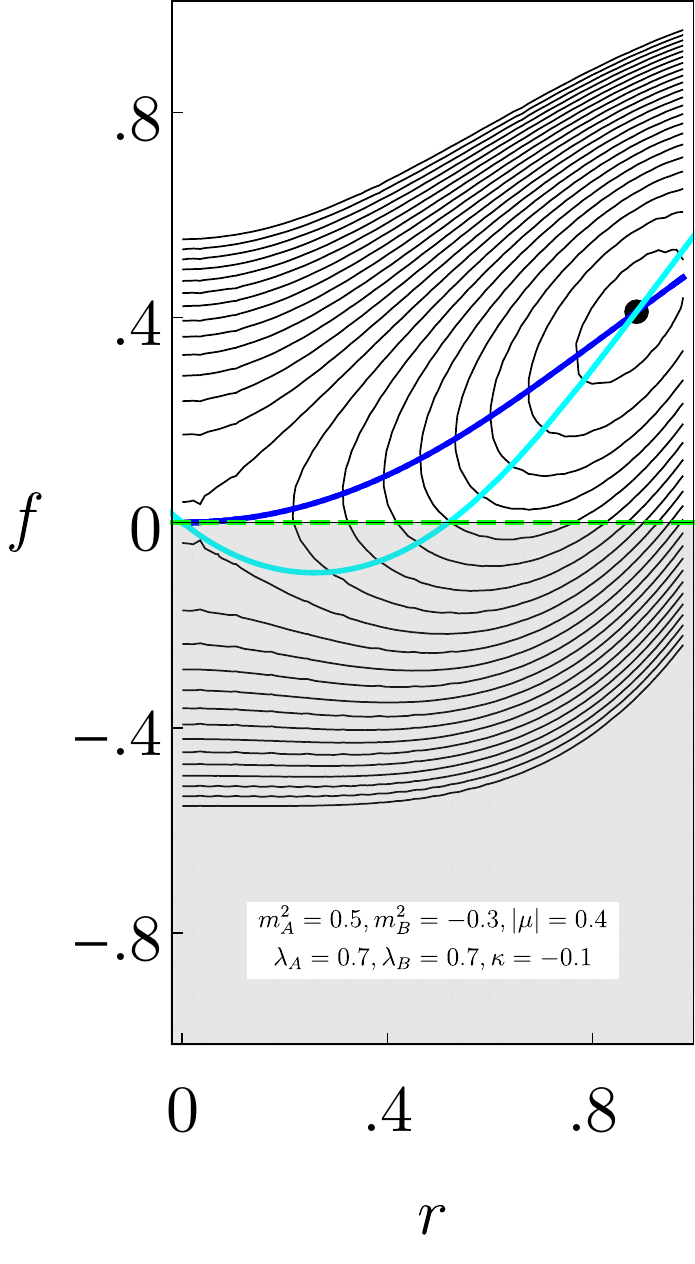}
  \hspace*{3ex}
  \includegraphics[width=0.3\textwidth]{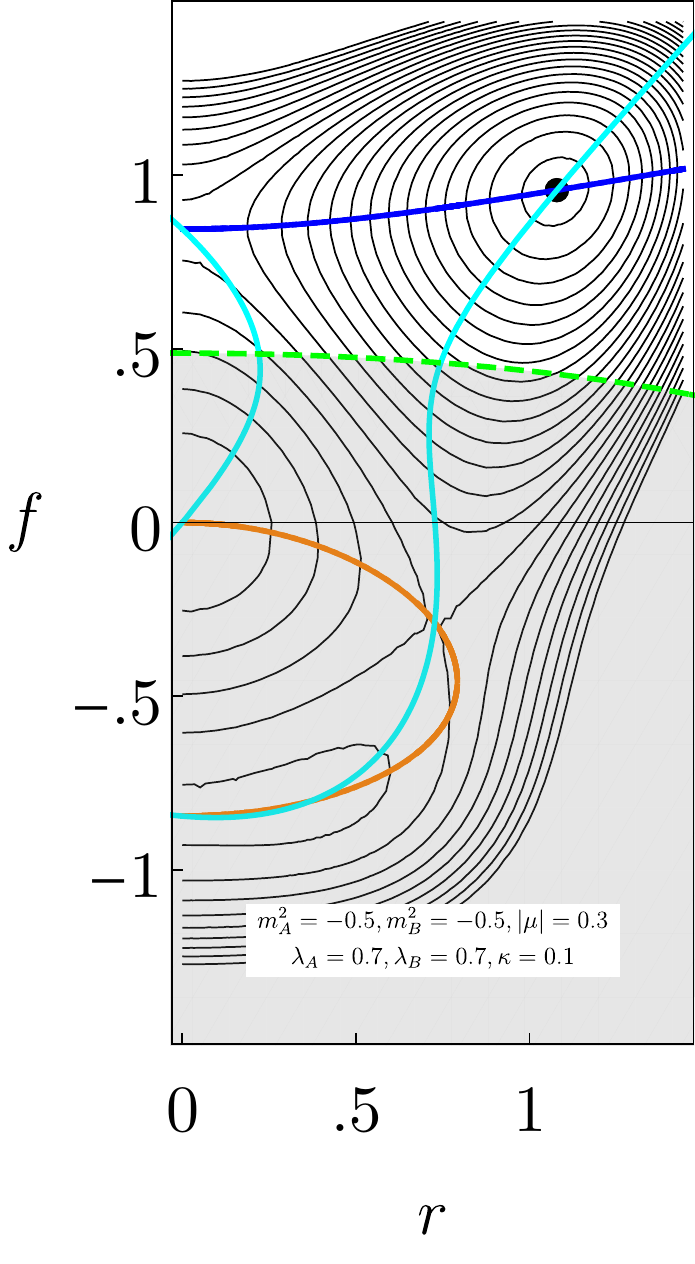}
\caption{EFT submanifolds in the Abelian toy model obtained by matching in the unitary basis and the mass basis for two representative sets of parameters, one with $m_A^2 > 0$ (left panel) and the other with $m_A^2 < 0$ (right panel). 
In each panel, contours of the potential \cref{eqn:abeluvVparam} are shown in the $(r,f)$ plane in arbitrary mass units with the indicated choice of parameters. We fix $\beta = -\arg \mu + \pi$. 
The location of the global minimum is denoted by a black dot, while the fixed point corresponds to $(r,f) = (0,0)$.  
The EFT submanifold branches obtained by matching in the \emph{unitary basis} are shown in blue and orange; blue corresponds to solution(s) of $\frac{\partial V}{\partial f}=0$ in regions where the UV fluctuations (in $f$ and $\beta$) have positive definite mass matrix, and orange is for the opposing case. 
The region in which the UV fluctuations do not have positive definite mass matrix is shown in light gray, with a dashed green boundary.
The EFT submanifolds obtained by matching in the \emph{mass basis} are shown in cyan, corresponding to the solutions $\frac{\partial V}{\partial H}=0$. 
In the left panel, the choice of parameters (namely $m_A^2 > 0$) admits a SMEFT, as illustrated by the existence of an EFT submanifold connecting the UV fixed point to the global minimum through a region where the mass matrix is positive definite. 
In the right panel, the choice of parameters (namely $m_A^2 < 0$) requires a HEFT, as none of the EFT submanifolds connect the fixed point to the global minimum.}
\label{fig:AbelEx}
\end{figure}

Explicitly, we start with the potential in the gauge eigenbasis~\cref{eqn:abeluvfbetaparam}, and substitute
\begin{equation}
  r= v_B + h_B, \qquad \text{and} \qquad f = v_A + h_A\, .
\end{equation}
Here $(r,f)=(v_B,v_A)$ are the coordinates of the global minimum (when $\beta = -\arg \mu + \pi$) and $h_A$ and $h_B$ are further rotated into the light and heavy (CP even) mass eigenstates $h$ and $H$, defined as
\begin{subequations}\label{eqn:ChargeToMass}
\begin{align}
{h_A} &= h \s {\cos \alpha } - H \s {\sin \alpha } \,, \\[6pt]
{h_B} &= h \s {\sin \alpha } + H \s {\cos \alpha } \,,
\end{align}
\end{subequations}
for some appropriate value of $\alpha$. 
To derive the effective Lagrangian, we again solve for the EOMs using $\frac{\partial V}{\partial \beta} = \frac{\partial V}{\partial H} = 0$. 
The $\beta$ EOM has the same solution as in the unitary basis, $\beta = -\arg \mu + \pi$, and upon substitution into $\frac{\partial V}{\partial H}=0$ we find a bivariate cubic equation involving $h$ and $H$, which we solve for $H$. 
This yields the EFT submanifold in the mass basis, which is generically not the same as the EFT submanifold obtained when matching in the unitary basis, see~\cref{fig:AbelEx}. 
Consequently, the EFT obtained in the mass basis generically possesses a different tree-level two-derivative Lagrangian, and hence has different curvature invariants. 
This is consistent because, although the two bases are related by non-derivative field redefinitions in the UV theory, transforming between them requires derivative field redefinitions in the IR EFTs, see~\cref{app:UVFieldRedefs}; the apparent difference in curvature invariants is an artifact of the fact that the geometric quantities are only defined up to two derivatives. 

\clearpage

\subsubsection{The EFT Submanifold}
The salient features of this Abelian toy model  -- and the general discussion of EFT submanifolds in \cref{sec:eftsubmani} -- are nicely illustrated by studying the theory in the $(r,f)$ plane for fixed $\beta$. 
In \cref{fig:AbelEx} we depict two example points in parameter space for the Abelian toy model, one with $m_A^2 > 0$ and one with $m_A^2 < 0$.
We plot the EFT submanifolds obtained in both the unitary (blue and orange) and mass bases (cyan). 
When $m_A^2 > 0$ (left panel), the EFT submanifolds obtained in both the unitary and mass bases connect the UV fixed point to the global minimum (the equivalent of the observed vacuum in this toy model). 
In the unitary basis, the EFT submanifold given by solutions of $\frac{\partial V}{\partial f}=0$ is entirely contained within a connected region in which the UV fluctuations have positive definite mass matrix, ensuring that the EFT can be written as a SMEFT.

On the other hand, when $m_A^2 < 0$ (right panel), both bases admit more than one EFT submanifold branch.
Importantly, the branch that includes the global minimum fails to reach the UV fixed point. 
This is as expected from our general discussion in \cref{sec:eftsubmani}, since the mass matrix of the UV fluctuations is not positive definite at the fixed point. 
Although the EFT submanifolds in the unitary and mass bases differ, their qualitative behavior remains the same: when the fluctuations being integrated out are tachyonic at the $U(1)$ invariant point, the EFT that includes the global minimum does not connect to the invariant point, and thus is described by a HEFT.

\subsection{Two Higgs Doublet Model}%
\label{sec:the_two_higgs_doublet_model}
Having worked through the details of the toy model, we now turn to an example of broad phenomenological interest.
In particular, we will argue under what conditions it is possible to match the tree-level custodially-symmetric two Higgs doublet model (2HDM) onto SMEFT. 
Such a ``two Higgs quadruplet model'' (2HQM) is a special case of the scalar sector of the full 2HDM. 
We reserve an in-depth treatment of the full 2HDM for a companion paper \cite{twohdm}; for previous EFT studies, see \eg~\cite{Gunion:2002zf, deBlas:2014mba, Gorbahn:2015gxa, Chiang:2015ura, Brehmer:2015rna, Egana-Ugrinovic:2015vgy, Belusca-Maito:2016dqe, Faro:2020qyp}.

Following the approach taken for the Abelian toy model, we will match in the unitary basis.
However, the 2HQM requires keeping track of some additional subtleties. 
First, the quadruplets generically can mix, and so we must pick a basis prior to integrating one of them out -- this field redefinition  does not affect the dynamics of the model but does, in general, change the tree-level EFT at any given derivative order, see the discussion in \cref{app:UVFieldRedefs}. 
Second, there exist configurations such that the vevs of the two quadruplets are misaligned, thereby breaking the heretofore assumed $O(3)$ symmetry. 
This is the analogue of the charge breaking vacua in the 2HDM. 
An EFT manifold which includes $O(3)$ breaking configurations cannot be described by the SMEFT or HEFT of \cref{sec:Define}, which assume custodial symmetry persists to low energies.

Nevertheless, we will show that the intuition developed by studying the Abelian toy model is validated by the matching calculations that follow. 
When there are extra sources of symmetry breaking, \ie, when the quadruplet being integrated out is tachyonic at the UV $O(4)$ invariant point, HEFT results; otherwise, we can match onto SMEFT.
\clearpage

\subsubsection{UV Model}
The UV model includes two quadruplets $\vec \phi_a$, which are distinguished using the index $a=1,2$. 
The most general renormalizable Lagrangian takes the form
  \begin{equation}
    \mathcal{L} = \frac{1}{2}\s \partial \vec \phi_a \cdot \partial \vec \phi_a  - V \label{eq:2HDMuvlag} \, ,
  \end{equation}
where
\begin{equation}
V = \frac{1}{2}\s m^2_{ab}\s \vec \phi_a \cdot \vec \phi_b + \frac{1}{4}\s \lambda_{abcd}\s \big(\vec \phi_a \cdot \vec \phi_b\big) \big(\vec \phi_c \cdot \vec \phi_d\big)
  \label{eq:2HDMuvV} \, .
\end{equation}
Note that repeated indices indicate sums over $a,b,c,d \in \{1,2\}$, and ``$\s\cdot\s$'' denotes contraction of the $O(4)$ indices.
The parameters are real and have the symmetry properties
\begin{equation}
m^2_{ab} = m^2_{(ab)}, \qquad \lambda_{abcd} = \lambda_{(ab)(cd)}, \qquad \text{and} \qquad \lambda_{abcd} = \lambda_{cdab}\,,
\end{equation}
where the parenthesis in the subscripts implies symmetrization over the indices as usual.
This implies that the mass-squared matrix $m^2_{ab}$ contains 3 independent real parameters,
\begin{equation}
   m^2_{11}\,,\quad m^2_{12}\,,\quad m^2_{22} \, ,
   \label{eq:mcomp}
\end{equation}
while $\lambda_{abcd}$ contains 6 independent real parameters,\footnote{Up to rational factors, these correspond to the familiar parameters
\begin{equation}
   \lambda_1\,,\quad \Re \lambda_6\,,\quad \lambda_3\,, \quad \lambda_4\,,\quad \Re \lambda_7\,,\quad \lambda_2 \,,
\end{equation}
that appear in parametrizations of the full 2HDM, see~\eg~\cite[App.\ A]{Davidson:2005cw}. }
\begin{equation}
    \lambda_{1111}\,,\quad \lambda_{1112}\,, \quad\lambda_{1122}\,,\quad \lambda_{1212}\,,\quad \lambda_{1222}\,,\quad \lambda_{2222} \, .
   \label{eq:lamcomp}
\end{equation}
The potential misalignment of the two quadruplets is accounted for by an angle $\beta$, defined via
\begin{equation}
    \cos \beta = \frac{\vec \phi_1 \cdot \vec \phi_2}{\sqrt{\vec\phi_1 \cdot \vec\phi_1} \sqrt{\vec\phi_2 \cdot \vec\phi_2}} \, ,
\end{equation}
Insisting that the potential be bounded from below implies
\begin{equation}
    \forall \theta\,,\beta \in [0,2\pi)\,,\qquad \bar \lambda_{abcd}(\beta) \, n_a n_b n_c n_d \geq 0\,,
\end{equation}
where
\begin{equation}
    n_a = \begin{pmatrix} \cos \theta \\ \sin \theta \end{pmatrix}\qquad \text{ and }\qquad
    \begin{pmatrix} \bar \lambda_{1111} \\ \bar \lambda_{1112} \\ \bar \lambda_{1122} \\ \bar \lambda_{1212} \\ \bar \lambda_{1222} \\ \bar \lambda_{2222} \end{pmatrix} =
    \begin{pmatrix} \lambda_{1111} \\ \lambda_{1112} \cos \beta \\ \lambda_{1122} \\ \lambda_{1212} \cos^2 \beta \\ \lambda_{1222} \cos \beta \\ \lambda_{2222} \end{pmatrix}
    \, .
\end{equation}
Finally, we note that one parameter is unphysical, since it can be removed using the field redefinition that mixes the quadruplets, namely
\begin{equation}
    \vec \phi_a \to O_{ab}\s \vec \phi_b;  \qquad O_{ab} = \begin{pmatrix}
      1 & 0 \\ 0 & \pm 1
    \end{pmatrix} \begin{pmatrix} \cos \alpha & \sin \alpha \\ -\sin \alpha & \cos \alpha \end{pmatrix} \,\in\, O(2) \, .
    \label{eq:UVrot}
\end{equation}

  \subsubsection{Matching in the Unitary Basis}
Now we have everything we need to integrate out $\vec \phi_2$ in order to build an EFT for $\vec \phi_1$. 
As per the above discussion of the Abelian toy model, we will perform this calculation in the unitary basis.
For the 2HQM, this basis can be parameterized by
  \begin{equation}
    \vec \phi_2 = \frac{f}{r} \exp \begin{pmatrix}
        0 & 0 & 0 & \beta_1 \\
        0 & 0 & 0 & \beta_2 \\
        0 & 0 & 0 & \beta_3 \\
        -\beta_1 & -\beta_2 & -\beta_3 & 0
    \end{pmatrix} 
    \vec \phi_1    \qquad \text{with} \qquad
    \begin{array}{l}  
    f \,\in\, \rr\\[5pt] 
    \beta_i \in [0,2\pi)
    \end{array} \,,
    \label{eq:phi2param}
  \end{equation}
as an arbitrary rotation and scaling of $\vec \phi_1$, where $r \equiv \sqrt{\vec \phi_1 \cdot \vec \phi_1} \geq 0$ and $|f| = \sqrt{\vec \phi_2 \cdot \vec \phi_2}$. 
The unitary basis takes a particularly simple form in the 2HQM, \ie, $\vec \phi_2 \propto \vec \phi_1$, because the BSM state transforms in the same representation as the Higgs.
In this parametrization,
  \begin{equation}
    \vec \phi_1 \cdot \vec \phi_2 = r\s f\s \cos \beta + \frac{f}{r}\s \frac{(1- \cos\beta)}{\beta^2} \Big[ \phi_1^i\s \phi_1^i\s \beta^2 - \big(\phi_1^i \s\beta^i\big)^2 \Big] \, ,
    \label{eq:onedottwo}
  \end{equation}
where $\beta \equiv \sqrt{\beta_1^2 + \beta_2^2 + \beta_3^2}$ and $\phi^i_1$, $i=1,2,3$ are the top three components of the quadruplet that is being retained in the EFT.\footnote{See, \eg~\cite[(A.2)]{Alonso:2016oah} for the closed form of the exponential in \cref{eq:phi2param}.}

Next, we express the potential in the unitary basis, which to second order is
  \begin{align}
    V\Big(\vec \phi_1, \vec \phi_2\Big) &= V\big(\s r, \phi_1^i,f, \beta^i\s \big) \nonumber \\[5pt]
    &= \frac{1}{2}\s \Big( m_{11}^2\s r^2 + 2\s m_{12}^2\s r\s f + m_{22}^2\s f^2 \Big) \nonumber \\[5pt]
    &\hspace{15pt}+\bigg[\frac{\lambda_{1111}}{4}\s r^4 + \lambda_{1112}\s r^3\s f + \big[\frac{\lambda_{1122}}{2} + \lambda_{1212}\big]\s r^2\s f^2 + \lambda_{1222}\s r\s f^3 + \frac{\lambda_{2222}}{4}\s f^4 \bigg] \nonumber \\[5pt]
    &\hspace{15pt} - \frac{1}{2}\s rf\! \left[ m_{12}^2 + \begin{pmatrix}
        r & f
    \end{pmatrix}\! \begin{pmatrix}
        \lambda_{1112} & \lambda_{1212} \\
        \lambda_{1212} & \lambda_{1222}
    \end{pmatrix}\! \begin{pmatrix}
      r \\ f
    \end{pmatrix} \right]\! \left[ \beta^2 - \frac{1}{r^2} \Big(\phi_1^i\s \phi_1^i\s \beta^2 - \big(\phi_1^i\s \beta^i\big)^2\Big) \right] \nonumber \\[5pt]
    &\hspace{15pt}+ \mathcal{O}(\beta^4)
    \label{eq:2HDMpotparam}
  \end{align}
Note that it is an even function of the $\beta^i$.
The UV Lagrangian up to two derivatives is then
  \begin{align}
    \mathcal{L} &= \frac{1}{2}\s \partial \vec \phi_1 \cdot \partial \vec \phi_1 + \frac{1}{2}\s (\partial f)^2 + \frac{1}{2}\s \frac{f^2}{r^2} \Bigg[ \big(\partial \vec \phi_1\big)^2 - \frac{\big(\vec \phi_1 \cdot \partial \vec \phi_1\big)^2}{\vec \phi_1 \cdot \vec \phi_1} \Bigg] +\big[\partial \beta \text{  terms}\big] - V \, .
  \end{align}

To find the EFT manifold(s), we solve the zero-derivative EOMs, %
\begin{subequations}
  \begin{align}
    \frac{\partial V}{\partial f}\big(r, \phi_1^i, f, \beta^i\big) &= 0 \, ,\\[5pt]
    \frac{\partial V}{\partial \beta^i} \big(r, \phi_1^i, f, \beta^i\big) &= 0 \, .
    \label{eq:eom}
  \end{align}
\end{subequations}
Obviously, $\beta^i = 0$ is always a solution of the EOMs $\frac{\partial V}{\partial \beta^i}=0$, since the potential is an even function of $\beta^i$.\footnote{Other classes of solutions yield charge-breaking vacua, and thus we do not consider them here.} 
Such solutions imply $\vec \phi_2 \propto \vec \phi_1$: the two quadruplets are aligned on the EFT submanifold defined using these solutions. 
We can then solve for $f$ as a function of $r$ via the cubic equation
  \begin{align}
    \frac{\partial V}{\partial f} \Big\vert_{\beta^i=0} = m_{12}^2\s r + m_{22}^2\s f + \lambda_{1112}\s r^3 + \big(\lambda_{1122} + 2\s \lambda_{1212}\big)\s r^2\s f + 3\s \lambda_{1222}\s r\s f^2 + \lambda_{2222}\s f^3 = 0\,, \notag\\
  \label{eq:o3symeom}
  \end{align}
resulting in the effective Lagrangian
\begin{align}
\mathcal{L}_\text{Eff}  &= \frac{1}{2}\s \partial \vec \phi_1 \cdot \partial \vec \phi_1\s \bigg[ 1 + \frac{f^2}{r^2} \bigg] + \frac{1}{2}\s \big(\vec \phi_1 \cdot \partial \vec \phi_1\big)^2 \Bigg[ \frac{\big(r\s f^\prime - f\big)\big(r\s f^\prime + f\big)}{r^4} \Bigg] - V(r,f)\, .
 \label{eq:eft} 
  \end{align}
Equivalently, we can rewrite this effective Lagrangian using polar coordinates $\vec \phi_1 = r \s \vec n$ for the Higgs, which yields
  \begin{equation}
    \mathcal{L} = \frac{1}{2}\s (\partial r)^2\s \bigg[ 1 + \big(f^\prime\big)^2 \bigg] + \frac{1}{2}\s (\partial \vec n)^2 \big[ r^2 + f^2 \big] - V \, . \label{eq:eftpolar}
  \end{equation}
In this form, it is clear that the $\beta^i = 0$ solutions of the EOMs preserve $O(3)$ at low energies, reflected in the fact that the Lagrangian does not single out any component of $\vec n$.

As in the Abelian toy model, $f(r)$ must be analytic and $f(0) = 0$ in order to be able to match onto SMEFT. 
The invariance of the potential under $(r,f) \to (-r,-f)$ guarantees that any solution of the cubic satisfying $f(0)=0$ is odd in $r$, and therefore the terms in square brackets in \cref{eq:eft} will admit a SMEFT-like expansion in $r^2 = \vec \phi_1 \cdot \vec \phi_1$.
Conversely solutions with $f(0) \neq 0$ are HEFT-like due to the lack of a zero in the coefficient of $(\partial \vec n)^2$ in \cref{eq:eftpolar}. 
These are manifest as the EFT submanifold either includes, or does not include, the $O(4)$ fixed point respectively.

\begin{figure}[t]
  \centering
  \includegraphics[height=0.6\textwidth]{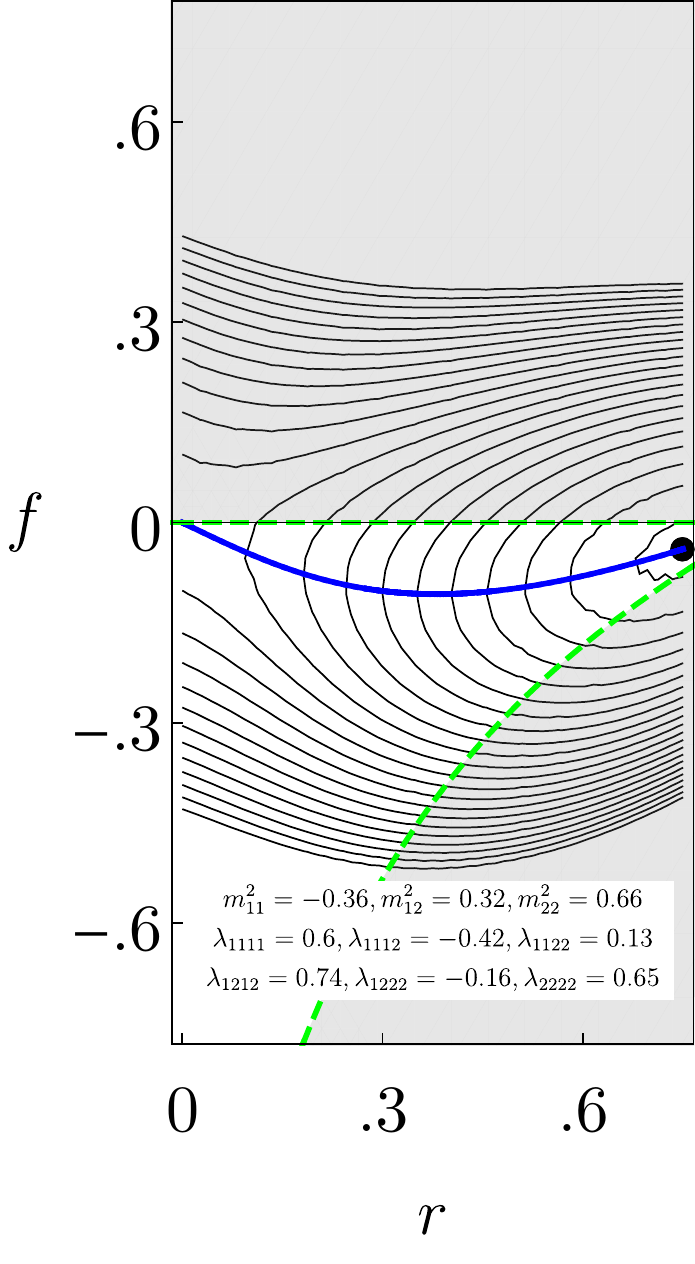}
  \hspace*{3ex}
  \includegraphics[height=0.6\textwidth]{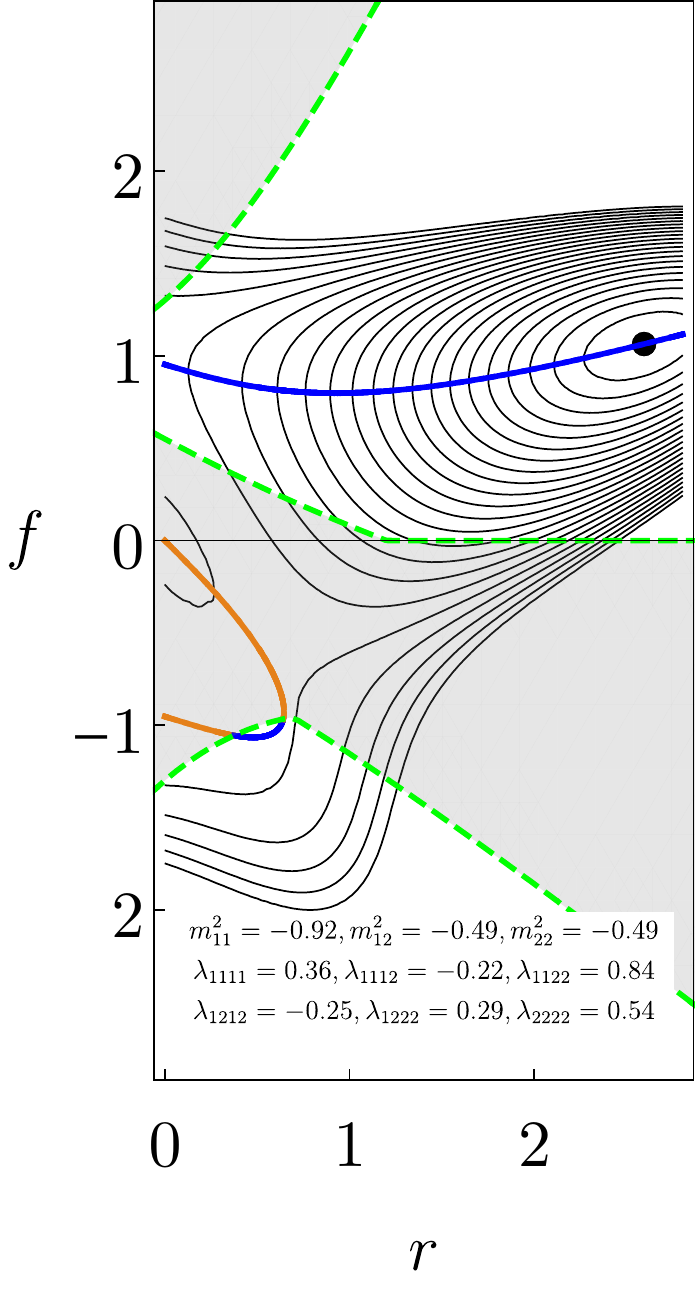}
  \caption{EFT submanifolds  in the 2HQM obtained by matching in the unitary basis for two representative sets of parameters, one with $m_{22}^2 > 0$ (left panel) and the other with $m_{22}^2 < 0$ (right panel). 
In each panel, contours of the potential \cref{eq:2HDMpotparam} are shown in the $(r,f)$ plane in arbitrary mass units with the indicated choice of parameters. 
We have fixed $\beta^i = 0$. 
The location of the global minimum is denoted by a black dot, while the fixed point corresponds to $(r,f) = (0,0)$.  
The EFT submanifolds are shown in blue and orange; blue corresponds to solution(s) of $\frac{\partial V}{\partial f}=0$ in regions where the UV fluctuations (in $f$ and $\beta^i$) have positive definite mass matrix, and orange otherwise. 
The region in which the UV fluctuations do not have positive definite mass matrix is shown in light gray, with a dashed green boundary.
In the left panel, the choice of parameters (namely $m_{22}^2 > 0$) admits SMEFT, as illustrated by the existence of an EFT submanifold connecting the UV fixed point to the global minimum through a region of positive definite mass matrix. 
In the right panel, the choice of parameters (namely $m_{22}^2 < 0$) requires HEFT, as none of the EFT submanifolds connect the fixed point to the global minimum.
\label{fig:2HDMEx}}
\end{figure}

\subsubsection{The EFT submanifold}

In \cref{fig:2HDMEx}, we depict two example points in the UV parameter space.
The UV potentials in the $(r,f)$ plane when $\beta^i =0$ are shown as contours, and are overlaid with the EFT submanifolds corresponding to the solutions of the cubic equation~\cref{eq:o3symeom}. 
When $m_{22}^2 > 0$ (left panel), there is EFT submanifold solution which connects the global minimum (here the observed vacuum) to the UV invariant point, and thus yields SMEFT.
Alternatively, the potential with $m_{22}^2 < 0$ does not have such a submanifold, and thus must be matched onto HEFT.

Such behavior can be understood generally as follows, see~\cref{sec:scalarsmeft}.
When $m_{22}^2 < 0$, $\frac{ \partial^2 V}{\partial f^2}$ is positive at the global minimum and negative at the UV invariant point. 
The evolution of the solution(s) of $\frac{\partial V}{\partial f}=0$ is given by
  \begin{equation}
    0 = \dd \frac{\partial V}{\partial f} =
    \frac{\partial^2 V}{\partial r \s \partial f} \s \dd r + \frac{\partial^2 V}{\partial f^2} \s \dd f = 0 \, ,
\end{equation}
implying that $\dd r$ changes sign whenever $\frac{\partial^2 V}{\partial f^2}$ does. 
For example, a solution that sets out towards the $f$ axis at the global minimum ($\dd r <0$) must turn away from the $f$ axis in the neighborhood of the UV invariant point ($\dd r > 0$). 
We can conclude that when $m_{22}^2 <0$, a solution passing through the global minimum cannot connect to the UV invariant point.
In the other case with $m_{22}^2 > 0$, a solution may connect the global minimum and UV invariant point, and using our freedom to rotate the fields in the UV, \cref{eq:UVrot}, we can always find a UV basis that results in a SMEFT~\cite{twohdm}.

\clearpage

\subsection{Higgs Triplet Extension and Custodial Symmetry Violation}
\label{appsec:Triplet}
For our final concrete example of BSM symmetry breaking, we explore the implications for tree-level matching when the SM is extended to include a triplet scalar $\Phi$; see~\eg~\cite{Skiba:2010xn, Khandker:2012zu, deBlas:2014mba, Chiang:2015ura, Ellis:2016enq, Dawson:2017vgm} for previous EFT studies of this scenario.
However, since custodial symmetry is not respected by the UV couplings, the EFT will only have manifest $SU(2)_L \times U(1)_Y$ invariance, instead of the larger custodial $O(4)$ symmetry.
This implies that our simplifying assumptions defined in \cref{sec:Define} and used throughout do not hold here,~\eg~we cannot use our custodially symmetric HEFT parametrization in \cref{eqn:LHEFT}, our expressions for the curvature invariants \cref{eqn:Rgeneral}, and so forth.
Nevertheless, we will find it interesting to analyze the effective Lagrangian that results from integrating out $\Phi$.

In particular, we will see that there are two classes of solutions to the EOMs, depending on relations among the parameters.
For the first type, corresponding to phases 1 and 3 in \cref{tbl:Branches}, the $\Phi$ vev $v_\Phi$ is driven by the Higgs vev $v_0$, namely $v_\Phi \to 0$ as $v_0\to 0$.
In this case, the theory can be matched onto SMEFT, such that all BSM effects decouple as the triplet mass is taken to be large, even though the triplet vev is non-zero in the physical vacuum.
In the second type, corresponding to phases 2 and 4 in \cref{tbl:Branches}, the triplet vev is driven by an instability in its own potential, even in the limit $v_0\to 0$.
In this case, the triplet produces an independent source of electroweak symmetry breaking.
After integrating out the BSM physics, a non-decoupling effect remains behind even when the triplets become heavy.
The resulting low energy description must be HEFT.

\subsubsection{UV Model}
We start with the most general renormalizable Lagrangian for the scalar sector 
\begin{equation}
\Lag = {\left| {\partial H} \right|^2} + \frac{1}{2}\big(\partial \Phi\big)^2 - V \,,
\label{eqn:LUVTriplet2d}
\end{equation}
with the potential
\begin{equation}
V = -\mu_H^2\s{|H|^2} + {\lambda_H}{|H|^4} + \frac{1}{2}\s {m^2}\s \Phi^2 - \mu\s {H^\dag }\s{t_L^a}\s H\s {\Phi_a} + \kappa\s |H|^2 \s\Phi^2 + \frac{1}{4}\s\lambda_\Phi\s {\Phi^4} \,,
\label{eqn:LUVTriplet0d}
\end{equation}
where $H$ is the Higgs doublet, $\Phi_a$ with $a=1,2,3$ is a real Higgs triplet of $SU(2)_L$, and $t_L^a\equiv\sigma^a/2$ are generators in the fundamental representation of $SU(2)_L$.
We use the freedom to redefine $\Phi_a \to - \Phi_a$ to set $\mu > 0$ without loss of generality.

Note that the $\mu$ coupling between $H$ and $\Phi$ breaks the custodial $O(4)$ symmetry.
We can see this explicitly by expressing the Higgs field as a custodial bilinear in the usual way:
\begin{equation}
\Sigma \equiv \mqty( \tilde{H} & H ) \,,
\end{equation}
with $\tilde{H}\equiv i\s\sigma^2 H^*$.
Then the interaction can be written as
\begin{equation}
{H^\dag }{t_L^a}H{\Phi_a} =  - \tr \Big( {{\Sigma^\dag }\s{t_L^a}\s\Sigma\s t_R^3} \Big){\Phi_a} \,.
\end{equation}
where $t_L^a$ are the generators of $SU(2)_L$ and $t_R^3$ is the third generator of $SU(2)_R$.
Clearly, this violates custodial symmetry since it singles out a particular direction in the $SU(2)_R$ space. 

\subsubsection{Phases of the UV Moduli Space}
\label{appsubsec:vevStructure}
Our goal is first to derive the coupled equations that determine the two vevs $v_0$ and $v_\Phi$.
We will then explore the properties of their solutions, and will discover that these solutions can be classified as belonging to one of four phases.
Ultimately we wish to understand the connection between the various allowed solutions and the low energy description in terms of SMEFT or HEFT.
In particular, we will show that some regions of the parameter space lead to non-decoupling effects that force one to match onto HEFT.

We parameterize the components of our fields as
\begin{equation}
H = \frac{1}{\sqrt{2}}\s \mqty( \phi_1 + i\s\phi_2 \\ v_0+h + i\s\phi_4 )  \qquad\text{and} \qquad  \Phi = \mqty( \Phi_1 \\ \Phi_2 \\ v_\Phi + h_\Phi)  \,,
\end{equation}
where the notation for $H$ parallels \cref{eqn:Doublethn} with the identification $\phi_3 = v_0+h$.
We then plug these into the potential in \cref{eqn:LUVTriplet0d} and extract the linear terms.
Minimizing the potential is equivalent to requiring that these linear terms vanish:%
\begin{subequations}\label{eqn:vevEquations}
\begin{align}
\Big( {-\mu_H^2 + {\lambda_H}\s v_0^2 + \kappa\s v_\Phi^2 - \frac{1}{2}\s\mu\s {v_\Phi }} \Big)\s{v_0} &= 0 \label{eqn:hvevEq} \\[7pt]
\Big( {{m^2} + \lambda_\Phi\s v_\Phi^2 + \kappa v_0^2} \Big){v_\Phi } - \frac{1}{4} \s \mu v_0^2 &= 0 \,. \label{eqn:phivevEq}
\end{align}
\end{subequations}

This set of two cubic equations should yield nine solutions.\footnote{To see that there are nine solutions requires allowing the vevs to be complex. However, the physically interesting solutions should of course be real valued.  Ultimately the Lagrangian parameters determine which solutions are accessible.}
We will now argue that they can be categorized into the four phases summarized in \cref{tbl:Branches}.
There are two ways to satisfy \cref{eqn:hvevEq}:
\begin{align}
v_0 = 0  \qquad \text{or} \qquad -\mu_H^2 + \lambda_H\s v_0^2 + \kappa\s v_\Phi^2 - \frac{1}{2}\s\mu\s v_\Phi = 0 \,.
\label{eqn:vhBranches}
\end{align}
Intuitively, the global minimum of the potential occurs for $v_0=0$ when the Higgs quadratic term is positive.
When the Higgs quadratic term is negative, the second condition in \cref{eqn:vhBranches} must be satisfied.

Similarly, there are two ways to satisfy \cref{eqn:phivevEq}.
To see this, consider the limit $\mu\to 0$:%
\begin{equation}
\Big( {{m^2} + \lambda_\Phi\s v_\Phi^2+ \kappa\s v_0^2} \Big){v_\Phi } = \frac{1}{4}\s\mu\s v_0^2 \, \to\, 0 \,.
\end{equation}
Clearly, there are two options
\begin{align}
v_\Phi \sim \mu \,\to\, 0 \qquad \text{or} \qquad m^2+\lambda_\Phi\s v_\Phi^2+ \kappa\s v_0^2 \sim \mu \,\to\, 0 \,,
\label{eqn:vphiBranches}
\end{align}
independent of which solution was chosen for \cref{eqn:hvevEq}.
In summary, each of the two vev equations implies two possible classes of solutions, yielding four phases in total:
\vspace{-30pt}
\begin{center}
\begin{equation}
\renewcommand{\arraystretch}{1.8}
\setlength{\arrayrulewidth}{.3mm}
\setlength{\tabcolsep}{2 em}
\begin{tabular}{c|c|c}
Phases    & Sourced $v_\Phi$ & Tachyon $v_\Phi$ \\\hline
Sourced $v_0$   & 1                & 2 \\
Tachyon $v_0$ & 3                & 4 \\
\end{tabular}
\label{tbl:Branches}
\end{equation}
\end{center}
where the ``sourced'' ($m^2>0$) means that a non-zero vev is driven by the vev of the other field, and ``tachyon'' ($m^2<0$) implies that the vev would be non-zero due to a negative quadratic term in the potential.
As we will argue next, which $v_\Phi$ phase we are in determines the nature of the EFT.
We will show that when $v_\Phi$ is sourced by $v_0$ (phases 1 and 3), there is no electroweak symmetry breaking when $v_0 = 0$.
This implies that the theory which results from integrating out the triplet can be matched onto SMEFT.
On the other hand, if the parameter space implies that $v_\Phi$ is non-zero due to $\Phi$ being tachyonic (phases 2 or 4), then there is a non-decoupling extra source of electroweak symmetry breaking.
Integrating out the triplet results in an EFT that must be written as HEFT.

\subsubsection{Matching in the Unitary Basis}

To integrate out the triplet, we parametrize it in the unitary basis as
\begin{equation}
    \Phi_a = \frac{4\s f}{r^2} \exp \begin{pmatrix}
        0 & 0 & \beta_1 \\
        0 & 0 & \beta_2 \\
        -\beta_1 & -\beta_2 & 0
    \end{pmatrix}
    \begin{pmatrix}
    H^\dagger t^1 H \\
    H^\dagger t^2 H \\
    H^\dagger t^3 H
    \end{pmatrix}
     \qquad \text{with} \qquad
    \begin{array}{l}  
    f \,\in\, \rr\\[5pt] 
    \beta_i \in [0,2\pi)
    \end{array} \,,
    \label{eq:tripparam}
\end{equation}
where $r \equiv \sqrt{2\s H^\dagger H}$. 
In this case, $\Phi_a$ is proportional to the appropriate function of $H$ such that it transforms as an $SU(2)_L$ triplet. 
The UV Lagrangian in unitary basis is 
\begin{equation}
  \Lag = \left[ 1 + \frac{4\s f^2}{r^2} \right] |\partial H|^2 + \frac{1}{2}\s (\partial f)^2 + \frac{2\s f^2}{r^4} \bigg[ \big(\partial(H^\dagger H)\big)^2 + \Big(H^\dagger \overset{\leftrightarrow}{\partial} H\Big)^2 \bigg] + \big[\partial \beta\text{ terms}\big] - V
  \label{eq:tripUVlag}
\end{equation}
where the potential takes the form
\begin{align}
  V &= -\frac{1}{2}\s \mu_H^2\s r^2  + \frac{1}{2}\s m^2\s f^2 + \frac{1}{4}\s {\lambda_H}\s r^4 + \frac{1}{2}\s \kappa\s r^2\s f^2 +  \frac{1}{4}\s\lambda_\Phi\s f^4 \nonumber \\[5pt]
    &\hspace{15pt} - \frac{1}{4}\s \mu\s f\s r^2 + 4\s \frac{\mu\s f}{r^2} (1- \cos \beta) \Bigg[ \frac{\big(H^\dagger t^i H\s \beta^i\big)^2}{\beta^2} + \big(H^\dagger t^3 H\big)^2 \beta^2 \Bigg]  \,,
  \label{eq:TripletPotParam}
\end{align}
where the repeated index $i=1,2$ is summed over, and $\beta = \sqrt{\beta_1^2 + \beta_2^2}$.

The EFT branches for the light field $H$ are derived by solving 
\begin{subequations}
\begin{align}
  \frac{\partial V}{\partial f} &= 0 \,, \\[8pt]
  \frac{\partial V}{\partial \beta} &= 0 \,,
\end{align}
\end{subequations}
for $f$ and $\beta^i$.
As for the 2HQM, the EOMs admit the charge-preserving solution $\beta^i=0$, which is always a local minimum in the $\beta^i$ directions when $f>0$. 
Enforcing this condition, the remaining EOM becomes
\begin{equation}
  \frac{\partial V}{\partial f} \Big\vert_{\beta^i = 0} =  -\frac{1}{4}\s \mu\s r^2 + \big(m^2+\kappa\s r^2\big)\s f + \lambda_\Phi\s f^3 = 0\,,
\end{equation}
which is a depressed cubic equation identical to the form considered in the Abelian toy model~\cref{eqn:depcubabelian}. 

\subsubsection{The EFT Submanifold}

We recall the result that, when $m^2 > 0$, an analytic even solution $f(r)$ satisfying $f(0)=0$ connects the global minimum to the invariant point. In the present model, this corresponds to the ``sourced'' phases for $\Phi$, and can be matched onto SMEFT. 
When $m^2<0$, an analytic even solution $f(r)$ connects the global minimum to the point $f(0) = \sqrt{-m^2/\lambda_\Phi}$. 
This corresponds to the ``tachyon'' phases for $\Phi$, and must be matched onto HEFT.

\begin{figure}[h!]
  \centering
  \includegraphics[width=0.3\textwidth]{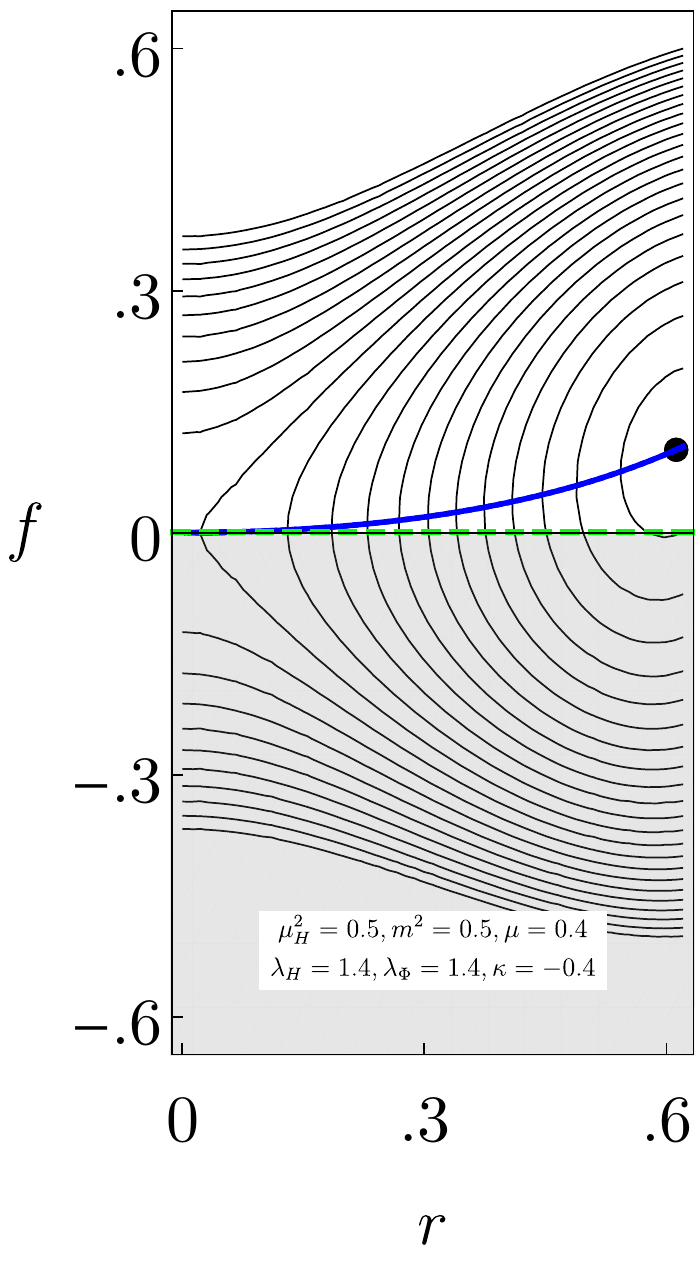}
  \hspace*{3ex}
  \includegraphics[width=0.3\textwidth]{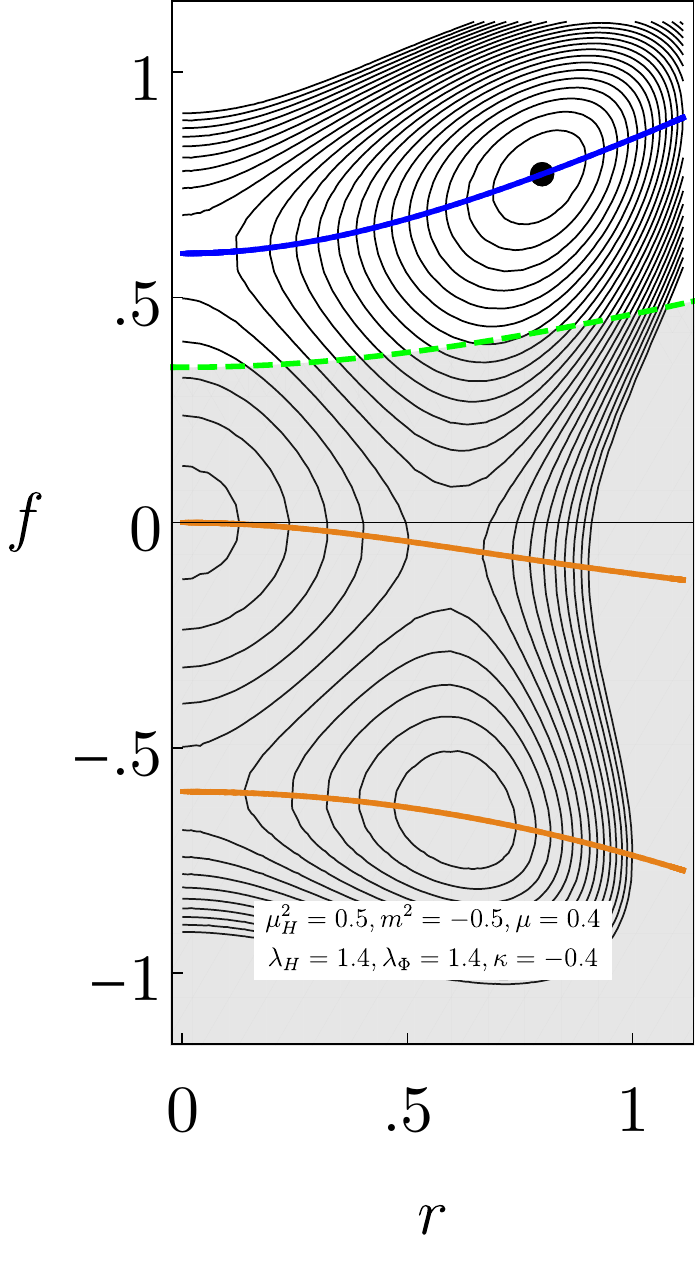}
  \caption{EFT submanifolds  in the triplet model obtained by matching in the unitary basis for two representative sets of parameters, one with $m^2 > 0$ (left panel) and the other with $m^2 < 0$ (right panel). In each panel, contours of the potential \cref{eq:TripletPotParam} are shown in the $(r,f)$ plane in arbitrary mass units with the indicated choice of parameters. We fix $\beta^i = 0$. The location of the global minimum is denoted by a black dot, while the fixed point corresponds to $(r,f) = (0,0)$.  The EFT submanifolds are shown in blue and orange; blue corresponds to solution(s) of $\frac{\partial V}{\partial f}=0$ in regions where the UV fluctuations (in $f$ and $\beta^i$) have a positive definite mass matrix, and the curve is orange otherwise.
The region in which the UV fluctuations do not have a positive definite mass matrix is shown in light gray, with a dashed green boundary.
In the left panel, the choice of parameters (namely $m^2 > 0$) admits a SMEFT, as illustrated by the existence of an EFT submanifold connecting the UV fixed point to the global minimum through a region of positive definite mass matrix. In the right panel, the choice of parameters (namely $m^2 < 0$) requires a HEFT, as none of the EFT submanifolds connect the fixed point to the global minimum. \label{fig:TripletEx}}
\end{figure}

Upon substitution of $f(r)$ into the UV Lagrangian \cref{eq:tripUVlag}, we obtain the effective Lagrangian
\begin{equation}
  \Lag_\text{Eff} = \bigg[ 1 + \frac{4\s f^2}{r^2} \bigg]\s |\partial H|^2 + \frac{1}{2}\s \Bigg[ \frac{\big(f^\prime\big)^2}{r^2} + \frac{4\s f^2}{r^4} \Bigg] \Big(\partial\big(H^\dagger H\big)\Big)^2 + \frac{2\s f^2}{r^4} \Big(H^\dagger \overset{\leftrightarrow}{\partial} H\Big)^2 - V \, ,
\end{equation} 
Note the presence of the custodial symmetry-violating third term, which is not included within the scope of our custodially-symmetric framework.
However, the qualitative conclusions remain the same, see~\cref{fig:TripletEx}. 
The two derivative terms admit a SMEFT expansion when $m^2 > 0$ (left panel) as the coefficients of all three terms are analytic at the invariant point, while HEFT results when $m^2 < 0$ (right panel). 
The simple intuition that new sources of electroweak symmetry breaking imply we must match the theory onto HEFT has been demonstrated by each of our example UV models.

\vspace{40pt}

\clearpage

\section{All Orders Versus Truncated EFT Expansions}
\label{sec:Truncate}
Thus far we have focused entirely on in-principle distinctions between HEFT and SMEFT. 
However, equally important are the practical considerations for when one should match onto HEFT:
\begin{center}
\vspace{3mm}
\begin{tcolorbox}[colback=light-gray]
\begin{minipage}{5.5in}
\textbf{Practical Criterion:}  One should match onto HEFT when integrating out a state whose mass is near (or below) the electroweak scale.
\end{minipage}
\end{tcolorbox}
\vspace{3mm}
\end{center}
This practical criterion follows by simply considering the radius of convergence of a given EFT expansion. 
Even when the scalar manifold formally admits SMEFT, one may encounter issues with convergence when this description is evaluated at our physical vacuum, in that the predictions may converge much more slowly than the corresponding HEFT, or not at all. 
Either situation favors the use of HEFT when intepreting data.

To illustrate these considerations, we will focus on the set of zero-derivative bosonic operators in SMEFT,
\begin{align}
\Lag &\supset  \sum_k C_k\s \frac{\lambda}{\Lambda^{2\s k-4}} |H|^{2\s k} \notag \\
&= \sum_k \frac{1}{2^{k}}\s C_k\s \frac{\lambda}{\Lambda^{2\s k-4}}\s\Bigg[ h^{2\s k} + \begin{pmatrix} 2\s k\\ 1 \end{pmatrix} v_0\,h^{2\s k-1} + \dots + \begin{pmatrix} 2\s k\\ 2\s k-1 \end{pmatrix} v_0^{2\s k-1}\,h + v_0^{2\s k}\Bigg] \,,
\label{eq:ToyEFTExpan}
\end{align}
where $k$ is an integer, $\Lambda$ is a dimensionful scale, we have extracted a coupling $\lambda$ with units of the Higgs quartic in order to make the Wilson coefficients $C_k$ dimensionless (when $\hbar \neq 1$), and in the second line we have expanded the Higgs around its vev. 
Although we are focusing on the zero-derivative sector here, similar considerations pertain to the 2-derivative sector for momenta of order $p^2 \sim \lambda\s v_0^2$.

To highlight issues of convergence, we would like to extract a prediction from this EFT for an observable.  
For the sake of definiteness, we will compute the effective dimensionful Higgs cubic coupling parameter $\lambda_{h^3}$ (which of course is only a proxy for an observable):
\begin{align}
\mathcal{L} \supset \lambda_{h^3}\,h^3 = \sum_k \frac{2^{1-k}}{3} \, k\,(k-1)\,(2\s k-1)\,   C_k \left( \frac{ v_0}{\Lambda} \right)^{2\s k-4} \, \lambda \,v_0\,h^3 \,.
\end{align}
Clearly the series converges quickly for $v_0 \ll \Lambda$, but not for $v_0 \sim \Lambda$.
This exemplifies the familiar statement that SMEFT is not a useful description when integrating out states whose mass is at or below the electroweak scale, since one must include the infinite tower of operators to extract predictions. 
What is more surprising, however, is the rate of convergence of the SMEFT expansion for $v_0 \lesssim \Lambda$, in that the SMEFT expansion exists and converges, but may do so much more slowly than HEFT. 
In what follows, we sharpen these statements with a concrete example.

\subsection{On EFT Convergence}

To highlight the issue of convergence, we return to the $\mathbb{Z}_2$ symmetric singlet scalar model presented in \cref{sec:IntOutDoublet}, in the phase where the nontrivial matching arises at one loop.
The resulting EFT potential to all orders in the coupling was given in \cref{eqn:LEFT022HDM}, which we reproduce here (the $H$ kinetic term is canonically normalized):
\begin{align}
{V^\text{Full}_{{\text{Eff}}}}\left( H \right) &=  -\mu_H^2{|H|^2} + \lambda_H {|H|^4} - \frac{1}{{{64\s\pi^2}}}{\left( {m^2 + {\kappa}\s{{|H|}^2}} \right)^2}\left( {\ln \frac{{{\mu^2}}}{{m^2 + {\kappa}\s{{|H|}^2}}} + \frac{3}{2}} \right) \,.
\label{eqn:VEFT2HDM}
\end{align}
Here $m$ is the mass parameter for the scalar singlet and $\kappa$ is the cross quartic coupling that connects the singlet to the Higgs doublet; see \cref{sec:IntOutDoublet} for details.
Extracting EFT Wilson coefficients from this potential requires Taylor expanding in terms of the Higgs field and truncating.
From this point of view, SMEFT and HEFT can be simply thought of as two different choices for how to expand \cref{eqn:VEFT2HDM}.

In SMEFT, we work with the full Higgs doublet $H$, so that the EFT is derived by expanding in powers of
\begin{align}
X_\text{SMEFT} = \frac{\kappa\s|H|^2}{m^2} = \frac{\kappa\s v_0^2}{2\s m^2}\s\bigg(1+\frac{h}{v_0}\bigg)^2 \,.
\label{eq:XSMEFTDef}
\end{align}
The resulting EFT potential takes the form
\begin{align}
{V_{{\text{SMEFT}}}}( H ) &= -\mu_H^2\s|H|^2 + \lambda_H \s|H|^4 - \frac{{m^4}}{{{64\s\pi^2}}}\Bigg[
 {\ln \frac{{{\mu^2}}}{{m^2}} + \frac{3}{2}}  + \left( \ln\frac{\mu^2}{m^2} + 1 \right) 2X_\text{SMEFT}  \notag\\[6pt]
&\hspace{64pt}  + \bigg(\ln \frac{\mu^2}{m^2}\bigg) X_\text{SMEFT}^2 + \sum\limits_{k=3}^{k_\text{max}} \frac{2\s(-1)^k}{k\s(k-1)(k-2)} X_\text{SMEFT}^k
 \Bigg] \,, \label{eqn:VEFT2HDMSMEFT}
\end{align}
where we see that an infinite tower of operators have been generated, as in the toy expressions given in \cref{eq:ToyEFTExpan}. In this case, $\Lambda^2 = m^2 / \kappa$, $\lambda = \kappa^2/(16 \pi^2)$, and the Wilson coefficients are all generated at one loop.

On the other hand, if our goal is to describe this theory using HEFT, we first express $H$ in terms of the vev, the physical Higgs boson, and the Goldstones following \cref{sec:MapBetween}.
Given the form of \cref{eqn:VEFT2HDM}, it is convenient to define the expansion parameter
\begin{equation}
 X_\text{HEFT} \equiv \frac{\kappa\s \Big(|H |^2-\frac{1}{2}\s v_0^2 \Big)}{m^2 + \frac{1}{2}\s \kappa\s v_0^2} =  \frac{\kappa\s v_0^2}{2\s m^2 +\kappa\s v_0^2} \Bigg[2\s \frac{h}{v_0} + \bigg(\frac{h}{v_0}\bigg)^2\s\Bigg] \,,
\label{eq:XHEFTDef}
\end{equation}
so that
\begin{equation}
m^2 + {\kappa}\s{|H|^2} = \bigg(m^2 + \frac{1}{2}\s{\kappa}\s{v_0^2}\bigg)\s (1+X_\text{HEFT}) \,.
\end{equation}
It is then straightforward to expand in powers of $X_\text{HEFT}$:
\begin{align}
\hspace{-3pt}{V_{{\text{HEFT}}}}\left( h \right) =\,& -\frac{1}{2}\s \mu_H^2\s(v_0+h)^2 + \frac{1}{4}\s \lambda_H\s(v_0+h)^4 \notag\\[6pt]
& - \frac{{{{\big( {m^2 + \frac{1}{2}\s{\kappa}\s{v_0^2}} \big)}^2}}}{{{64\s\pi^2}}}\Bigg[ \ln \frac{\mu^2}{m^2 + \frac{1}{2}\s \kappa \s v_0^2} + \frac{3}{2} +\bigg( {\ln \frac{{{\mu^2}}}{{m^2 + \frac{1}{2}\s{\kappa}\s{v_0^2}}} + 1} \bigg)\s 2\s X_\text{HEFT}   \notag\\[6pt]
&\hspace{30pt} + \bigg( {\ln \frac{{{\mu^2}}}{{m^2 + \frac{1}{2}\s{\kappa}\s{v_0^2}}}} \bigg)\s X_\text{HEFT}^2 + \sum\limits_{k = 3}^{k_\text{max}}  {\frac{{2\s{{( { - 1} )}^k}}}{{k\s ( {k - 1} )\s( {k - 2} )}}{\s X_\text{HEFT}^k}}
 \Bigg] \,. \label{eqn:VEFT2HDMHEFT}
\end{align}

\begin{figure}
\begin{center}
  \begin{tikzpicture}[smeftstyle/.style={orange},heftstyle/.style={blue}]
\draw[thick,->] (-3,0) -- (6,0);
\draw[thick,->] (0,-3.6) -- (0,3.6);
\coordinate (smeft) at (0,0);
\coordinate (heft) at (2,0);
\coordinate (sing) at (-1.5,0);
\draw (sing) node[cross=4pt] {};
\draw[decorate,decoration={snake}] (sing) -- (-3,0);
\node[label=-135:{$-\frac{m^2}{\kappa}$}] at (sing) {};
\filldraw[smeftstyle] (smeft) circle (2pt);
\filldraw[heftstyle,label={[label distance=50pt]90:{$\frac{1}{2} v_0^2$}}] (heft) circle (2pt);
\node[smeftstyle] at (1.9,1.2) {SMEFT};
\node[heftstyle] at (4.7,3.0) {HEFT};
\node[above] at (heft) {$\frac{1}{2} v_0^2$};
\node at (-2.6,2.8) {$\left|H\right|^2$};
\draw[rounded corners] (-3,2.3) -- (-2.1,2.3) -- (-2.1,3.2);
\draw[smeftstyle] (smeft) circle (1.5);
\draw[heftstyle] (heft) circle (3.5);
\draw[Latex-Latex] (smeft) -- (-30:1.5) node[midway,label={[smeftstyle,label distance=-12pt]-135:{\scriptsize$\frac{1}{2} v_0^2 r$}}] {};
\draw[Latex-Latex] (heft) -- +(-30:3.5) node[midway,label={[heftstyle,label distance=-12pt]-135:{\scriptsize$\frac{1}{2} v_0^2 (r+1)$}}] {};
\end{tikzpicture}
\caption{Radii of convergence of the SMEFT (orange) and HEFT (blue) expansions in the complex plane of $\left|H\right|^2$ as a function of $r = m^2 / (\kappa\s v_0^2/2)$. The full potential has a branch point at $\left|H\right|^2 = -m^2/\kappa$. \label{fig:potsingularities}}
\end{center}
\end{figure}
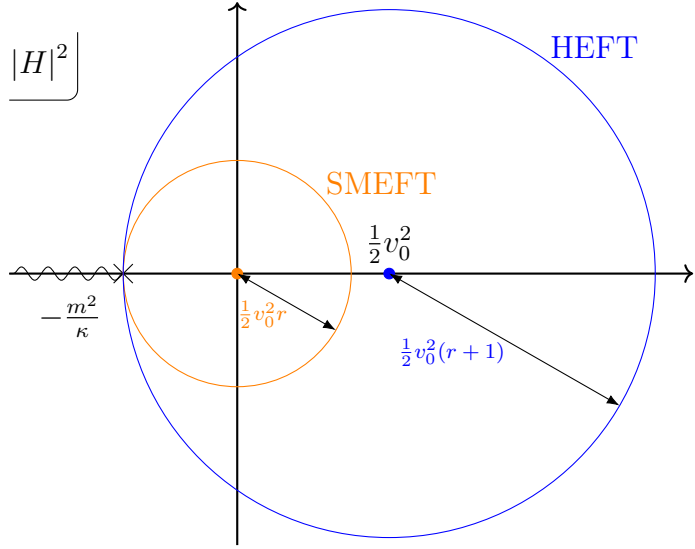

Now we are set up to explore the convergence properties of the two EFTs. In the complex plane of $\left|H\right|^2$, depicted in \cref{fig:potsingularities}, the full potential \cref{eqn:VEFT2HDM} has a branch point at $\left|H\right|^2 = -m^2/\kappa$. 
The SMEFT and HEFT potentials are Taylor expansions about $\left|H\right|^2 = 0$ and $\left|H\right|^2=v_0^2/2$ respectively. 
We introduce the ratio between the two mass scales
\begin{align}
r\equiv\frac{m^2}{\frac{1}{2}\s \kappa\s v_0^2} \,,
\label{eq:rDef}
\end{align}
where $r \rightarrow \infty$ corresponds to the $m^2 \rightarrow \infty$ decoupling limit for the singlet scalar, such that the new physics contributions to observables vanish. 
In terms of $r$, the SMEFT and HEFT expansions have respective radii of convergence $v_0^2\s r/2$ and $v_0^2\s (r+1)/2$, the latter being strictly greater.

Typically, we want to evaluate the expansions somewhere at or between the invariant point ($|H|^2 = 0$) and our low energy vacuum ($|H|^2= v_0^2/2$), typically about the latter when making predictions for experiments. 
Depending on the size of $r$, there are three different scenarios:
\begin{itemize}
 \item $r > 1$: Both expansions converge to reproduce \cref{eqn:VEFT2HDM}.  
 \item $0<r\le1$: This corresponds to the situation where the majority of the mass of the scalar singlet is due to the vev of the SM-like Higgs. As one moves along the $|H|^2$ axis, the HEFT expansion converges slowly since $\frac{1}{2}\le\frac{1}{r+1}<1$, while SMEFT may not converge at all. Importantly, SMEFT does not converge at our low energy vacuum, as it lies outside the expansion's radius of convergence.
 \item $r=0$: This corresponds to the situation where the mass of the scalar singlet is entirely due to the vev of the SM-like Higgs.  Therefore, the SMEFT expansion \emph{does not exist}, by the criteria established in preceding sections. HEFT converges slowly in the region of interest about the physical vacuum, but not at the fixed point.
\end{itemize}
Thus HEFT is the appropriate EFT in two of the three scenarios, including one in which the SMEFT expansion exists but does not converge at our vacuum.

However, it is interesting to focus on $r \gtrsim 1$, where both HEFT and SMEFT expansions exist and converge at our physical vacuum. 
In this situation, the rate of convergence of the HEFT expansion can be much faster, making it the preferred EFT parametrization of Higgs data despite the formal validity of the SMEFT expansion. 
To make this discussion concrete, we can explore the convergence of the SMEFT expansion at the low energy vacuum as quantified by its impact on the Higgs trilinear and quartic couplings. 
As we will show, the HEFT expansion is effectively exact here.

To make the comparison, we expand the EFT potential $V(H)$ as a function of $h$, and map it onto the parametrization
\begin{equation}
{V_{{\text{EFT}}}}\left( H \right) = {c_0} + {c_1}\s v_0\s h + \frac{1}{2}\s{c_2}\s{h^2} + {c_3}\s v_0\s {h^3} + \frac{1}{4}\s {c_4}\s{h^4} + \cdots \,.
\end{equation}
The expansion coefficients $c_i$ are then given by
\begin{subequations}\label{eqn:ExpansionCoefficients}
\begin{align}
{c_1} &= -\mu_H^2 + \lambda_H\s v_0^2 + \frac{1}{{{16\s\pi^2}}}{\s\delta_1}  \\
{c_2} &= -\mu_H^2 + 3 \lambda_H\s v_0^2 + \frac{1}{{{16\s\pi^2}}}{\s\delta_2}  \\
{c_3} &= \lambda_H + \frac{1}{{{16\s\pi^2}}}{\s\delta_3}  \\
{c_4} &= \lambda_H + \frac{1}{{{16\s\pi^2}}}{\s\delta_4} \,,
\end{align}
\end{subequations}
where the $\delta_i$ are the one-loop corrections.
We can then calculate the $\delta_i$ for our three cases: all-orders using \cref{eqn:VEFT2HDM}, SMEFT using \cref{eqn:VEFT2HDMSMEFT}, and HEFT using \cref{eqn:VEFT2HDMHEFT}.
All three potentials are the same at tree level, but they differ at one loop.

In this toy model, we have four independent Lagrangian parameters $\mu_H^2$, $m^2$, $\lambda_H$, and $\kappa$. 
We take the vev condition $c_1=0$ and the mass condition $c_2=m_h^2$ to eliminate $\mu_H^2$ and $\lambda_H$ in favor of the more physical parameters vev $v_0$ and $m_h^2$. 
This has the benefit that the scheme dependence of the corrections to $c_3$ and $c_4$ is eliminated, and differences between EFT expansions are contained entirely in their corrections to the cubic and quartic couplings at this order. 
We parameterize this using
\begin{subequations}
\begin{align}
{c_3} &= \frac{{m_h^2}}{{2\s{v_0^2}}}\bigg[ {1 + \frac{1}{{{16\s\pi^2}}}\s\delta {c_3}} \bigg]  \\[5pt]
{c_4} &= \frac{{m_h^2}}{{2\s{v_0^2}}}\bigg[ {1 + \frac{1}{{{16\s\pi^2}}}\s\delta {c_4}} \bigg] \,,
\end{align}
\end{subequations}
where the one-loop corrections can be computed from the expansion coefficients in Eq.~\eqref{eqn:ExpansionCoefficients} as
\begin{subequations}\label{eqn:CorrectionRelation}
\begin{align}
\delta {c_3} &= \frac{{{\delta_1} - {\delta_2} + 2\s{v_0^2}\s{\delta_3}}}{{m_h^2}} \\[5pt]
\delta {c_4} &= \frac{{{\delta_1} - {\delta_2} + 2\s{v_0^2}\s{\delta_4}}}{{m_h^2}} \,.
\end{align}
\end{subequations}

Given this formalism, it is then straightforward to expand a given potential and derive expressions for these corrections to the Higgs cubic and quartic.
For concreteness, we will provide the resulting expressions derived from the all-orders potential, see \cref{eqn:VEFT2HDM}:
\begin{subequations}\label{eqn:c3c4Full}
\begin{align}
\delta {c_3^\text{Full}} &= \frac{{\kappa^3\s{v_0^4}}}{{3\s m_h^2\big( {2\s m^2 + {\kappa}{v_0^2}} \big)}}  \\[5pt]
\delta {c_4^\text{Full}} &= \frac{{4\s \kappa^3\s{v_0^4}\big( {3\s m^2 + {\kappa}\s{v_0^2}} \big)}}{{3\s m_h^2\s {{\big( {2\s m^2 + {\kappa}\s{v_0^2}} \big)}^2}}} \,\,,
\end{align}
\end{subequations}
the SMEFT potential, see \cref{eqn:VEFT2HDMSMEFT}:
\begin{subequations}\label{eqn:c3c4SMEFT}
\begin{align}
\delta {c_3^\text{SMEFT}} &= \sum\limits_{k=3}^{k_\text{max}}\frac{4\s m^4\s (-1)^{k+1}}{3\s v_0^2\s m_h^2} \bigg(\frac{\kappa \s v_0^2}{2\s m^2}\bigg)^k  \\[10pt]
\delta {c_4^\text{SMEFT}} &= \sum\limits_{k=3}^{k_\text{max}}\frac{8\s k\s m^4\s (-1)^{k+1}}{3\s v_0^2\s m_h^2} \bigg(\frac{\kappa \s v_0^2}{2\s m^2}\bigg)^k \,\,,
\end{align}
\end{subequations}
and the HEFT potential, see \cref{eqn:VEFT2HDMHEFT}:
\begin{subequations}\label{eqn:c3c4HEFT}
\begin{align}
\delta {c_3^\text{HEFT}} &= \left\{\begin{array}{ll}
0 &\hspace{32pt}\text{for } k_\text{max} < 3 \\[8pt]
\dfrac{\kappa^3\s v_0^4}{3\s m_h^2\s \big(2\s m^2 + \kappa v_0^2\big)}&\hspace{32pt}\text{for } k_\text{max} \geq 3 \end{array} \right.  \\[10pt]
\delta {c_4^\text{HEFT}} &= \left\{\begin{array}{ll}
0 &\qquad \text{for } k_\text{max} < 3 \\[8pt]
\dfrac{2 \s \kappa^3\s v_0^4}{m_h^2\s \big(2\s m^2+ \kappa\s v_0^2\big)}&\qquad\text{for } k_\text{max} = 3\\[20pt]
\dfrac{4\s \kappa^3\s v_0^4\s \big(3\s m^2+\kappa \s v_0^2\big)}{3\s m_h^2\s \big(2\s m^2+\kappa \s v_0^2\big)^2}&\qquad\text{for } k_\text{max} > 3 \end{array} \right. \,\,.
\end{align}
\end{subequations}
Note that when we take a truncation order $k_\text{max}\le2$, namely up to renormalizable interactions, both SMEFT and HEFT expansions of our EFT yield no corrections to $\delta c_3$ and $\delta c_4$, due to the unbroken $\mathbb{Z}_2$ symmetry in our UV model.

\begin{figure}[t!]
\centering
\includegraphics[width=0.9\textwidth]{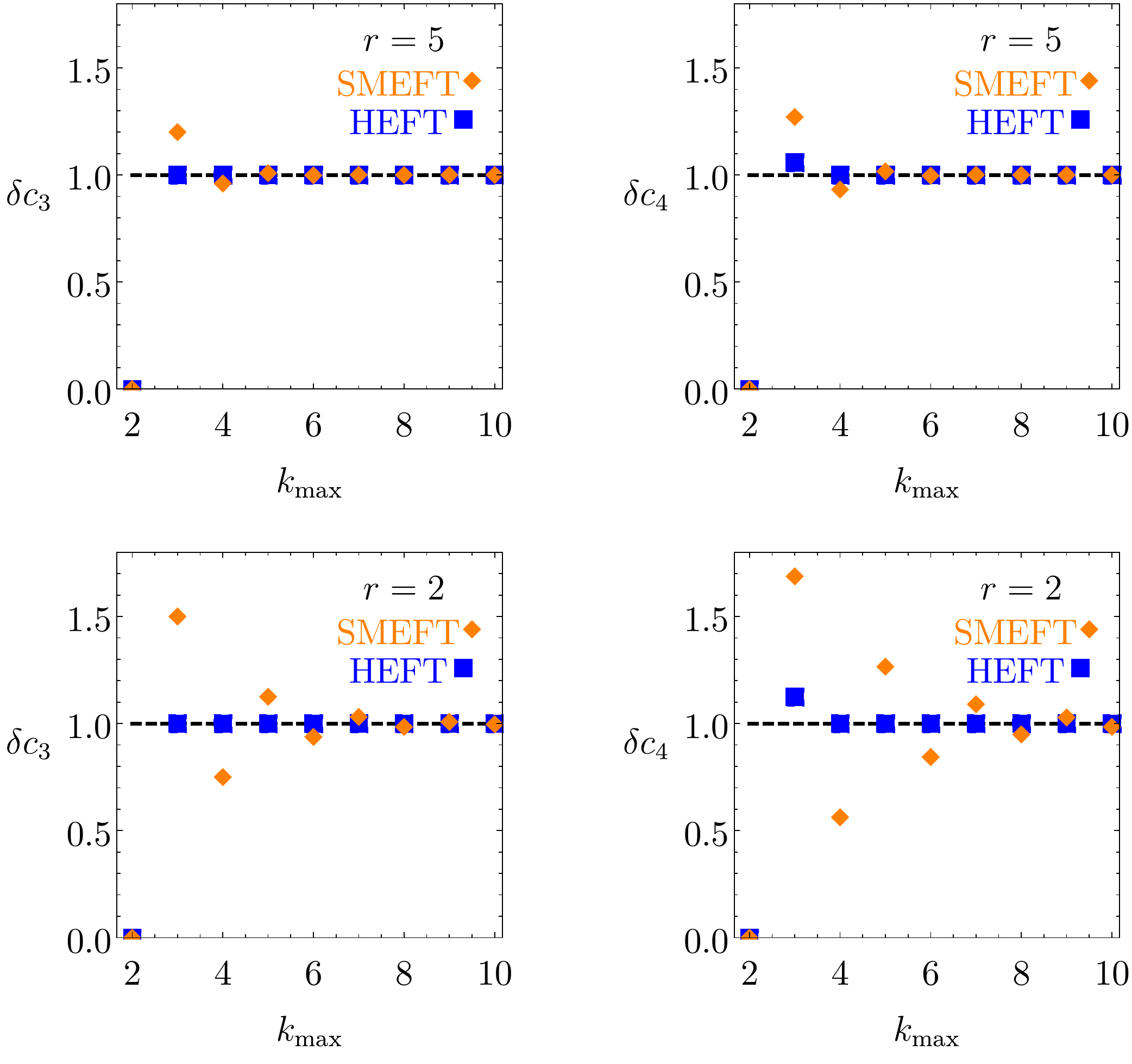}
\caption{Corrections to the Higgs trilinear (left panels) and quartic (right panels) couplings extracted from truncated SMEFT (orange diamond) and HEFT (blue square) expansions, normalized to the actual all-order prediction in \cref{eqn:c3c4Full}. The upper (lower) panels are for $r=5$ ($r=2$), see \cref{eq:rDef} for the definition.}\label{fig:c3c4}
\end{figure}

We see from \cref{eqn:c3c4SMEFT} that observables computed with the SMEFT expansion converge as $\Big(\frac{\kappa\s v_0^2}{2\s m^2}\Big)^k=\Big(\frac{1}{r}\Big)^k$, in accordance with our previous discussion on the convergence radius. To visualize this convergence rate and contrast it with the HEFT, \cref{fig:c3c4} shows the SMEFT and HEFT predictions for $\delta c_3$ and $\delta c_4$ normalized to the all-order potential predictions, where the horizontal axis $k_\text{max}$ is the EFT truncation order as defined in \cref{eqn:VEFT2HDMSMEFT,eqn:VEFT2HDMHEFT}.
There are two scenarios shown in \cref{fig:c3c4} with $r \gtrsim 1$, namely $r=5$ and $r=2$. Although both SMEFT and (trivially) HEFT converge given that $r>1$ in these cases, the truncation order $k_\text{max}$ required for SMEFT to provide a reasonable approximation of the all-order result increases rapidly as $r \rightarrow 1$. 
This highlights the sense in which HEFT can provide the optimal fit to data even in situations where the SMEFT expansion exists and converges in our vacuum, a consideration above and beyond the formal criteria explored in preceding sections.
Clearly these lessons generalize beyond this example, hence the Practical Criterion stated at the beginning of this section.

The slow convergence of the SMEFT expansion when $r \sim 1$ was previously noted and extensively studied in \cite{Englert:2014uua,Brehmer:2015rna}. 
These authors showed that the numerical agreement between a finite truncation of SMEFT and exact predictions in a perturbative UV completion could be significantly improved by defining the scale $\Lambda$ as the physical mass of new particles in the broken phase, including contributions from the Higgs vev, a prescription called ``$v$-improved matching.'' 
For this prescription to be effective, the operators retained in the finite truncation must span the observables of interest. 
In terms of the geometric picture developed in this paper, more conventional matching to a finite truncation of SMEFT (using a scale $\Lambda$ defined by the masses of the new particles in the {\it unbroken} phase) amounts to constructing a simplified EFT manifold that is locally ``tangent'' to the true EFT manifold at the fixed point. 
In contrast, $v$-improved matching to a finite truncation of SMEFT constructs a simplified EFT manifold that is ``tangent'' to the true EFT manifold at the observed vacuum. 
When $r \gg 1$ the true EFT manifold has small and slowly-varying curvature, so the simplified EFT manifolds obtained from conventional and $v$-improved matching are similar. 
But when $r \sim 1$ they may differ substantially, the latter providing a better fit to measurements made in the observed vacuum.

From a practical perspective, $v$-improved matching is carried out by first matching in HEFT (invariably the case when matching in the broken phase) and subsequently writing all occurrences of $h, \vec n$ in terms of $H$. 
This is always possible in the regime of interest $r \gtrsim 1$ since both SMEFT and HEFT parametrizations are well defined. 
For example, in the singlet case explored above, $v$-improved matching is obtained by working in terms of the HEFT expansion in \eqref{eq:XHEFTDef} parameterized by the field $H$. 
In this sense, the improved convergence for $r \sim 1$ observed here using the HEFT parametrization coincides with $v$-improved matching in \cite{Englert:2014uua,Brehmer:2015rna}.

When one is in a situation where the convergence of SMEFT is suspect, our point of view is that it is most transparent to make predictions and derive constraints on the HEFT parameter space directly. Of course, in the event of observed deviations from the SM, the slow (or non-) convergence of a SMEFT parameterization is likely to be among the first indications that the deviation is due to new physics near the weak scale. 

\section{Conclusions}
\label{sec:Conclusions}
In this paper, we have explored the utility of the HEFT description by asking ``What characterizes a UV theory that \emph{cannot} be matched onto SMEFT?''
While this is an easy question to state, this paper clearly shows that it is not a simple question to answer in practice.
A major source of complication is inherent to quantum field theory itself: physical observables are unchanged by analytic field redefinitions.
Therefore, we found it advantageous to frame our answer using geometric quantities that are invariant under coordinate redefinitions, following in the footsteps of Alonso, Jenkins, and Manohar  \cite{Alonso:2015fsp, Alonso:2016oah} and connecting UV and IR geometries through the notion of the EFT scalar submanifold.
This led us to formulate and prove a set of basis independent curvature criteria that circumvent field redefinition ambiguities by relying on the Ricci scalar and scalar potential (as well as covariant derivatives thereof) defined on the EFT scalar submanifold.
We additionally examined the possible effects that integrating out heavy physics could have on different EFT submanifold branches.
This led us to a compelling picture for what differentiates SMEFT from the more general space of possibilities realized by HEFT.
The electroweak symmetric point must exist on the EFT submanifold, and the constraints induced from integrating out the heavy states must yield a smooth curve that connects this fixed point to the observed vacuum where electroweak symmetry is broken.
When one encounters intersections between branches, or no branch can be found that smoothly connects these two special points, then HEFT is the only possible low energy description.

We then related these observations to a variety of new physics scenarios, from which two general lessons emerged.
First, if the UV model contains a state whose mass is entirely determined by electroweak symmetry breaking, then the theory must be matched onto HEFT.
This is intuitive since this BSM state is a massless fluctuation when the theory is expanded about the fixed point, and thus yields non-analytic contributions to the effective action upon being integrated out.
The second scenario arises when the UV model provides new sources of electroweak symmetry breaking.
Then the fixed point is disconnected from the physical vacuum, and again HEFT emerges in the IR.
Finally, we argued that when integrating out a state whose mass is near the weak scale, one would find that although it might be possible to write the theory as SMEFT, the convergence of the EFT expansion will be significantly improved by matching onto HEFT instead.
Taken together, we have provided a compelling set of reasons to study the phenomenological consequences of HEFT.

We encountered many subtleties along the way.
In particular, we provided cases where field redefinitions in terms of $h$ could produce ``fake'' non-analyticities in the effective Lagrangian expressed in terms of $H$, and we showed how the geometric approach is insensitive to such pathological issues.
We showed that applying curvature criteria (even in leading-order form) requires working to all orders in the light fields when performing matching calculations. This motivated utilizing functional methods to derive some new general one-loop formulas for matching onto the two derivative terms in the effective Lagrangian.
We additionally explored the impact of basis changes on the branches that define different EFT submanifolds, which led us to discover an optimal choice one can make when matching, that we call the ``unitary basis.''
This approach paved the way to a simple characterization of the EFTs that emerge from the two Higgs doublet and Higgs triplet extensions of the SM.

Looking forward, there are many directions that would be exciting to explore.
It would be interesting to systematically apply the geometric language to EFT scattering amplitudes, expressing them in a way that was manifestly invariant under non-derivative field redefinitions \cite{amplitudes}.
The simplicity of performing matching calculations in the unitary basis suggests that it would be fruitful to apply to other extensions of the SM scalar sector.
Since all of the work presented here (other than the Higgs triplet model) assumed fundamental $O(4)$ invariance at the fixed point, it would be interesting to explore extending the geometric framework and curvature criteria developed here to non-custodially symmetric examples \cite{twohdm}.
More speculatively, the role of derivative field redefinitions motivates extending the geometric formulation to theories with four or more derivatives.
Finally, there is important phenomenological work to do \cite{higglons}; it would be valuable to identify measurements one could make that would test if our low energy EFT could be SMEFT or must be HEFT.
This paper makes it clear that HEFT covers an interesting class of BSM models, and is broadly relevant to describing extensions of the SM lying at the edge of experimental reach.

\acknowledgments

We thank Zackaria Chacko, Spencer Chang, Isabel Garcia Garcia, Seth Koren, Graham Kribs, Ian Lewis, Markus Luty, and Matthew McCullough for helpful discussions.
TC and XL are supported by the U.S. Department of Energy, under grant number DE-SC0011640.
NC is supported by the U.S. Department of Energy under the grant DE-SC0011702.
DS has received funding from the European Union's Horizon 2020 research and innovation programme under the Marie Skłodowska-Curie grant agreement No.\ 754496.
We are grateful to the KITP and support from the National Science Foundation under Grant No.\ NSF PHY-1748958.

\appendix
\section*{Appendices}
\addcontentsline{toc}{section}{\protect\numberline{}Appendices}%
\addtocontents{toc}{\setcounter{tocdepth}{1}}
\section{Scalar Curvature in HEFT}
\label{appsec:ScalarCurveature}
In this appendix, we evaluate the Ricci scalar $R$ relevant for the general parametrization of the HEFT Lagrangian in \cref{eqn:LHEFT}.
The metric is determined from the two-derivative term, which we reproduce here
\begin{equation}
\Lag_\text{HEFT}^{(2)} = \frac{1}{2}{\left[ {K\left( h \right)} \right]^2}{\left( {\partial h} \right)^2} + \frac{1}{2}{\left[v\s{F\left( h \right)} \right]^2}{( {\partial \vec n} )^2} \,.
\label{eqn:LHEFTKE}
\end{equation}
Recall that $\vec n\left(\pi\right)\in S^3$ is a four-component unit vector, which satisfies $\vec n \cdot \vec n = 1$:
\begin{equation}
\vec n = \left( {\begin{array}{*{20}{c}}
{{n_1}}\\
{{n_2}}\\
{{n_3}}\\
{\sqrt {1 - n_1^2 - n_2^2 - n_3^2} }
\end{array}} \right) \,.
\end{equation}
In order to identify the metric, we choose the coordinates $\mqty(h, n_1, n_2, \cdots, n_{N_\varphi})$ with $N_\varphi=3$.
Using
\begin{equation}
{( {\partial \vec n} )^2} = \left( {{\delta_{ij}} + \frac{{{n_i}{n_j}}}{{1 - {n^2}}}} \right)\left( {{\partial^\mu }{n_i}} \right)\left( {{\partial_\mu }{n_j}} \right) \,,
\end{equation}
we can rewrite the HEFT Lagrangian into the form
\begin{equation}
\Lag_\text{HEFT}^{(2)} = \frac{1}{2}{{\left[ {K\left( h \right)} \right]}^2}{{\left( {\partial h} \right)}^2} + \frac{1}{2}{{\left[ v\s F\left( h \right) \right]}^2}\left( {{\delta_{ij}} + \frac{{{n_i}{n_j}}}{{1 - {n^2}}}} \right)\left( {{\partial^\mu }{n_i}} \right)\left( {{\partial_\mu }{n_j}} \right) \,.
\end{equation}
\clearpage
Here $i,j$ runs from $1$ to $N_\varphi=3$.
Following Eq. (4.1) in~AJM~\cite{Alonso:2016oah}, we identify the non-zero elements of the metric to be
\begin{equation}
\left\{ \begin{array}{l}
{g_{hh}} = {K^2}\\[10pt]
{g_{ij}} = {v^2}{F^2}\left( {{\delta_{ij}} + \dfrac{{{n_i}{n_j}}}{{1 - {n^2}}}} \right)
\end{array} \right.  \,,\quad  \left\{ \begin{array}{l}
{g^{hh}} = \dfrac{1}{K^2}\\[10pt]
{g^{ij}} = \dfrac{1}{v^2 F^2} \left( {{\delta_{ij}} - {n_i}{n_j}} \right)
\end{array} \right. \,.
\label{eq:DefMetric}
\end{equation}
Next, we use these to compute the Christoffel symbols:
\begin{subequations}
\begin{align}
{\Gamma^h}_{hh} &= \frac{1}{2}{g^{hh}}\left( {{\partial_h}{g_{hh}}} \right) = \frac{1}{K}\left( {{\partial_h}K} \right) \,, \\[5pt]
{\Gamma^h}_{hi} &= \frac{1}{2}{g^{hh}}\left( {{\partial_i}{g_{hh}}} \right) = 0 \,, \\[5pt]
{\Gamma^h}_{ij} &=  - \frac{1}{2}{g^{hh}}\left( {{\partial_h}{g_{ij}}} \right) =  - \frac{{{v^2}F}}{{{K^2}}}\left( {{\partial_h}F} \right)\left( {{\delta_{ij}} + \frac{{{n_i}{n_j}}}{{1 - {n^2}}}} \right) \,, \\[5pt]
{\Gamma^i}_{hh} &=  - \frac{1}{2}{g^{ij}}\left( {{\partial_j}{g_{hh}}} \right) = 0 \,, \\[5pt]
{\Gamma^i}_{hj} &= \frac{1}{2}{g^{ik}}\left( \partial_h g_{kj} \right) = \frac{1}{F}\left( {{\partial_h}F} \right){\delta_{ij}} \,, \\[5pt]
{\Gamma^i}_{jk} &=  - \frac{1}{2}{g^{il}}\left( {{\partial_l}{g_{jk}} - {\partial_j}{g_{kl}} - {\partial_k}{g_{jl}}} \right) = {n_i}\left( {{\delta_{jk}} + \frac{{{n_j}{n_k}}}{{1 - {n^2}}}} \right) \,.
\end{align}
\end{subequations}
These can then be used to compute the non-zero components of the Riemann tensor:
\begin{subequations}
\begin{align}
{R^h}_{ihj} &= {\partial_h}{\Gamma^h}_{ij} + {\Gamma^h}_{hh}{\Gamma^h}_{ij} - {\Gamma^h}_{jk}{\Gamma^k}_{hi} \notag\\[5pt]
 &= -\frac{{v^2 F}}{{{K^2}}}\left[ {\left( {\partial_h^2F} \right) - \left( {{\partial_h}K} \right)\left( {\frac{1}{K}{\partial_h}F} \right)} \right]\left( {{\delta_{ij}} + \frac{{{n_i}{n_j}}}{{1 - {n^2}}}} \right) \,, \\[8pt]
{R^i}_{hhj} &= {\partial_h}{\Gamma^i}_{hj} + {\Gamma^i}_{hk}{\Gamma^k}_{hj} - {\Gamma^h}_{hh}{\Gamma^i}_{hj} \notag\\[5pt]
 &= \frac{1}{F}\left[ {\left( {\partial_h^2F} \right) - \left( {{\partial_h}K} \right)\left( {\frac{1}{K}{\partial_h}F} \right)} \right]{\delta_{ij}} \,, \\[8pt]
{R^k}_{ijk} &= {\partial_j}{\Gamma^k}_{ik} - {\partial_k}{\Gamma^k}_{ij} + {\Gamma^k}_{hj}{\Gamma^h}_{ik} + {\Gamma^k}_{jm}{\Gamma^m}_{ik} - {\Gamma^k}_{hk}{\Gamma^h}_{ij} - {\Gamma^k}_{km}{\Gamma^m}_{ij} \notag\\[5pt]
 &=  - \left( {{N_\varphi } - 1} \right)\left( {{\delta_{ij}} + \frac{{{n_i}{n_j}}}{{1 - {n^2}}}} \right)\left[ {1 - {{\left( {\frac{{{v}}}{K}{\partial_h}F} \right)}^2}} \right] \,.
\end{align}
\end{subequations}
These in turn yield the following non-zero components of the Ricci tensor:
\begin{subequations}
\begin{align}
{R_{hh}} &=  - {R^i}_{hhi} =  - \frac{{{N_\varphi }}}{F}\left[ {\left( {\partial_h^2F} \right) - \left( {{\partial_h}K} \right)\left( {\frac{1}{K}{\partial_h}F} \right)} \right] \,, \\[8pt]
{R_{ij}} &= {R^h}_{ihj} - {R^k}_{ijk} = \Bigg\{ - \frac{{v^2F}}{{{K^2}}}\left[ {\left( {\partial_h^2F} \right) - \left( {{\partial_h}K} \right)\left( {\frac{1}{K}{\partial_h}F} \right)} \right] \notag\\
&\hspace{110pt} + \left( {{N_\varphi } - 1} \right)\left[ {1 - {{\left( {\frac{{{v}}}{K}{\partial_h}F} \right)}^2}} \right] \Bigg\}\left( {{\delta_{ij}} + \frac{{{n_i}{n_j}}}{{1 - {n^2}}}} \right) \,.
\end{align}
\end{subequations}
Finally, we obtain the Ricci scalar curvature:
\begin{align}
R &= {g^{hh}}{R_{hh}} + {g^{ij}}{R_{ij}} \notag\\[5pt]
&= - \frac{{2{N_\varphi }}}{{{K^2}F}}\left[ {\left( {\partial_h^2F} \right) - \left( {{\partial_h}K} \right)\left( {\frac{1}{K}{\partial_h}F} \right)} \right] + \frac{{{N_\varphi }\left( {{N_\varphi } - 1} \right)}}{{v^2{F^2}}}\left[ {1 - {{\left( {\frac{{{v}}}{K}{\partial_h}F} \right)}^2}} \right] \,.
\label{eqn:Rgeneral}
\end{align}
This expression is used extensively in the main text.

\section{Proving the Curvature Criteria}
\label{appsec:CurvatureCriterionK1}

In this appendix, we provide the detailed proof that the Curvature Criteria presented in \cref{subsec:CurvatureCriterion} are a basis independent generalization of the Fixed Basis Criteria provided in \cref{subsubsec:K1Criterion}.
Specifically, we will show that in the basis where the $h$ kinetic term is canonical, and the HEFT Lagrangian takes the form given in \cref{eqn:LHEFTK1}, the following two sets of conditions for when HEFT can be expressed as SMEFT are equivalent
\begin{equation}\renewcommand\arraystretch{1.6}
\left. \begin{array}{l}
{F^{\left( {2k} \right)}}\s\left( {{h_*}} \right) = 0, \, \forall k \in \mathbb{N} \\
v\s F'\s\left( {{h_*}} \right) = 1\\
{V^{\left( {2k + 1} \right)}}\s\left( {{h_*}} \right) = 0, \, \forall k \in \mathbb{N}
\end{array} \right\}
\Longleftrightarrow
\left\{ \begin{array}{l}
F\s\left( {{h_*}} \right) = 0\\
{\left. {{\nabla^{{\mu_1}}}{\nabla_{{\mu_1}}} \cdots {\nabla^{{\mu_n}}}{\nabla_{{\mu_n}}}R} \right|_{{h_*}}} < \infty , \, \forall n \in \mathbb{N} \\
{\left. {{\nabla^{{\mu_1}}}{\nabla_{{\mu_1}}} \cdots {\nabla^{{\mu_n}}}{\nabla_{{\mu_n}}}V} \right|_{{h_*}}} < \infty , \, \forall n \in \mathbb{N}
\end{array} \right. \,. \label{eqn:EquivalenceApp}
\end{equation}
Throughout, we will additionally assume that the functions $F(h)$ and $V(h)$ are analytic in $h$ about the point $h=h_*$.
Following the discussion in the main body, if either $F(h)$ or $V(h)$ are non-analytic at $h_*$, then the theory is not SMEFT.

We begin by defining some notation.
First, we introduce the variable
\begin{equation}
\phi \equiv h - h_* = v_0 + h = \sqrt{\vec\phi \cdot \vec\phi} \,\,,
\end{equation}
which follows from \cref{eqn:HTohnPotential}.
Then $F = F(\phi)$ and $V = V(\phi)$ are both single-argument functions of $\phi$.
Evaluating a function at the $O(4)$ invariant fixed point $h=h_*$ is equivalent to taking $\phi=0$.
Next we introduce a new function $Y(\phi)$:
\begin{equation}
 Y(\phi) \equiv \frac{v^2F^2}{\phi^2} -1  \,,\quad  v\s F = \phi\sqrt{1+Y} \,\,.
 \label{eqn:Ydef}
\end{equation}
Note that the metric derived using \cref{eq:DefMetric} from the Lagrangian in \cref{eqn:LHEFTK1} is flat when $Y(\phi)=0$, so $Y$ can be interpreted as a function that encodes the curvature of the manifold (it is straightforward to see that $R = 0$ by evaluating \cref{eqn:Rgeneral} with $K=1$).

We note a few simple relations between $F$ and $Y$.
First, when $F$ has a well-defined Taylor expansion about $\phi=0$ ($h=h_*$) with vanishing leading term $F\left( \phi=0 \right)=0$, then equivalently $Y$ has a well-defined Taylor expansion about $\phi=0$:
\begin{equation}
 F = \sum\limits_{k=1}^\infty \frac{1}{k!} F_k \phi^k  \qquad\Longleftrightarrow\qquad  Y = \sum\limits_{k=0}^\infty \frac{1}{k!} Y_k \phi^k \,,
 \label{eqn:Y0}
\end{equation}
were we have adopted the abbreviation
\begin{equation}
 f_k \equiv \left. \frac{\text{d}^k f}{\text{d}\phi^k} \right|_{\phi=0} \,,
\end{equation}
which will be used across this Appendix, for all single-argument functions $f(\phi)$.
Furthermore, if $v\s F$ is an \emph{odd power} Taylor series about $\phi=0$ with $v\s F_1=1$, then \cref{eqn:Ydef} implies that $Y$ is an \emph{even power} Taylor series about $\phi=0$ with vanishing leading term:
\begin{equation}
\hspace{-8pt}
F = \sum\limits_{k=0}^\infty \frac{1}{(2k+1)!} F_{2k+1} \phi^{2k+1}  \quad\text{and}\quad  v\s F_1=1  \quad\Longleftrightarrow\quad  Y = \sum\limits_{k=1}^\infty \frac{1}{(2k)!} Y_{2k} \phi^{2k} \,.
\end{equation}
We can therefore rewrite \cref{eqn:EquivalenceApp} as
\begin{equation}\renewcommand\arraystretch{1.6}
\left\{ \begin{array}{l}
Y_0 = Y_{2k+1} = 0\\
V_{2k+1} = 0
\end{array} \right.
\qquad\Longleftrightarrow\quad
\left\{ \begin{array}{l}
{\left. {{\nabla^{{\mu_1}}}{\nabla_{{\mu_1}}} \cdots {\nabla^{{\mu_n}}}{\nabla_{{\mu_n}}}R} \right|_{\phi=0}} < \infty \\
{\left. {{\nabla^{{\mu_1}}}{\nabla_{{\mu_1}}} \cdots {\nabla^{{\mu_n}}}{\nabla_{{\mu_n}}}V} \right|_{\phi=0}} < \infty
\end{array} \right. \,. \label{eqn:EquivalenceRewrite}
\end{equation}
Again, we emphasize the implicit condition that $Y(\phi)$ and $V(\phi)$ are both analytic about $\phi=0$.

To prepare for the proof, we will write down some useful curvature invariants in terms of $\phi$ and $Y$.
From the definition of $Y$ in \cref{eqn:Ydef}, we find
\begin{subequations}
\begin{align}
{v}{\partial_h}F &= \frac{\phi }{{2\sqrt {1 + Y} }}\frac{{\dd Y}}{{\dd\phi }} + \sqrt {1 + Y} \,, \\
{v}\partial_h^2F &= \frac{\phi }{{2\sqrt {1 + Y} }}\frac{{{\dd^2}Y}}{{\dd{\phi^2}}} - \frac{\phi }{{4{{\left( {1 + Y} \right)}^{3/2}}}}{\left( {\frac{{\dd Y}}{{\dd\phi }}} \right)^2} + \frac{1}{{\sqrt {1 + Y} }}\frac{{\dd Y}}{{\dd\phi }} \,.
\end{align}
\end{subequations}
The scalar curvature $R$ derived in \cref{eqn:Rgeneral} can be rewritten as
\begin{align}
R &=  - \frac{{2{N_\varphi }}}{F}\left( {\partial_h^2F} \right) + \frac{{{N_\varphi }\left( {{N_\varphi } - 1} \right)}}{{v^2{F^2}}}\left[ {1 - {{\left( {{v}{\partial_h}F} \right)}^2}} \right] \nonumber \\
 &= - {N_\varphi }\left[ {\frac{1}{{1 + Y}}\frac{{{\dd^2}Y}}{{\dd{\phi^2}}} - \frac{1}{2}\frac{1}{{{{\left( {1 + Y} \right)}^2}}}{{\left( {\frac{{\dd Y}}{{\dd\phi }}} \right)}^2} + \frac{2}{{1 + Y}}\frac{1}{\phi }\frac{{\dd Y}}{{\dd\phi }}} \right] \nonumber \\
 &\quad - {N_\varphi }\left( {{N_\varphi } - 1} \right)\left[ {\frac{1}{4}\frac{1}{{{{\left( {1 + Y} \right)}^2}}}{{\left( {\frac{{\dd Y}}{{\dd\phi }}} \right)}^2} + \frac{1}{{1 + Y}}\frac{1}{\phi }\frac{{\dd Y}}{{\dd\phi }} + \frac{Y}{{1 + Y}}\frac{1}{{{\phi^2}}}} \right] \,,
\label{eqn:RYphi}
\end{align}
where we have taken $K=1$ to simplify these expressions, as is appropriate for the fixed basis HEFT Lagrangian in \cref{eqn:LHEFTK1}.
Next, we can evaluate the action of $\nabla^2$ on a scalar function $X$:
\begin{align}
{\nabla^2}X &= {g^{hh}}{\nabla_h}{\nabla_h}X + {g^{ij}}{\nabla_i}{\nabla_j}X = \left[ {\partial_h^2 + {N_\varphi }{F^{ - 1}}\left( {{\partial_h}F} \right){\partial_h}} \right]X \nonumber \\[5pt]
 &= \left[ {\frac{{{\dd^2}}}{{\dd{\phi^2}}} + {N_\varphi }\left( {\frac{1}{\phi } + \frac{1}{2}\frac{1}{{1 + Y}}\frac{{\dd Y}}{{\dd\phi }}} \right)\frac{\dd}{{\dd\phi }}} \right]X \nonumber \\[5pt]
 &= \left[ \frac{\dd^2}{\dd\phi^2} + \left( N_\varphi + A \right) \frac{1}{\phi} \frac{\dd}{\dd\phi} \right] X \,.
\label{eqn:nabla2XYphi}
\end{align}
Here we have introduced another function for convenience
\begin{equation}
A\left( \phi  \right) \equiv \frac{{{N_\varphi }}}{2}\frac{\phi }{{1 + Y}}\frac{{\dd Y}}{{\dd\phi }} \,.
\label{eqn:Adef}
\end{equation}
For higher scalar derivatives, we have the obvious generalization:
\begin{equation}
\nabla^{2n} X = \nabla^2 \left(\nabla^{2n-2} X\right) = \left[ \frac{\dd^2}{\dd\phi^2} + \left( N_\varphi + A \right) \frac{1}{\phi} \frac{\dd}{\dd\phi} \right] \left(\nabla^{2n-2} X\right) \,.
\end{equation}
This expression will be critical to the inductive proof of \cref{eqn:EquivalenceRewrite} that follows.

\subsection{\boldmath Fixed Basis Criteria $\Longrightarrow$ Curvature Criteria}
\label{appsubsec:Forward}
First, we prove that the Fixed Basis Criteria imply the Curvature Criteria, \ie, the ``$\Rightarrow$'' direction in \cref{eqn:EquivalenceRewrite}.
We start with the assumption that the function $Y(\phi)$ is an even power Taylor series at $\phi=0$ with vanishing leading term
\begin{equation}
{Y_0} = {Y_{2k + 1}} = 0  \qquad\Longrightarrow\qquad  Y = \sum\limits_{k = 1}^\infty  {\frac{1}{{\left( {2k} \right)!}}{Y_{2k}}{\phi^{2k}}} \,.
\end{equation}
This implies that all the functions
\begin{equation}
\frac{1}{{1 + Y}}  \,,\quad  \frac{{{\dd^2}Y}}{{\dd{\phi^2}}}  \,,\quad  {\left( {\frac{{\dd Y}}{{\dd\phi }}} \right)^2}  \,,\quad  \frac{1}{\phi }\frac{{\dd Y}}{{\dd\phi }}  \,,\quad  \frac{Y}{{1 + Y}}\frac{1}{{{\phi^2}}} \,,
\end{equation}
are analytic in $\phi$ about $\phi=0$ and that each can be expressed as an even power series.
Therefore, the scalar curvature $R$ derived in \cref{eqn:RYphi} also has a well-defined Taylor expansion about $\phi=0$ and is an even power series:
\begin{equation}
{\left. R \right|_{\phi  = 0}} < \infty  \,,\qquad  R = \sum\limits_{k = 0}^\infty  {\frac{1}{{\left( {2k} \right)!}}{R_{2k}}{\phi^{2k}}} \,.
\end{equation}
We conclude that the Ricci scalar is finite.

Next, we turn to derivatives of the scalar curvature $\nabla^{2n}R$, which we analyze by applying the result in \cref{eqn:nabla2XYphi}.
Note that the function $A$ defined in \cref{eqn:Adef} is an even power series (with vanishing leading term):
\begin{equation}
A\left( \phi  \right) \equiv \frac{{{N_\varphi }}}{2}\frac{\phi }{{1 + Y}}\frac{{\dd Y}}{{\dd\phi }} = \sum\limits_{k = 1}^\infty  {\frac{1}{{\left( {2k} \right)!}}{A_{2k}}{\phi^{2k}}} \,.
\label{eq:ASeries}
\end{equation}
Therefore, \cref{eqn:nabla2XYphi} tells us that if $R$ is an even power Taylor series about $\phi=0$, then $\nabla^2 R$ will be as well:
\begin{equation}
R = \sum\limits_{k = 0}^\infty  {\frac{1}{{\left( {2k} \right)!}}{R_{2k}}{\phi^{2k}}}  \qquad\Longrightarrow\qquad  {\nabla^2}R = \sum\limits_{k = 0}^\infty  {\frac{1}{{\left( {2k} \right)!}}{{\left( {{\nabla^2}R} \right)}_{2k}}{\phi^{2k}}} \,.
\label{eq:RSeries}
\end{equation}
This can be used as the starting point for an inductive proof; through the successive application of \cref{eqn:nabla2XYphi} augmented by \cref{eq:ASeries}, it is clear that $\nabla^{2n}R$ can each be written as even power Taylor series at $\phi=0$ for all $n\in\mathbb{N}$.
Therefore, they are all finite at $\phi=0$:
\begin{align}
\left. \nabla^{2n}R \right|_{\phi=0}<\infty \,,
\end{align}
for all $n\in\mathbb{N}$.

The same analysis holds for the potential.
Following the same inductive logic, when $V$ is an even power Taylor series about $\phi=0$, then even power covariant derivatives of the potential will be finite:
\begin{align}
\left. \nabla^{2n}V \right|_{\phi=0}<\infty \,,
\end{align}
for all $n\in\mathbb{N}$ .
This completes our proof of the forward direction in \cref{eqn:EquivalenceRewrite}.

\subsection{\boldmath Curvature Criteria $\Longrightarrow$ Fixed Basis Criteria}
\label{appsubsec:Backward}
In this section, we prove that the Curvature Criteria imply the Fixed Basis Criteria, \ie, the ``$\Leftarrow$'' direction in \cref{eqn:EquivalenceRewrite}.
Our starting point is the assumption that $Y$ has a well-defined Taylor expansion about $\phi=0$:
\begin{equation}
 Y(\phi) = \sum\limits_{k=0}^\infty \frac{1}{k!} Y_k \phi^k \,,
 \label{eq:YTaylorExp}
\end{equation}
Let us first focus on proving that $Y_0 = Y_{2k+1} = 0$.
We will demonstrate this inductively in two steps:
\begin{enumerate}
  \item Base step: The finiteness of $\left.R\right|_{\phi=0}$ implies that $Y_0=Y_1=0$.
  \item Induction step:  For any $n\ge1$, assuming $Y_0=Y_1=\cdots=Y_{2n-1}=0$ and the finiteness of $\left. \nabla^{2n} R \right|_{\phi=0}$ implies that $Y_{2n+1}=0$.
\end{enumerate}
Then we will show that $V_{2n+1} = 0$ follows from the finiteness of $\left. \nabla^{2n+2} V \right|_{\phi=0}$.
This will prove that the Curvature Criteria $\Longrightarrow$ the Fixed Basis Criteria by induction.

\subsubsection*{\boldmath Base Step: Proving  $Y_0 = Y_1 = 0$}
The goal of step 1 is to prove that $Y_0 = Y_1 = 0$.
Since this will rely on the fact that the function $\frac{1}{1+Y}$ has a well-defined Taylor expansion about $\phi=0$, we first need to argue that $Y_0\ne -1$ is a consequence of $\left. R \right|_{\phi=0} < \infty$.
We will show this by contradiction; if we assume that $Y_0 = -1$, we can show that the scalar curvature is infinite when it is evaluated at $\phi = 0$.

If $Y$ is a constant, we can see that $Y = -1$ yields infinite curvature by plugging $Y = -1 + \epsilon$ into \cref{eqn:RYphi} to find
\begin{align}
Y = -1 + \epsilon \qquad \Longrightarrow \qquad R = -\frac{N_\phi (N_\phi-1)(-1+\epsilon)}{\epsilon \phi^2} \,,
\end{align}
so the curvature clearly diverges as $\epsilon \rightarrow 0$, for any $\phi$.\footnote{Throughout this appendix, we assume that the dimension of the scalar manifold is greater than two, namely that $N_\varphi>1$.}
Next, we can generalize this to the situation where $Y$ takes the form
\begin{equation}
Y = -1 + \frac{1}{k!}Y_k \phi^k + \cdots \,,
\end{equation}
where we have assumed that there exists a smallest non-zero power $k$, and ``$+\cdots$'' represents higher power terms.
Plugging this into \cref{eqn:RYphi} yields
\begin{align}
R = - N_\varphi \left[ \frac{k(k+2)}{2\phi^2} + \cdots \right] - N_\varphi\left(N_\varphi-1\right) \left[ -\frac{k!}{Y_k\phi^{k+2}} +\cdots \right] \,.
\end{align}
In each of these brackets, we only explicitly show the leading divergent terms in the limit $\phi\to0$, assuming $k\ge1$.
Clearly, $\left. R \right|_{\phi=0}$ is divergent as long as $N_\varphi>0$.
We conclude that $Y_0 = -1$ is incompatible with the finite curvature requirement, and hence
\begin{equation}
\left.R\right|_{\phi=0} < \infty  \qquad\Longrightarrow\qquad  Y_0 \ne -1  \qquad\Longrightarrow \qquad  \left.\frac{1}{1+Y}\right|_{\phi=0} < \infty \,.
\end{equation}

Armed with the fact that $\frac{1}{1+Y}$ has a well-defined Taylor expansion about $\phi=0$, we can reanalyze \cref{eqn:RYphi} assuming \cref{eq:YTaylorExp} to identify the leading divergent terms for the calculation of $R$:
\begin{align}
R &\supset  - {N_\varphi }\frac{2}{{1 + Y}}\frac{1}{\phi }\frac{{\dd Y}}{{\dd\phi }} - {N_\varphi }\left( {{N_\varphi } - 1} \right)\left( {\frac{1}{{1 + Y}}\frac{1}{\phi }\frac{{\dd Y}}{{\dd\phi }} + \frac{Y}{{1 + Y}}\frac{1}{{{\phi^2}}}} \right) \notag\\[5pt]
 &\supset  - {N_\varphi }\frac{{2{Y_1}}}{{1 + {Y_0}}}\frac{1}{\phi } - {N_\varphi }\left( {{N_\varphi } - 1} \right)\left[ {\frac{{{Y_0}}}{{1 + {Y_0}}}\frac{1}{{{\phi^2}}} + \frac{{{Y_1}}}{{1 + {Y_0}}}\frac{1}{\phi } + \frac{{{Y_1}}}{{{{\left( {1 + {Y_0}} \right)}^2}}}\frac{1}{\phi }} \right] \,.
\end{align}
We see that for $N_\varphi>1$, $\left. R \right|_{\phi=0} < \infty$ requires that $Y_0=Y_1=0$.\footnote{For $N_\varphi=1$, $Y_0$ is allowed to be nonzero.}

\subsubsection*{\boldmath Induction Step: Proving  $Y_{2n+1} = 0$}
At this point, we have shown that $Y_0=Y_1=0$.
Now we will prove that all odd powers of $Y$ vanish.
To do so, we will relate the requirement that $\nabla^{2n}R$ are finite to the need for the odd term $Y_{2n+1}$ to vanish, see \cref{eqn:YoddVanish} below.

We begin with the assumption that $Y$ is an analytic even power series up to some arbitrary order $\phi^{2n}$, with $n\geq 1$, and has no constant term:
\begin{align}
Y_0 = Y_1 = \dots = Y_{2n-1} = 0 \,.
\label{eqn:YCoeffsZero}
\end{align}
Then from the definition of the function $A$ in \cref{eqn:Adef}, we see that $A$ is also an analytic even power series up to the same order, and has no constant term:
\begin{equation}
A_0=A_1=\cdots=A_{2n-1}=0 \,.
\label{eqn:ACoeffsZero}
\end{equation}
For the scalar curvature $R$, we can now apply \cref{eqn:YCoeffsZero,eqn:ACoeffsZero} to \cref{eqn:RYphi}. It is straightforward to see that $R$ has a well-defined Taylor expansion at $\phi=0$
\begin{equation}
R = \sum\limits_{k=0}^\infty {\frac{1}{k!}{R_k} \phi^k}  \,,
\label{eqn:Ranalytic}
\end{equation}
and that there could be an odd power in the expansion of $R$, where the first non-zero coefficient would start at order $\phi^{2n-1}$:
\begin{equation}
 R_1=\cdots=R_{2n-3}=0 \,,\quad  {R_{2n - 1}} =  -{N_\varphi }\frac{{2n + 2}}{{2n + 1}}\left( {1 + \frac{{{N_\varphi }}}{{2n}}} \right) {Y_{2n + 1}} \,.
 \label{eqn:RCoeffsZero}
\end{equation}

Before working through the argument for general $n\ge1$, it is instructive to explain how this works for the special case of $n=1$.
Specifically, we will show that the finiteness of $\nabla^2 R$ requires $Y_3=0$.
To analyze $\nabla^2R$, we apply \cref{eqn:nabla2XYphi}. Using $A_0=0$ derived above in \cref{eqn:ACoeffsZero} and the analyticity of $R$ derived above in \cref{eqn:Ranalytic}, we see that the only potentially divergent term is proportional to $R_1$:
\begin{equation}
\nabla^2 R \supset {N_\varphi }\frac{1}{\phi }{R_1} \,.
\label{eqn:R1Vanish}
\end{equation}
Furthermore, from \cref{eqn:RCoeffsZero} we know that $R_1$ is proportional to $Y_3$:
\begin{equation}
R_1= - Y_3 N_\varphi \frac{4}{3}\left( 1 + \frac{N_\varphi}{2} \right) \,.
\label{eqn:R1Y3}
\end{equation}
Therefore, we see that $Y_3=0$ is required by the finiteness of $\left.\nabla^2 R\right|_{\phi=0}$:
\begin{equation}
\left.\nabla^2 R\right|_{\phi=0} < \infty  \qquad\Longrightarrow\qquad  R_1=0  \qquad\Longrightarrow\qquad  Y_3=0 \,.
\label{eqn:Y3Vanish}
\end{equation}

Before moving on, let us quickly summarize what we have learned from this $n=1$ case.
The key point is that \cref{eqn:R1Vanish} applies to not only $R$ but also any scalar function $X$ that has a well defined Taylor expansion about $\phi=0$:
\begin{equation}
X = \sum\limits_{k=0}^\infty {\frac{1}{k!}{X_k} \phi^k}  \,.
\label{eqn:Xanalytic}
\end{equation}
The only potentially divergent term in $\nabla^2 X$ is given by the coefficient $X_1$:
\begin{equation}
\nabla^2 X \supset {N_\varphi }\frac{1}{\phi }{X_1} \,.
\end{equation}
Therefore, with \cref{eqn:Xanalytic} assumed, we have
\begin{equation}
\nabla^2 X|_{\phi = 0} <\infty  \qquad\Longrightarrow\qquad  X_1 = 0 \,.
\label{eqn:X1Vanish}
\end{equation}

Next we can generalize to cases with $n\ge 1$.
Our goal is to prove that the following is true:
\begin{equation}
\hspace{-8pt} \left.\nabla^{2n} R\right|_{\phi=0} < \infty  \quad\Rightarrow\quad  \left(\nabla^{2n-2}R\right)_1=0  \quad\Rightarrow\quad  R_{2n-1}=0  \quad\Rightarrow\quad  Y_{2n+1}=0 \,.
\label{eqn:YoddVanish}
\end{equation}
The first implication is true by simply using $X=\nabla^{2n-2}R$ in \cref{eqn:X1Vanish}.\footnote{Note that to completely justify using $X=\nabla^{2n-2}R$ in \cref{eqn:X1Vanish}, we also need \cref{eqn:Xanalytic} to hold first, namely that $\nabla^{2n-2}R$ has a well defined Taylor expansion about $\phi=0$. We will see that this is true during our proof of the middle step in \cref{eqn:YoddVanish}. In particular, see the discussion around \cref{eqn:nabla2n2R}.}
The last implication is true by the connection between $R_{2n-1}=0$ and $Y_{2n+1}=0$ given in \cref{eqn:RCoeffsZero}.
So the only non-trivial aspect of the generalization to $n\ge1$ is the middle step in \cref{eqn:YoddVanish}.
This step was trivial for the special case $n=1$, as the two expressions are the same, but requires some effort to generalize for $n>1$ as we will now show.

The rest of this section is devoted to showing that $\left(\nabla^{2n-2}R\right)_1=0  \Rightarrow  R_{2n-1}=0 $ is true for $n\ge2$.
The key observation is that the coefficient $R_{2n-1}$ is responsible for the lowest odd power contribution to the Taylor expansion of $\nabla^{2k}R$ for all $0\le k\le n-1$, as summarized in \cref{tbl:nabla2kR}.
To see this, we recall that for any scalar function $X$, the Laplacian acts as
\begin{equation}
\nabla^2 X = \left[ \frac{\dd^2}{\dd\phi^2} + \left(N_\varphi+A\right)\frac{1}{\phi}\frac{\dd}{\dd\phi} \right] X \,,
\label{eqn:nabla2X}
\end{equation}
see~\cref{eqn:nabla2XYphi}.
Therefore, if the lowest odd power in $X$ is $\phi^{2s+1}$, then the lowest odd power in $\nabla^2X$ will be $\phi^{2s-1}$, because the differential operators $\frac{\dd^2}{\dd\phi^2}$ and $\frac{1}{\phi}\frac{\dd}{\dd\phi}$ will reduce the power of a given term by two.
However, for this argument to be valid, we need two constraints on $s$:
\begin{enumerate}[(i)]
\item $\nabla^2X$ must be also non-singular (in addition to $X$ itself) at $\phi=0$. From \cref{eqn:X1Vanish}, we see that this requires $X$ having a well defined Taylor expansion, with its lowest odd power $2s+1>1 \,\,\Rightarrow\,\, s\geq 1$.
\item Recall that the function $A$ contains odd power terms starting from order $2n+1$, see \cref{eqn:ACoeffsZero}.
For these terms to be irrelevant for the lowest odd term in $\nabla^2 X$, we need the restriction $2s-1<2n+1\,\, \Rightarrow\,\, s\leq n$.
\end{enumerate}
We conclude that if the lowest odd power coefficient in $X$ is $X_{2s+1}$ with $1\le s \le n$, then $\nabla^2 X$ has a well defined Taylor expansion about $\phi=0$, with its lowest odd power coefficient $\left(\nabla^2 X\right)_{2s-1}$ given by (this is straightforward to derive from \cref{eqn:nabla2X})
\begin{equation}
\left(\nabla^2 X\right)_{2s-1} = \left(1+\frac{N_\varphi}{2s}\right) X_{2s+1} \,.
\label{eqn:oddPass}
\end{equation}
In other words, each time we apply a Laplacian, the lowest odd power is reduced by two, and the coefficient depends on the power, \ie, it depends on $s$.

Now, we simply iterate this logic and apply it to $R$.
Taking $X_{2s+1}=R_{2n-1}$ and acting on it with $\nabla^2$ using \cref{eqn:oddPass} repeatedly, we get that $\nabla^{2k} R$ has well defined Taylor expansion about $\phi=0$, with the lowest odd power coefficient given by
\begin{equation}
\left(\nabla^{2k} R \right)_{2n-1-2k} = R_{2n-1} \prod\limits_{r=1}^k \left(1+\frac{N_\varphi}{2n-2r}\right) \,,
\label{eq:Nabla2kROddPower}
\end{equation}
for all $1\le k\le n-1$, as summarized in \cref{tbl:nabla2kR}; the trivial case $k=0$ is also included in the table for completeness.
Finally, taking $k=n-1$ in \cref{eq:Nabla2kROddPower}, which corresponds to the last line in \cref{tbl:nabla2kR}, we see that $\nabla^{2n-2}R$ has a well defined Taylor expansion, and
\begin{equation}
\left(\nabla^{2n-2} R \right)_1 = R_{2n-1} \prod\limits_{r=1}^{n-1} \left(1+\frac{N_\varphi}{2n-2r}\right) \,.
\label{eqn:nabla2n2R}
\end{equation}
This proves the middle step in \cref{eqn:YoddVanish}.

Since we have already shown the other two steps in \cref{eqn:YoddVanish}, the inductive proof is complete.
We conclude that if $\left. \nabla^{2n}R \right|_{\phi=0}<\infty$ for all $n\in\mathbb{N}$, then $Y$ is an even power series in $\phi$ to all orders.

\begin{table}[t!]
\renewcommand{\arraystretch}{2.2}
\setlength{\arrayrulewidth}{.3mm}
\setlength{\tabcolsep}{1 em}
\begin{center}
\begin{tabular}{c|c|c}
$k$      & Lowest odd power in $\nabla^{2k}R$ & Coefficient $\left(\nabla^{2k}R\right)_{2n-1-2k}$                                         \\\hline
$0$      & $2n-1$                             & $R_{2n-1}$                                                                                 \\
$1$      & $2n-3$                             & $\left(1+\frac{N_\varphi}{2n-2}\right)R_{2n-1}$                                            \\
$\vdots$ & $\vdots$                           & $\vdots$                                                                                   \\
$k$      & $2n-1-2k$                          & $\left(1+\frac{N_\varphi}{2n-2k}\right)\cdots\left(1+\frac{N_\varphi}{2n-2}\right)R_{2n-1}$\\
$\vdots$ & $\vdots$                           & $\vdots$                                                                                   \\
$n-1$    & $1$                                & $\left(1+\frac{N_\varphi}{2}\right)\cdots\left(1+\frac{N_\varphi}{2n-2}\right)R_{2n-1}$    \\
\end{tabular}\vspace{0.5cm}
\caption{A summary of the relation between the lowest odd power coefficients in the Taylor expansion for $\nabla^{2k}R$ in the range $0\le k\le n-1$, and the relation to $R_{2n-1}$. There is a similar table for $\nabla^{2k}V$, which can be obtained by making the replacement $R\to V$ and $n\to n+1$ in the above.}
\label{tbl:nabla2kR}
\end{center}
\end{table}


\subsubsection*{\boldmath Proving $V_{2n+1} = 0$}
To complete our proof that the Fixed Basis Criteria imply the Curvature Criteria, \ie, the ``$\Rightarrow$'' direction in \cref{eqn:EquivalenceRewrite}.
We will now show that if $\nabla^{2n+2}V|_{\phi=0} < \infty$ for all $n \in \mathbb{N}$, then $V$ is an even power Taylor expansion in $\phi$.
We will follow the same logic as the proof that was just presented for $Y$.
Even better, we now have the additional advantage that the function $A$ is an even power series to all orders, as was proven previously, see \cref{eqn:ACoeffsZero}.

We start with the assumption that $V$ has a well-defined Taylor expansion
\begin{equation}
V\left( \phi  \right) \equiv \sum\limits_{k = 0}^\infty  {\frac{1}{{k!}}{V_k}{\phi^k}} \,,
\label{eq:VTaylorExp}
\end{equation}
and then use induction to investigate the consequence of assuming $\left. \nabla^{2n+2} V \right|_{\phi=0}<\infty$.

First we prove the base step $n=0$:
\begin{equation}
\left. \nabla^{2} V \right|_{\phi=0}<\infty \qquad \Longrightarrow \qquad V_1 = 0\,.
\end{equation}
This is true simply by using $X=V$ in \cref{eqn:X1Vanish}. Next, we move to the inductive step.
Our goal is to prove that for any $n>0$,
\begin{align}
V_1 = V_3 = \cdots = V_{2n-1} = 0 \quad \text{and} \quad \left. \nabla^{2n+2} V \right|_{\phi=0}<\infty \quad\Longrightarrow \quad V_{2n+1}=0\,.
\label{eq:VProofGoal}
\end{align}
Follow the exact same logic that led us to \cref{eq:Nabla2kROddPower}, we can now derive that $\nabla^{2k} V$ have well defined Taylor expansion for all $1 \leq k \leq n$, with the lowest odd power coefficient
\begin{align}
{\left( {{\nabla^{2k}}V} \right)_{2n + 1 - 2k}} &= {V_{2n + 1}}\prod\limits_{r = 0}^{k - 1} {\left( {1 + \frac{{{N_\varphi }}}{{2n - 2r}}} \right)} \,.
\end{align}
Then taking $k=n$ yields
\begin{equation}
\left( \nabla^{2n}V \right)_1 = {V_{2n + 1}}\prod\limits_{r = 0}^{n - 1} {\left( {1 + \frac{{{N_\varphi }}}{{2n - 2r}}} \right)} \,.
\label{eq:Nabla2nV1Relation}
\end{equation}
Therefore, we get
\begin{equation}
\left. \nabla^{2n+2}V \right|_{\phi=0} < \infty  \qquad\Longrightarrow\qquad  \left( \nabla^{2n}V \right)_1 = 0  \qquad\Longrightarrow\qquad  V_{2n+1} = 0 \,,
\end{equation}
where the first implication follows from plugging $\nabla^{2n} V$ into \cref{eqn:X1Vanish}, and the second implication follows from \cref{eq:Nabla2nV1Relation}.
This completes our proof that $V$ is an even Taylor series in $\phi$ to all orders.

\section{Why Massless States Cause Submanifold Singularities}
\label{appsec:SubmanifoldSingularities}
This appendix provides support for the discussion in~\cref{sec:scalarsmeft}, by considering what happens to the EFT submanifold at a point where $\frac{\partial^2 V}{\partial \Phi^a\s \partial \Phi^b}$ is not invertible.
Consider the tangent plane to the EFT submanifold:
\begin{equation}
\dd  \frac{\partial V}{\partial\Phi^a} = \frac{\partial^2 V}{\partial \Phi^a\s \partial \phi^i}\, \dd \phi^i + \frac{\partial^2 V}{\partial \Phi^a\s \partial \Phi^b}\, \dd \Phi^b_\mathbf{c} = 0 \,,
\label{eq:tanplane}
\end{equation}
and in particular the rank $r$ of the matrix $\left(\frac{\partial^2 V}{\partial \Phi^a\s \partial \phi^i} ,\frac{\partial^2 V}{\partial \Phi^a\s \partial \Phi^b}\right)$, which has $n$ rows and $4+n$ columns; this matrix has rank at most $r = n$.
Assuming that the sub-matrix $\frac{\partial^2 V}{\partial \Phi^a\s \partial \Phi^b}$ is \emph{not} invertible, such that its rank is less than $n$, there are two cases to consider:
\begin{enumerate}
  \item $r<n$. We interpret \cref{eq:tanplane} as a linear equation for the $(4+n)$-vector $\left(\dd\phi^i, \dd\Phi^a\right)$. The dimension of its solution space is $4+n-r>4$, indicating that the EFT tangent plane is not uniquely defined. \emph{Often} multiple EOM solution branches join at this point, where each branch \emph{may} exhibit some non-analyticity at the point.\footnote{Due to the restriction that only real solutions (for real field components) count as EOM solutions, there are cases that only a single EOM branch exists at this point with no uniquely defined tangent plane. In these cases, although singularities do not show up at tree-level, they are generically expected at loop order.}  It implies that there exists at least one linear combination of the $\Phi$ modes which is a null eigenvector of the full UV mass matrix.
    
    \item $r = n$.
In this case, we can interpret \cref{eq:tanplane} as a linear equation for the $n$-vector $\dd\Phi^a$. 
The \emph{coefficient matrix} of this linear equation, is nothing but the sub-mass matrix $\frac{\partial^2 V}{\partial\Phi^a\s \partial\Phi^b}$, whose rank is smaller than $n$ by assumption, implying that it is non-invertible.
On the other hand, the \emph{augmented matrix} has a rank equal to $n$ for generic values of $\dd\phi^i$:
\begin{equation}
\rank \left( \frac{\partial^2 V}{\partial \Phi^a\s \partial \phi^i}\, \dd \phi^i , \frac{\partial^2 V}{\partial\Phi^a\s \partial\Phi^b}\right) = n > \rank \left(\frac{\partial^2 V}{\partial\Phi^a\s \partial\Phi^b}\right) \,.
\end{equation}
No solution (for all $\dd \Phi^a$) exists for generic values of $\dd\phi^i$. The only way to find solutions is by setting particular components of the vector $\dd\phi^i$ to zero in order to reduce the rank of the augmented matrix to match the coefficient matrix. This implies that there will be a divergent derivative $\dd\Phi^a/\dd\phi^j$ at the putative $O(4)$ fixed point, and hence the resulting EFT is non-analytic at this point.
\end{enumerate}

Non-analyticities at the $O(4)$ fixed point are usually of the case 1 kind. This is because, when the fields are arranged into a 4-plet $\phi^i$ and a set of $O(4)$ irreps $\Phi^a$, $\frac{\partial^2 V}{\partial \Phi^a\s \partial \phi^i} = 0$ by symmetry, unless $\Phi^a$ contains another 4-plet. Non-analyticities at generic points on the path between the $O(4)$ fixed point and our observed vacuum are usually of the case 2 variety. Both kinds of non-analyticity require a HEFT description of the EFT. A sketch of the two cases is provided in~\cref{fig:singularEFTSubmanifold} in \cref{sec:eftsubmani}.

\section{Computing the Effective Lagrangian}
\label{appsec:TwoDerUniversal}
In this appendix, we use functional methods to derive new general expressions for the two derivative contributions to the one-loop effective Lagrangian that include all orders in the fields.
Specifically, we will evaluate the functional determinant of an elliptic (functional) operator of the form
\begin{equation}
\mathrm{O} = \partial^2 + M^2 + U \,,
\end{equation}
where $M^2$ is a degenerate mass square parameter that is proportional to the identity matrix $M^2\propto I$, and $U$ is a local spacetime operator, such that $U \left| x \right\rangle = \left| x \right\rangle U(x)$.
Then the one-loop contribution to the effective action from integrating out the physics associated with $\mathrm{O}$ is
\begin{align}
T(U) &\equiv i\ln \det \left( \partial^2 + M^2 + U \right) = i\Tr \ln \left( \partial^2 + M^2 + U \right) \,,
\label{eqn:TUdef}
\end{align}
which we organize as a derivative expansion\footnote{All of the $T^{(2k+1)}$ are zero due to the fact that \cref{eqn:TUdef} is even under $\partial_\mu \to -\partial_\mu$.}
\begin{align}
T(U) = T^{(0)}( U ) + T^{(2)}( U ) + \mathcal{O}\big(\partial^4\big) \,.
\label{eqn:TDerExpan}
\end{align}
The leading contribution $T^{(0)}(U)$ collects terms with no explicit derivatives acting on $U$.\footnote{Here we are using the superscript to track the number of explicit derivatives that appear.  There can additionally be implicit derivatives contained within $U$.}
This includes the famous Coleman-Weinberg potential~\cite{Coleman:1973jx}, and generalizes it to allow $U$ to contain derivatives.
The next term in the series is $T^{(2)}(U)$, which collects terms with two explicit derivatives acting on $U$.

We begin the evaluation of the functional trace as follows (see Eq.~(A.1) in~\cite{Henning:2016lyp}):
\begin{align}
T(U) &\equiv i\ln \det \left( {{\partial^2} + {M^2} + U} \right) = i \Tr \ln \left( {{\partial^2} + {M^2} + U} \right) \notag\\[3pt]
 &= i\int {\frac{{{\dd^4}p}}{{{{\left( {2\pi } \right)}^4}}} \mel**{p}{\tr \ln \left( -{\hat p^2} + \hat{M}^2 + \hat{U} \right)}{p} } \notag\\[3pt]
 &= i\int {\dd^4 x \int {\frac{{{\dd^4}p}}{{{{\left( {2\pi } \right)}^4}}} \bra{p}\ket{x} \mel**{x}{\tr \ln \left( -{\hat p^2} + \hat{M}^2 + \hat{U} \right)}{p} } } \notag\\[3pt]
 &= i\int {\dd^4 x \int {\frac{{{\dd^4}p}}{{{{\left( {2\pi } \right)}^4}}}\, {e^{ipx}}\, \tr \ln \left[ { - {{\left( {i\partial } \right)}^2} + {M^2} + U} \right]{e^{ - ipx}}} } \notag\\[3pt]
 &= i\int {\dd^4 x \int {\frac{{{\dd^4}p}}{{{{\left( {2\pi } \right)}^4}}} \tr \ln \left[ { - {{\left( {i\partial + p} \right)}^2} + {M^2} + U} \right]} } \notag\\[3pt]
 &= i\int {\dd^4 x \int {\frac{{{\dd^4}p}}{{{{\left( {2\pi } \right)}^4}}} \tr \ln \left[ { - {{\left( {i\partial - p} \right)}^2} + {M^2} + U} \right]} } \notag\\[3pt]
 &= i\int {\dd^4 x \int {\frac{{{\dd^4}p}}{{{{\left( {2\pi } \right)}^4}}} \tr \ln \left[ { - {p^2} + {M^2} + U + \left( {2ip \cdot \partial  + {\partial^2}} \right)} \right]} } \,,
\label{eqn:TraceStart}
\end{align}
where in going from the first line to the second we have explicitly evaluated the ``Tr'' trace, leaving only a ``tr'' trace which acts on internal indices.
We have also added explicit ``hats'' on operators at the relevant intermediate steps to make clear at what step these operators become $c$-numbers.

We simplify the evaluation by first taking the derivative of $T(U)$ with respect to $M^2$:
\begin{equation}
  \frac{\partial T(U)}{\partial M^2} = i \int \dd^4 x \int \frac{\dd^4 p}{(2\pi)^4} \tr \left[ \frac{1}{(-p^2 + M^2 + U) + (2 i p \cdot \partial + \partial^2)} \right] \,. \label{eqn:DerivativeAssistance}
\end{equation}
Our goal is to evaluate this expression to second order in derivatives.  To this end we write $\Delta = -p^2 + M^2 + U$ and only keep the contributions to the integrand that do not vanish under $p \to - p$:
\begin{align}
\frac{\partial T(U)}{\partial M^2} &= i \int \dd^4 x \int \frac{\dd^4 p}{(2\pi)^4} \tr \Big[ \Delta^{-1} - \Delta^{-1}  \partial^2 \Delta^{-1} \notag\\
&\hspace{120pt} + \Delta^{-1} (2 i p \cdot \partial ) \Delta^{-1}  (2 i p \cdot \partial) \Delta^{-1} + \mathcal{O}\left(\partial^4\right) \Big] \,,
\end{align}
where the partial derivatives that appear in these products are understood to act on everything to their right. Integration by parts then gives
\begin{align}
\frac{\partial T(U)}{\partial M^2} &= i \int \dd^4 x \int \frac{\dd^4 p}{(2\pi)^4} \tr \Big[ \Delta^{-1} + \left(\partial_\mu \Delta^{-1}\right)  \left(\partial^\mu \Delta^{-1}\right) \notag\\
&\hspace{120pt} + 4 p^\mu p^\nu \left(\partial_\mu \Delta^{-1}\right) \Delta^{-1} \left(\partial_\nu \Delta^{-1}\right) + \mathcal{O}\left(\partial^4\right) \Big] \,.
\end{align}
The only $x$ dependent part of $\Delta$ is $U(x)$, and thus $\partial_\mu \Delta^{-1} = -\Delta^{-1} \left(\partial_\mu U\right) \Delta^{-1}$. Then using the cyclic property of the trace, we find
\begin{align}
\frac{\partial T(U)}{\partial M^2} &= i \int \dd^4 x \int \frac{\dd^4 p}{(2\pi)^4} \tr \Big[ \Delta^{-1} + \left(\partial_\mu U\right) \Delta^{-2} \left(\partial^\mu U\right) \Delta^{-2} \notag\\
&\hspace{120pt} + p^2 \left(\partial_\mu U\right) \Delta^{-2} \left(\partial^\mu U\right) \Delta^{-3} + \mathcal{O}\left(\partial^4\right) \Big] \,,
\label{eqn:ptpmsqrOld}
\end{align}
where we have also substituted $p^\mu p^\nu \to \frac14 \eta^{\mu\nu} p^2$. Now note that the following combination is a total derivative of $p$:
\begin{align}
&\tr \left[ \left(\partial_\mu U\right) \Delta^{-2} \left(\partial^\mu U\right) \Delta^{-2} + 2p^2 \left(\partial_\mu U\right) \Delta^{-2} \left(\partial^\mu U\right) \Delta^{-3} \right] \notag\\[3pt]
&\hspace{50pt} = \frac{\partial}{\partial p^\rho} \tr \left[\frac{1}{4} p^\rho \left(\partial_\mu U\right) \Delta^{-2} \left(\partial^\mu U\right) \Delta^{-2} \right] \,,
\end{align}
which will yield zero upon performing the loop integral. We can therefore simplify \cref{eqn:ptpmsqrOld} to yield
\begin{align}
\frac{\partial T(U)}{\partial M^2} &= i \int \dd^4 x \int \frac{\dd^4 p}{(2\pi)^4} \tr \left[ \Delta^{-1} + \frac{1}{2} \left(\partial_\mu U\right) \Delta^{-2} \left(\partial^\mu U\right) \Delta^{-2} + \mathcal{O}\left(\partial^4\right) \right] \,.
\label{eqn:ptpmsqr}
\end{align}
At this point, we can identify the contributions which belong to $T^{(0)}(U)$ and $T^{(2)}(U)$ respectively. What follows is the evaluation of each in turn.

First, we integrate the zero-derivative part of \cref{eqn:ptpmsqr} with respect to $M^2$ to regain
\begin{equation}
  {T^{(0)}}( U ) = i\int\dd^4 x\, \int {\frac{\dd^4 p}{(2\pi)^4} \tr \ln \left( { - {p^2} + {M^2} + U} \right)}  \,.
\end{equation}
The evaluation of $T^{(0)}$ is treated in many textbooks, and we will briefly reproduce the key steps here for completeness (we have numbered the equal signs for ease of reference):
\begin{align}
{T^{(0)}}( U ) &\stackrel{1}{=} i\int {\dd^4 x\int {\frac{\dd^4 p}{(2\pi)^4} \tr \ln \left( { - {p^2} + {M^2} + U} \right)} } \notag\\[5pt]
 &\stackrel{2}{=}  - i\int {\dd^4 x\, \tr \left\{ {{{\left. {\frac{\dd}{{\dd\alpha }}\left[ {\int {\frac{\dd^4 p}{(2\pi)^4}\frac{1}{{{{\left( -p^2 + M^2 + U \right)}^\alpha }}}} } \right]} \right|}_{\alpha  = 0}}} \right\}} \notag\\[5pt]
 &\stackrel{3}{=} \int {\dd^4 x\, \tr \left\{ {{{\left. {\frac{\dd}{{\dd\alpha }}\left[ {\frac{{{\mu^{4 - d}}}}{{{{\left( {4\pi } \right)}^{d/2}}}}\frac{{\Gamma \left( {\alpha  - d/2} \right)}}{{\Gamma \left( \alpha  \right)}}\frac{1}{{{{\left( {{M^2} + U} \right)}^{\alpha  - d/2}}}}} \right]} \right|}_{\alpha  = 0}}} \right\}} \notag\\[5pt]
 &\stackrel{4}{=} \int {\dd^4 x\, \tr \left\{ {{{\left. {\left[ {\frac{{{\mu^{4 - d}}}}{{{{\left( {4\pi } \right)}^{d/2}}}}\frac{{\Gamma \left( {\alpha  - d/2} \right)}}{{\alpha \Gamma \left( \alpha  \right)}}\frac{1}{{{{\left( {{M^2} + U} \right)}^{\alpha  - d/2}}}}} \right]} \right|}_{\alpha  = 0}}} \right\}} \notag\\[5pt]
 &\stackrel{5}{=} \int {\dd^4 x\, \tr \left[ {\frac{{{\mu^{4 - d}}}}{{{{\left( {4\pi } \right)}^{d/2}}}}\frac{{\Gamma \left( { - d/2} \right)}}{{{{\left( {{M^2} + U} \right)}^{ - d/2}}}}} \right]} \notag\\[5pt]
 &\stackrel{6}{\to} \int {\dd^4 x\, \frac{1}{{{16\s\pi^2}}} \tr \left[ {\frac{1}{2}{{\left( {{M^2} + U} \right)}^2}\left( {\ln \frac{{{\mu^2}}}{{{M^2} + U}} + \frac{3}{2}} \right)} \right]} \,.
\label{eqn:T0Evaluation}
\end{align}
To clarify a few of the steps in \cref{eqn:T0Evaluation}, we begin with the second equality, where we have used
\begin{equation}
\ln x =  - {\left. {\frac{\dd}{{\dd\alpha }}{x^{ - \alpha }}} \right|_{\alpha  = 0}} \,,
\end{equation}
to rewrite $\ln\left(-p^2+M^2+U\right)$.
For the third equality, we shifted to $d$ dimensions and have performed the loop integral using dimensional regularization, with the requisite introduction of the subtraction scale $\mu$.
The fourth equality results from carrying out the derivative $\frac{\dd}{\dd\alpha}$ by explicitly evaluating
\begin{equation}
{\left. {\frac{{\dd f\left( \alpha  \right)}}{{\dd\alpha }}} \right|_{\alpha  = 0}} = \mathop {\lim }\limits_{\alpha  \to 0} \frac{{f\left( \alpha  \right) - f\left( 0 \right)}}{\alpha} \,.
\end{equation}
Finally, to derive the final result, we have subtracted the UV divergence using the $\overline{\text{MS}}$ renormalization scheme.

Next, we evaluate the two derivative piece of \cref{eqn:ptpmsqr}.  Note that the momentum integral is finite. First, we Wick rotate and make the change of variables $t = p_\text{E}^2 = - p^2$:
\begin{align}
\frac{\partial T^{(2)}(U)}{\partial M^2} &= - \int \dd^4 x \int_0^\infty \frac{t\, \dd t}{16\s\pi^2} \tr \left[ \frac{1}{2} \left(\partial_\mu U\right) \Delta^{-2} \left(\partial^\mu U\right) \Delta^{-2} \right] \,,
\end{align}
with $\Delta=t+M^2+U$. Now we can get rid of the $M^2$ derivative by recognizing the above as
\begin{align}
\frac{\partial T^{(2)}(U)}{\partial M^2} &= \int\dd^4 x \int_0^\infty \frac{\dd t}{16\s\pi^2} \left( -\frac{\partial}{\partial t}\frac{1}{4}t^2 \right) \tr \left[ \left(\partial_\mu U\right) \Delta^{-2} \left(\partial^\mu U\right) \Delta^{-2} \right] \notag\\[6pt]
&= \int\dd^4 x \int_0^\infty \frac{\dd t}{16\s\pi^2} \frac{1}{4}t^2 \frac{\partial}{\partial M^2} \tr \left[ \left(\partial_\mu U\right) \Delta^{-2} \left(\partial^\mu U\right) \Delta^{-2} \right] \,,
\end{align}
where in going from the first to the second line, we moved the $t$ derivative using integration by parts.
Finally, we evaluate
\begin{align}
T^{(2)}(U) &= \int \dd^4x \int_0^\infty \frac{\dd t}{16\s\pi^2} \frac14 t^2 \tr \left[ \left(\partial_\mu U\right) \Delta^{-2} \left(\partial^\mu U\right) \Delta^{-2} \right] \notag\\[6pt]
&= \int\dd^4x\, \frac{1}{16 \pi^2} \tr \Bigg[ \frac{1}{4 M^2} \int_0^1 \dd z (1 - z)^2 \left(\partial_\mu U\right) \left( 1 + \frac{z U}{M^2} \right)^{-2} \notag\\
&\hspace{200pt} \times \left(\partial^\mu U\right) \left( 1 + \frac{z U}{M^2} \right)^{-2} \Bigg] \,.
\end{align}
In the last line we obtain the final form of $T^{(2)}(U)$ by making the change of variable $t \to z \equiv \frac{M^2}{M^2+t}$, in order to arrive at a form that is amenable to being Taylor expanded in terms of $\frac{U}{M^2}$. 
Note that in the special case where $\left[U, \partial_\mu U\right]=0$, we can simplify the final result:
\begin{align}
{T^{(2)}}( U ) &= \int\dd^4 x\, \frac{1}{16\s\pi^2} \tr \left[ {\frac{1}{{4{M^2}}}\int_0^1 {\dd z{{\left( {1 - z} \right)}^2}{{\left( {1 + \frac{{zU}}{{{M^2}}}} \right)}^{ - 4}}{{\left( {\partial U} \right)}^2}} } \right] + \mathcal{O}\left(\left[U, \partial_\mu U\right]\right) \notag\\[5pt]
 &= \int\dd^4 x\, \frac{1}{16\s\pi^2} \tr \left[ {\frac{1}{{12}}\frac{1}{{{M^2} + U}}{{\left( {\partial U} \right)}^2}} \right] +  \mathcal{O}\left(\left[U, \partial_\mu U\right]\right) \,,
\end{align}
where we have used
\begin{equation}
\int_0^1 {\dd z\frac{{{{\left( {1 - z} \right)}^2}}}{{{{\left( {1 + zx} \right)}^4}}}}  = \frac{1}{3}\frac{1}{{1 + x}} \,.
\end{equation}
This form is applicable for the singlet scalar example provided in \cref{sec:IntOutDoublet} above; the more general formula is relevant to the vector-like fermion example in \cref{sec:VectorLikeFermion}.

\subsection{Summary and Heavy Mass Expansion}
In summary, we started by organizing $T$ as a derivative expansion
\begin{align}
T(U) &\equiv i \ln \det \left(\partial^2 + M^2 + U\right) = T^{(0)}(U) + T^{(2)}(U) + \mathcal{O}\big(\partial^4\big) \,, \label{eqn:TFinal}
\end{align}
and obtained the following results
\begin{subequations}\label{eqn:TraceFinal}
\begin{align}
T^{(0)}(U) &\to \int\dd^4 x\, \frac{1}{16\s\pi^2} \tr \left[ \frac{1}{2} \left( M^2 + U \right)^2 \left( \ln \frac{\mu^2}{M^2+U} + \frac{3}{2} \right) \right] \,, \label{eqn:T0Final} \\[15pt]
T^{(2)}(U) &= \int\dd^4 x\, \frac{1}{16\s\pi^2} \tr \Bigg[ \frac{1}{4M^2} \int_0^1 \dd z (1-z)^2 \left(\partial_\mu U\right) \left(1 + \frac{zU}{M^2}\right)^{-2} \notag\\[-3pt]
 &\hspace{200pt} \times \left(\partial^\mu U \right) \left( 1 + \frac{zU}{M^2} \right)^{-2} \Bigg] \notag\\[5pt]
 &= \int\dd^4 x\, \frac{1}{16\s\pi^2} \tr \left[ \frac{1}{12} \frac{1}{M^2+U} (\partial U)^2 \right] + \mathcal{O}\left(\comm{U}{\partial_\mu U} \right) \,. \label{eqn:T2Final}
\end{align}
\end{subequations}
These results provide a direct calculation of the all-orders dependence on the fields, couplings, and masses at one loop.
In the heavy mass limit, it can be useful to expand in inverse powers of $M^2$ and truncate in mass dimension, which yields
\begin{subequations}
\begin{align}
{T^{(0)}}( U )  &\supset \int\dd^4 x\, \frac{1}{16\s\pi^2} \tr \Bigg[  {M^2}\left( {\ln \frac{{{\mu^2}}}{{{M^2}}} + 1} \right)U + \left(\frac{1}{2}\, {\ln \frac{{{\mu^2}}}{{{M^2}}}} \right){U^2}- \frac{1}{6}\frac{1}{{{M^2}}}{U^3} \notag\\
 &\hspace{100pt}  + \frac{1}{{24}}\frac{1}{{{M^4}}}{U^4} - \frac{1}{{60}}\frac{1}{{{M^6}}}{U^5} + \frac{1}{{120}}\frac{1}{{{M^8}}}{U^6} \Bigg] \,, \\[15pt]
{T^{(2)}}( U )  &\supset \int\dd^4 x\, \frac{1}{16\s\pi^2} \tr \Bigg[ \frac{1}{{12}}\frac{1}{{{M^2}}}{\left( {\partial U} \right)^2} - \frac{1}{{12}}\frac{1}{{{M^4}}}U{\left( {\partial U} \right)^2}  \notag\\[-3pt]
 &\hspace{100pt} + \frac{1}{{20}}\frac{1}{{{M^6}}}{U^2}{\left( {\partial U} \right)^2} + \frac{1}{{30}}\frac{1}{{{M^6}}}U\left( {{\partial_\mu }U} \right)U\left( {{\partial^\mu }U} \right)  \Bigg] \,,
\end{align}
\end{subequations}
where we have not made any assumptions about $\comm{U}{\partial_\mu U}$ when expanding the all-orders result.
These expanded expressions have appeared in the literature; for example, they agree with Eq.~(2.54) in~\cite{Henning:2014wua}.
Again we emphasize that to our knowledge \cref{eqn:T2Final} had not previously appeared in the literature.

\clearpage

\section{Caveats for the Leading Order Criteria}
\label{sec:swampland}
In the main text, we explored various examples of perturbative UV models that must be matched onto HEFT.
The general lesson was that two physical scenarios result in HEFT: ($i$) integrating out a state whose mass is entirely sourced from electroweak symmetry breaking yields a non-analytic effective Lagrangian at the putative $O(4)$ fixed point (\cref{sec:Examples}), and ($ii$) there is an extra source of electroweak symmetry breaking, such that no $O(4)$ invariant point can be accessed on the manifold for $h$ (\cref{sec:BSMSymBreak}).
The question of whether or not HEFT was required for these examples was in accordance with our LO Criteria, which relied on the finiteness of the Ricci scalar, but neglected tests involving curvature invariants built from covariant derivatives acting on $R$.
Our goal here is to identify pitfalls associated with the restricted scope of this work, namely that the geometric quantities we study assume an effective Lagrangian truncated to two-derivative order, and the fact that our LO Criteria rely on a finite number of conditions.
Specifically, we identify two subtleties.
First, we consider UV theories that include non-renormalizable interactions, and second we explore the impact that performing non-derivative UV field redefinitions has on the effective Lagrangian.

\subsection{Non-renormalizable UV Theories \label{sec:scalarheft}}
Above in \cref{sec:SingletModel}, we encountered the expected behavior when we integrated out a BSM singlet state with a spontaneously broken $\mathbb{Z}_2$ symmetry that acquired mass from electroweak symmetry breaking. 
The resulting EFT was a SMEFT if the dynamical singlet field acquired at least some of its mass from terms independent of $H$. 
If the dynamical singlet field acquires all of its mass from electroweak symmetry breaking, however, the resulting tree-level effective potential was non-analytic and we were forced to match onto HEFT, in concert with expectations. 
In this section, we will discuss a related example where the theory includes a massless BSM state with non-renormalizable couplings; naively this should be matched onto HEFT at tree-level, but the LO Criteria fail.

For illustration, we introduce a simple model with a BSM singlet scalar $S$, whose Lagrangian is
\begin{equation}
  \mathcal{L}_\text{UV} \supset  \frac{1}{2} (\partial S)^2 - \frac{1}{2}\s \kappa\s S^2\s \big(\vec \phi \cdot \vec \phi\,\big) - \frac14\s \lambda_S\s S^4 - \frac{1}{\Lambda}\s S\s \big(\vec \phi \cdot \vec \phi\,\big)^2 \, ,
\end{equation}
where $\kappa$, $\lambda_S$, and $1/\Lambda$ are couplings, and we have only included the $S$ dependent terms.
We note that this form is non-generic, but it serves as a proxy for models where the BSM state forms a non-trivial irreducible representation of $O(4)$ such as the $25$-plet $\Phi^{(ijkl)}$, where the linear term $\Phi^{(ijkl)}\s \phi^i \phi^j \phi^k \phi^l$ naturally arises at dimension 5.
It is enlightening to investigate this Lagrangian from a geometric perspective.

First send $\Lambda \to \infty$, so that the model is renormalizable (and identical to the one studied in \cref{sec:SingletModel} above with $m^2 = 0$).
The equation of motion for $S$ has up to three real solutions 
\begin{equation}
  S_\textbf{c}[r] = 0,\,\, \pm\s r\s \sqrt{\frac{-\kappa}{\lambda_S}} \,
  \label{eq:singsol}
\end{equation}
where $r \equiv \sqrt{\vec \phi \cdot \vec \phi} = h + v_0$.
When $\kappa > 0$, only the solution $S_\textbf{c} = 0$ exists; the tree-level EFT is completely flat, but a non-trivial effective Lagrangian arises at one loop, as explored in \cref{sec:IntOutDoublet}. 
When $\kappa < 0$, however, the Higgs vev induces a singlet vev, and so there is effectively a linear singlet coupling which has non-trivial consequences at tree level, much as in \cref{sec:SingletModel}.
In this case, the global minimum lies on the latter two solution branches, $S_\textbf{c} \propto r$, and on the EFT manifold $R \to \infty$ at the fixed point due to a conical singularity. 
(The conical singularity can be visualized intuitively by noting that, when restricted to any 3d subspace $(\phi^i, \phi^j,S)$, $i,j \in \{1,2,3,4\}$, this $S_\textbf{c} \propto r$ a 2d cone.) 
The model must be matched onto HEFT at tree level. 
Operationally, the conical behavior $S_\textbf{c}(r) \propto r$ requires the interplay of two operators with opposite sign in the potential -- in this case $S^2 \s \big(\vec \phi \cdot \vec \phi\,\big)$ and $S^4$ --  that are equally relevant near the fixed point in the UV.

Next we proceed to investigate the implications of the non-renormalizable operator, \ie, we take $1/\Lambda$ finite, assuming $\kappa > 0$.
We will now demonstrate that it is possible to obtain a solution that is non-analytic at the $O(4)$ invariant fixed point, but which does not yield a conical singularity.
As one approaches the fixed point, there is one real solution to the EOM: 
\begin{align}
  S_\textbf{c}(r) &= \sqrt[3]{ - \left( \frac{r^4}{2\s \lambda_S\s \Lambda}\right) + \sqrt{ \left( \frac{r^4}{2\s \lambda_S\s \Lambda}\right)^2 + \left(  \frac{\kappa\s r^2}{3\s \lambda_S}  \right)^3} } \notag \\[5pt]
 & \hspace{12pt} + \sqrt[3]{ - \left( \frac{r^4}{2\s \lambda_S\s \Lambda}\right) - \sqrt{ \left( \frac{r^4}{2\s \lambda_S\s \Lambda}\right)^2 + \left(  \frac{\kappa\s r^2}{3\s \lambda_S}  \right)^3} } \ .
  \label{eq:singsoldeformed}
\end{align}
Still, $\frac{\dd S_\textbf{c}}{\dd r}(0)=0$, due to the added operator being less relevant at the fixed point. There is no conical singularity, $R$ is finite, and the deformed EFT satisfies the LO Criteria. However, the deformed EFT does not satisfy the full Curvature Criteria, as $S_\textbf{c}(r)$, and the resulting EFT metric, is only finitely differentiable at the fixed point.

This shows that even if the UV theory has massless fluctuations with linear couplings, it is possible to find an EFT manifold with finite curvature at the fixed point.
This can occur when integrating out a sufficiently large representations of $O(4)$, or if some terms in the Lagrangian are forbidden by a symmetry.
However, such theories, while satisfying our LO Criteria, face generic obstructions when attempting to match onto SMEFT.

First, there are often increasingly severe singularities that appear at higher derivative order in the effective Lagrangian, driven by the divergence at higher derivative order in the solution for $\Phi_\textbf{c}$.
For instance, from the two derivative piece of \cref{eqn:steepestdescent}, we solve for $\Phi_\textbf{c}^{(2)}$ via
\begin{equation}
  \frac{\delta^2 S^{(0)}}{\delta \Phi\s \delta \Phi}\big[\phi,\Phi^{(0)}_\textbf{c}\big]\, \Phi^{(2)}_\textbf{c} + \frac{\delta S^{(2)}}{\delta \Phi}\big[\phi,\Phi^{(0)}_\textbf{c}\big] = 0 \, .
\end{equation}
For example, if $\frac{\dif^2 V}{\dif \Phi\s \dif \Phi}(0,0) \propto \big(\vec \phi \cdot \vec \phi\,\big)$ near the fixed point, then $\Phi^{(2)}_\textbf{c}$ scales as
\begin{equation}
  \Phi^{(2)}_\textbf{c}  \sim \frac{1}{\big(\vec \phi \cdot \vec \phi\,\big)}\s \partial^2 \Phi^{(0)}_\textbf{c} \, ,
\end{equation}
which generically results in a $\Phi_\textbf{c}^{(2)}$ that grows faster than $\Phi^{(0)}_\textbf{c}$ as $r\to 0$.
By solving the derivative expansion iteratively, one can see that a term in $\Phi_\textbf{c}^{(2n)}$ will scale as
\begin{equation}
  \Phi^{(2n)}_\textbf{c}  \sim \left( \frac{1}{\big(\vec \phi \cdot \vec \phi\,\big)}  \partial^2 \right)^n \Phi^{(0)}_\textbf{c} \, ,
\end{equation}
which inevitably results in non-analyticities at some derivative order in the EFT Lagrangian about the fixed point (unless $\Phi^{(0)}_\textbf{c}\big(\vec \phi\,\big) \equiv 0$ everywhere).
A detailed study of these questions is beyond the scope of this paper.

Additionally, one-loop and higher corrections introduce non-analyticities in the EFT Lagrangian at all orders in the derivative expansion.
Specifically, the Coleman-Weinberg contribution,
\begin{equation}
\mathcal{L}_\text{Eff} \supset  \tr \left[ \left(\frac{\dif^2 V}{\dif \Phi\s \dif \Phi}\right)^2 \ln \left( \frac{\dif^2 V}{\dif \Phi\s \dif \Phi} \right) \right] \ ,
\end{equation}
guarantees that the EFT potential be only finitely differentiable if $\frac{\dif^2 V}{\dif \Phi \dif \Phi}(0,0)$ is singular about the fixed point.
If $\frac{\dif^2 V}{\dif \Phi \dif \Phi}(0,0) \propto \big(\vec \phi \cdot \vec \phi\,\big)$ near the fixed point, the Coleman-Weinberg contribution to the EFT potential is thrice differentiable.
Similarly the two derivative terms (using the simpler form of \cref{eqn:T2Final} that assumes the mass matrix commutes with its spatial derivative)
\begin{equation}
\mathcal{L}_\text{Eff} \supset  \tr \left[
    \bigg(\frac{\dif^2 V}{\dif \Phi\s \dif \Phi}\bigg)^{-1}
  \partial \bigg( \frac{\dif^2 V}{\dif \Phi\s \dif \Phi} \bigg)\,
  \partial \bigg( \frac{\dif^2 V}{\dif \Phi\s \dif \Phi} \bigg)
\right] \ ,
\end{equation}
guarantee that the one-loop EFT metric will only be differentiable a finite number of times when the mass matrix is singular.
For example, if $\frac{\dif^2 V}{\dif \Phi\s \dif \Phi}(0,0) \propto \big(\vec \phi \cdot \vec \phi\,\big)$ near the fixed point, then the metric is continuous but has no well defined derivatives about that point.
Thus $R$ is undefined, the hallmark of a singularity.

\subsection{UV Field Redefinitions}
\label{app:UVFieldRedefs}
A central goal of this study was to frame the question of HEFT versus SMEFT in terms of geometric quantities, so that the criteria would be invariant under (non-derivative) field redefinitions of the EFT fields.
As we will illustrate here using a simple example, there is a danger that coordinate choices in the UV theory can yield a singular HEFT-like chart prior to matching.
Working with the ``wrong'' coordinates can result in an EFT with divergent curvature invariants defined via the two-derivative action, even in the absence of massless BSM fluctuations.
This comes down to the fact that zero-derivative field redefinitions in the UV theory induce field redefinitions within the EFT that involve derivatives.
Since the geometric picture truncated at two-derivative order does not accommodate these redundancies, the value of curvature invariants computed within the EFT can change by an arbitrarily large amount in this case.

We can see this play out by working with a simple example.
Let the UV theory include an $O(4)$ fundamental multiplet $\vec \phi$ and a singlet $S$ with no interactions:
\begin{align}
  \Lag_\text{UV} \left[\vec \phi, S\right] &= \frac{1}{2}{\big(\s \partial {\vec \phi} \,\big)^2} + \frac{1}{2}{( {\partial S} )^2} - \frac{1}{2}{m^2} \big(\vec \phi \cdot \vec \phi\,\big)^2 - \frac{1}{2}\s{M^2}\s S^2 \notag \\[5pt]
  &= \frac{1}{2}\s{( {\partial r} )^2} + \frac{1}{2}\s r^2\s {( {\partial {\vec n}} )^2} + \frac{1}{2}\s{( {\partial S} )^2} - \frac{1}{2}\s{m^2}\s r^2 - \frac{1}{2}\s{M^2}\s S^2 \,,
\end{align}
with masses $m^2\ll M^2$, and in the second line we have introduced HEFT-like polar coordinates $\vec \phi = r\s \vec n$.
It is trivial to integrate out the heavy state $S$, yielding a free theory for $\vec \phi$: 
\begin{align}
\Lag_\text{EFT} \Big[\vec \phi \,\Big] &= \frac{1}{2}\s{\big(\s {\partial \vec {\phi}} \,\big)^2} - \frac{1}{2}\s{m^2}\s\big(\vec \phi \cdot \vec \phi\,\big)^2 \notag \\[5pt]
&= \frac{1}{2}\s{( {\partial r} )^2} + \frac{1}{2}\s r^2\s {( {\partial {\vec n}} )^2} - \frac{1}{2}\s{m^2}\s r^2 \, ,
\label{eqn:LEFTphi1}
\end{align}
which has constant curvature invariants
\begin{align}
R=0\qquad \text{and}\qquad \nabla^2 V = 4\s m^2 \, .
\label{eq:CurInvNoFieldRedef}
\end{align}

Next, we will apply the following zero-derivative field redefinition in the UV theory, which is simply to rotate the two massive scalars by an angle $\theta$:\footnote{This field redefinition serves as a proxy for the transformation needed to diagonalize the mass matrix at the global minimum in theories of extended scalar sectors with a non trivial potential.}
\begin{equation}\renewcommand\arraystretch{1.2}
\mqty( \tilde r \\ \tilde S) = \mqty( \cos \theta & -\sin \theta \\ \sin\theta & \cos\theta ) \mqty( r \\ S ) \,.
\label{eqn:UVRedef}
\end{equation}
Then the UV theory written in terms of $\tilde r$, $\vec n$, and $\tilde S$ appears to be slightly less trivial
\begin{equation}
  \Lag_\text{UV} \big[\tilde r, \vec n, \tilde S\big] = \frac{1}{2}\s{( {\partial \tilde r} )^2} + \frac{1}{2}\s{\big( {\partial \tilde S} \big)^2} - \frac{1}{2}\s \tilde m_1^2[\vec n]\, \tilde r^2 - \frac{1}{2}\s\tilde m_2^2[\vec n]\,\tilde S^2 - \tilde m_{12}^2[\vec n]\,{\tilde r}\s{\tilde S} \,,
\end{equation}
with
\begin{subequations}\label{eqn:MassMatrixParameters}
\begin{align}
  \tilde m_1^2[\vec n] &\equiv \big({m^2} - (\partial \vec n)^2\big)\s \cos^2\theta + {M^2}\s \sin^2\theta \ \\[5pt]
\tilde m_2^2[\vec n] &\equiv \big({m^2} - (\partial \vec n)^2\big)\s \sin^2\theta + {M^2}\s \cos^2\theta \ \\[5pt]
\tilde m_{12}^2[\vec n] &\equiv  - \left( {{M^2} - {m^2} + (\partial \vec n)^2} \right) \cos\theta \s \sin\theta  \,.
\end{align}
\end{subequations}
Although the new field $\tilde S$ is not a mass eigenstate, we can still formally integrate it out by solving its EOM:
\begin{equation}
\frac{\delta S_\text{UV}}{\delta\tilde S} = 0
\qquad\Longrightarrow\qquad
\tilde S \big[\tilde r, \vec n\big] = - \frac{1}{{\partial^2} + \tilde m_2^2[\vec n]}\s {\tilde m_{12}^2[\vec n]}\s {\tilde r} \,.
\label{eqn:EOMtildephi2}
\end{equation}
The resulting effective Lagrangian is then
\begin{align}
\Lag_\text{Eff} \left[\tilde r , \vec n\right] &= \frac{1}{2}{\left( {\partial {{\tilde r}}} \right)^2} - \frac{1}{2}\s\tilde m_1^2[\vec n]\,\tilde r^2 + \frac{1}{2}\s \tilde r\s {\tilde m_{12}^2[\vec n]} \s\frac{1}{{\partial^2} + \tilde m_2^2[\vec n]}\s   {\tilde m_{12}^2[\vec n]} \,\tilde r \notag\\[7pt]
                                               &= \frac{1}{2}\s k_R\s {( {\partial {{\tilde r}}} )^2}  + \frac{1}{2}\s k_N\s \tilde r^2\s {( {\partial {\vec n}} )^2} - \frac{1}{2}\s k_V\s \tilde r^2  + \mathcal{O}\big( \partial^4 \big) \ ,
\label{eqn:LEFTtildephi1}
\end{align}
where in the second line we have expanded up to two derivative order, and the coefficients are
\begin{subequations}
\begin{align}
k_R &=  1 + \frac{\tilde m_{12}^4[0]}{\tilde m_2^4[0]} \\[5pt]
k_N &= \left( \cos\theta - \sin\theta \frac{\tilde m_{12}^2[0]}{\tilde m_2^2[0]} \right)^2 \\[5pt]
k_V &= \tilde m_1^2[0] - \frac{\tilde m_{12}^4[0]}{\tilde m_{2}^2[0]}  \ ,
\end{align}
\end{subequations}
which can be used to compute the curvature invariants
\begin{align}
R = \frac{6}{k_R\s \tilde r^2} \left( \frac{k_R}{k_N} - 1 \right)\qquad \text{and} \qquad \nabla^2 V = 4\s \frac{k_V}{k_R} \, ,
\end{align}
which differ from the invariants computed in~\cref{eq:CurInvNoFieldRedef}.
Furthermore, $R$ diverges at the origin $(r\to 0)$ due to a conical singularity in the metric.

Of course, the two EFTs must yield identical predictions.
In fact, we can explicitly derive that they are equivalent, by transforming between them formally using the field redefinition
\begin{align}
  \tilde r &= \left[ (\partial^2 + \tilde m_1^2[\vec n]) - \tilde m_{12}^2[\vec n] (\partial^2 + \tilde m_{2}^2[\vec n])^{-1} \tilde m_{12}^2[\vec n] \right]^{-\frac{1}{2}} \left[\partial^2 + m^2 - (\partial \vec n)^2 \right]^\frac{1}{2} r \notag \\[5pt]
           &= \sqrt{\frac{\tilde m_2^2[0]}{M^2}} \left( 1 + \frac{ (M^2 - m^2) \s \sin^2\theta }{2\s \tilde m_2^2[0]\s M^2}\s \partial^2 - \frac{\sin^2\theta}{2\s \tilde m_2^2[0]} (\partial \vec n)^2 + \mathcal{O}\big(\partial^4\big) \right) r \, ,
\end{align}
where in the second line we have expanded to two-derivative order.
This field redefinition is singular at the origin, in the sense that $\vec {\tilde \phi} \equiv \tilde r \s\vec n$ cannot be written in terms of $\vec \phi \equiv r\s \vec n$ using expressions that are analytic in $\vec \phi$ at $\vec \phi = \vec 0$.
Furthermore, we see explicitly that a field redefinition without derivatives in the UV induces a field redefinition that includes derivatives in the EFT.
This explains why the curvature invariants differed between the two descriptions, and shows a limitation of the criteria presented in this paper.

Even so, this limitation is not fatal provided some care is taken to work with the ``right'' coordinates when matching to the UV theory.
If the goal is to match onto SMEFT when possible, then as long as the UV theory does not have any massless fluctuations at the $O(4)$ invariant point, or extra electroweak symmetry breaking, there must exist a field redefinition that can convert the resulting EFT into a SMEFT.
Specifically, we choose SMEFT-like coordinates $\vec \phi$ in the UV that can be used to match directly onto a benign analytic-to-all-orders SMEFT about the $O(4)$ invariant point, as in~\cref{sec:scalarsmeft}.
Note, however, that if ``spurious'' conical singularities can be defined away with derivative field redefinitions (as in the example provided here), then it is possible to remove ``real'' conical singularities that are the result of integrating out massless fluctuations.
The difference between these two cases is that when trying to define away a physical non-analyticity, one may be able to find coordinates such that the manifold is smooth, but there remains a tower of non-analytic terms at higher derivative order due to the massless fluctuations.
Hence, the Curvature Criteria would break down (which implies that the LO Criteria would also break down), although the pathology would be apparent in generalized curvature invariants defined beyond two-derivative order.

\end{spacing}


\begin{spacing}{1.09}
\addcontentsline{toc}{section}{\protect\numberline{}References}%
\bibliographystyle{utphys}
\bibliography{HEFT_vs_SMEFT}

\providecommand{\href}[2]{#2}\begingroup\raggedright\begin{thebibliography}{10}

\bibitem{Weinberg:1980wa}
S.~Weinberg, ``{Effective Gauge Theories},''
\href{http://dx.doi.org/10.1016/0370-2693(80)90660-7}{{\em Phys. Lett.} {\bf
  91B} (1980)  51--55}.

\bibitem{Polchinski:1992ed}
J.~Polchinski, ``{Effective field theory and the Fermi surface},'' in {\em
  {TASI 1992}}, pp.~0235--276.
\newblock 1992.
\newblock
\href{http://arxiv.org/abs/hep-th/9210046}{{\tt arXiv:hep-th/9210046
  [hep-th]}}.
\newblock

\bibitem{Georgi:1994qn}
H.~Georgi, ``{Effective field theory},''
\href{http://dx.doi.org/10.1146/annurev.ns.43.120193.001233}{{\em Ann. Rev.
  Nucl. Part. Sci.} {\bf 43} (1993)  209--252}.

\bibitem{Manohar:1995xr}
A.~V. Manohar, ``{Effective field theories},'' in {\em {Quarks and colliders
  1995}}, pp.~274--315.
\newblock 1995.
\newblock
\href{http://arxiv.org/abs/hep-ph/9508245}{{\tt arXiv:hep-ph/9508245
  [hep-ph]}}.
\newblock

\bibitem{Kaplan:1995uv}
D.~B. Kaplan, ``{Effective field theories},'' in {\em {Beyond the standard
  model 5}}.
\newblock 1995.
\newblock
\href{http://arxiv.org/abs/nucl-th/9506035}{{\tt arXiv:nucl-th/9506035
  [nucl-th]}}.
\newblock

\bibitem{Rothstein:2003mp}
I.~Z. Rothstein, ``{TASI lectures on effective field theories},''
\newblock 2003.
\newblock
\href{http://arxiv.org/abs/hep-ph/0308266}{{\tt arXiv:hep-ph/0308266
  [hep-ph]}}.
\newblock

\bibitem{Kaplan:2005es}
D.~B. Kaplan, ``{Five lectures on effective field theory},''
\newblock 2005.
\newblock
\href{http://arxiv.org/abs/nucl-th/0510023}{{\tt arXiv:nucl-th/0510023
  [nucl-th]}}.
\newblock

\bibitem{Skiba:2010xn}
W.~Skiba, \href{http://dx.doi.org/10.1142/9789814327183_0001}{``{Effective
  Field Theory and Precision Electroweak Measurements},''} in {\em {TASI
  2009}}, pp.~5--70.
\newblock 2011.
\newblock
\href{http://arxiv.org/abs/1006.2142}{{\tt arXiv:1006.2142 [hep-ph]}}.
\newblock

\bibitem{Petrov:2016azi}
A.~A. Petrov and A.~E. Blechman, \href{http://dx.doi.org/10.1142/8619}{{\em
  {Effective Field Theories}}}.
\newblock WSP,
2016.
\newblock

\bibitem{Manohar:2018aog}
A.~V. Manohar, ``{Introduction to Effective Field Theories},'' in {\em {Les
  Houches summer school 2017}}.
\newblock 2018.
\newblock
\href{http://arxiv.org/abs/1804.05863}{{\tt arXiv:1804.05863 [hep-ph]}}.
\newblock

\bibitem{Neubert:2019mrz}
M.~Neubert, ``{Les Houches Lectures on Renormalization Theory and Effective
  Field Theories},'' in {\em {Les Houches summer school 2017}}.
\newblock 2019.
\newblock
\href{http://arxiv.org/abs/1901.06573}{{\tt arXiv:1901.06573 [hep-ph]}}.
\newblock

\bibitem{Cohen:2019wxr}
T.~Cohen, ``{As Scales Become Separated: Lectures on Effective Field Theory},''
  {\em PoS: TASI 2018} (2019)  011,
\href{http://arxiv.org/abs/1903.03622}{{\tt arXiv:1903.03622 [hep-ph]}}.

\bibitem{Penco:2020kvy}
R.~Penco, ``{An Introduction to Effective Field Theories},''
  \href{http://arxiv.org/abs/2006.16285}{{\tt arXiv:2006.16285 [hep-th]}}.

\bibitem{Weinberg:1979sa}
S.~Weinberg, ``{Baryon and Lepton Nonconserving Processes},''
  \href{http://dx.doi.org/10.1103/PhysRevLett.43.1566}{{\em Phys. Rev. Lett.}
  {\bf 43} (1979)  1566--1570}.

\bibitem{Buchmuller:1985jz}
W.~Buchmuller and D.~Wyler, ``{Effective Lagrangian Analysis of New
  Interactions and Flavor Conservation},''
  \href{http://dx.doi.org/10.1016/0550-3213(86)90262-2}{{\em Nucl. Phys. B}
  {\bf 268} (1986)  621--653}.

\bibitem{Leung:1984ni}
C.~N. Leung, S.~Love, and S.~Rao, ``{Low-Energy Manifestations of a New
  Interaction Scale: Operator Analysis},''
  \href{http://dx.doi.org/10.1007/BF01588041}{{\em Z. Phys. C} {\bf 31} (1986)
  433}.

\bibitem{deFlorian:2016spz}
{\bf LHC Higgs Cross Section Working Group} Collaboration, D.~de~Florian {\em
  et al.}, ``{Handbook of LHC Higgs Cross Sections: 4. Deciphering the Nature
  of the Higgs Sector},'' \href{http://arxiv.org/abs/1610.07922}{{\tt
  arXiv:1610.07922 [hep-ph]}}.

\bibitem{Brivio:2017vri}
I.~Brivio and M.~Trott, ``{The Standard Model as an Effective Field Theory},''
  \href{http://dx.doi.org/10.1016/j.physrep.2018.11.002}{{\em Phys. Rept.} {\bf
  793} (2019)  1--98},
\href{http://arxiv.org/abs/1706.08945}{{\tt arXiv:1706.08945 [hep-ph]}}.

\bibitem{Feruglio:1992wf}
F.~Feruglio, ``{The Chiral approach to the electroweak interactions},''
  \href{http://dx.doi.org/10.1142/S0217751X93001946}{{\em Int. J. Mod. Phys.}
  {\bf A8} (1993)  4937--4972},
\href{http://arxiv.org/abs/hep-ph/9301281}{{\tt arXiv:hep-ph/9301281
  [hep-ph]}}.

\bibitem{Bagger:1993zf}
J.~Bagger, V.~D. Barger, K.-m. Cheung, J.~F. Gunion, T.~Han, G.~A. Ladinsky,
  R.~Rosenfeld, and C.~P. Yuan, ``{The Strongly interacting W W system: Gold
  plated modes},'' \href{http://dx.doi.org/10.1103/PhysRevD.49.1246}{{\em Phys.
  Rev.} {\bf D49} (1994)  1246--1264},
\href{http://arxiv.org/abs/hep-ph/9306256}{{\tt arXiv:hep-ph/9306256
  [hep-ph]}}.

\bibitem{Koulovassilopoulos:1993pw}
V.~Koulovassilopoulos and R.~S. Chivukula, ``{The Phenomenology of a
  nonstandard Higgs boson in W(L) W(L) scattering},''
  \href{http://dx.doi.org/10.1103/PhysRevD.50.3218}{{\em Phys. Rev.} {\bf D50}
  (1994)  3218--3234},
\href{http://arxiv.org/abs/hep-ph/9312317}{{\tt arXiv:hep-ph/9312317
  [hep-ph]}}.

\bibitem{Burgess:1999ha}
C.~P. Burgess, J.~Matias, and M.~Pospelov, ``{A Higgs or not a Higgs? What to
  do if you discover a new scalar particle},''
  \href{http://dx.doi.org/10.1142/S0217751X02009813}{{\em Int. J. Mod. Phys.}
  {\bf A17} (2002)  1841--1918},
\href{http://arxiv.org/abs/hep-ph/9912459}{{\tt arXiv:hep-ph/9912459
  [hep-ph]}}.

\bibitem{Grinstein:2007iv}
B.~Grinstein and M.~Trott, ``{A Higgs-Higgs bound state due to new physics at a
  TeV},'' \href{http://dx.doi.org/10.1103/PhysRevD.76.073002}{{\em Phys. Rev.}
  {\bf D76} (2007)  073002},
\href{http://arxiv.org/abs/0704.1505}{{\tt arXiv:0704.1505 [hep-ph]}}.

\bibitem{Alonso:2012px}
R.~Alonso, M.~B. Gavela, L.~Merlo, S.~Rigolin, and J.~Yepes, ``{The Effective
  Chiral Lagrangian for a Light Dynamical "Higgs Particle},''
  \href{http://dx.doi.org/10.1016/j.physletb.2013.04.037}{{\em Phys. Lett.}
  {\bf B722} (2013)  330--335},
\href{http://arxiv.org/abs/1212.3305}{{\tt arXiv:1212.3305 [hep-ph]}}.

\bibitem{Espriu:2013fia}
D.~Espriu, F.~Mescia, and B.~Yencho, ``{Radiative corrections to WL WL
  scattering in composite Higgs models},''
  \href{http://dx.doi.org/10.1103/PhysRevD.88.055002}{{\em Phys. Rev. D} {\bf
  88} (2013)  055002}, \href{http://arxiv.org/abs/1307.2400}{{\tt
  arXiv:1307.2400 [hep-ph]}}.

\bibitem{Buchalla:2013rka}
G.~Buchalla, O.~Cata, and C.~Krause, ``{Complete Electroweak Chiral Lagrangian
  with a Light Higgs at NLO},''
  \href{http://dx.doi.org/10.1016/j.nuclphysb.2016.09.010,
  10.1016/j.nuclphysb.2014.01.018}{{\em Nucl. Phys.} {\bf B880} (2014)
  552--573}, \href{http://arxiv.org/abs/1307.5017}{{\tt arXiv:1307.5017
  [hep-ph]}}.
[Erratum: Nucl. Phys.B913,475(2016)].

\bibitem{Brivio:2013pma}
I.~Brivio, T.~Corbett, O.~Eboli, M.~Gavela, J.~Gonzalez-Fraile,
  M.~Gonzalez-Garcia, L.~Merlo, and S.~Rigolin, ``{Disentangling a dynamical
  Higgs},'' \href{http://dx.doi.org/10.1007/JHEP03(2014)024}{{\em JHEP} {\bf
  03} (2014)  024}, \href{http://arxiv.org/abs/1311.1823}{{\tt arXiv:1311.1823
  [hep-ph]}}.

\bibitem{Alonso:2015fsp}
R.~Alonso, E.~E. Jenkins, and A.~V. Manohar, ``{A Geometric Formulation of
  Higgs Effective Field Theory: Measuring the Curvature of Scalar Field
  Space},'' \href{http://dx.doi.org/10.1016/j.physletb.2016.01.041}{{\em Phys.
  Lett.} {\bf B754} (2016)  335--342},
\href{http://arxiv.org/abs/1511.00724}{{\tt arXiv:1511.00724 [hep-ph]}}.

\bibitem{Alonso:2016oah}
R.~Alonso, E.~E. Jenkins, and A.~V. Manohar, ``{Geometry of the Scalar
  Sector},'' \href{http://dx.doi.org/10.1007/JHEP08(2016)101}{{\em JHEP} {\bf
  08} (2016)  101},
\href{http://arxiv.org/abs/1605.03602}{{\tt arXiv:1605.03602 [hep-ph]}}.

\bibitem{Buchalla:2017jlu}
G.~Buchalla, O.~Cata, A.~Celis, M.~Knecht, and C.~Krause, ``{Complete One-Loop
  Renormalization of the Higgs-Electroweak Chiral Lagrangian},''
  \href{http://dx.doi.org/10.1016/j.nuclphysb.2018.01.009}{{\em Nucl. Phys.}
  {\bf B928} (2018)  93--106},
\href{http://arxiv.org/abs/1710.06412}{{\tt arXiv:1710.06412 [hep-ph]}}.

\bibitem{Alonso:2017tdy}
R.~Alonso, K.~Kanshin, and S.~Saa, ``{Renormalization group evolution of Higgs
  effective field theory},''
  \href{http://dx.doi.org/10.1103/PhysRevD.97.035010}{{\em Phys. Rev.} {\bf
  D97} (2018) no.~3, 035010},
\href{http://arxiv.org/abs/1710.06848}{{\tt arXiv:1710.06848 [hep-ph]}}.

\bibitem{deBlas:2018tjm}
J.~de~Blas, O.~Eberhardt, and C.~Krause, ``{Current and Future Constraints on
  Higgs Couplings in the Nonlinear Effective Theory},''
  \href{http://dx.doi.org/10.1007/JHEP07(2018)048}{{\em JHEP} {\bf 07} (2018)
  048},
\href{http://arxiv.org/abs/1803.00939}{{\tt arXiv:1803.00939 [hep-ph]}}.

\bibitem{Falkowski:2019tft}
A.~Falkowski and R.~Rattazzi, ``{Which EFT},''
  \href{http://dx.doi.org/10.1007/JHEP10(2019)255}{{\em JHEP} {\bf 10} (2019)
  255},
\href{http://arxiv.org/abs/1902.05936}{{\tt arXiv:1902.05936 [hep-ph]}}.

\bibitem{Buchalla:2016bse}
G.~Buchalla, O.~Cata, A.~Celis, and C.~Krause, ``{Standard Model Extended by a
  Heavy Singlet: Linear vs. Nonlinear EFT},''
  \href{http://dx.doi.org/10.1016/j.nuclphysb.2017.02.006}{{\em Nucl. Phys. B}
  {\bf 917} (2017)  209--233}, \href{http://arxiv.org/abs/1608.03564}{{\tt
  arXiv:1608.03564 [hep-ph]}}.

\bibitem{Criado:2018sdb}
J.~Criado and M.~P\'erez-Victoria, ``{Field redefinitions in effective theories
  at higher orders},'' \href{http://dx.doi.org/10.1007/JHEP03(2019)038}{{\em
  JHEP} {\bf 03} (2019)  038}, \href{http://arxiv.org/abs/1811.09413}{{\tt
  arXiv:1811.09413 [hep-ph]}}.

\bibitem{Helset:2020yio}
A.~Helset, A.~Martin, and M.~Trott, ``{The Geometric Standard Model Effective
  Field Theory},'' \href{http://dx.doi.org/10.1007/JHEP03(2020)163}{{\em JHEP}
  {\bf 03} (2020)  163}, \href{http://arxiv.org/abs/2001.01453}{{\tt
  arXiv:2001.01453 [hep-ph]}}.

\bibitem{amplitudes}
T.~Cohen, N.~Craig, X.~Lu, and D.~Sutherland, ``{Linking Convergence and
  Unitarity in Scalar EFTs}.'' To appear.

\bibitem{twohdm}
T.~Cohen, N.~Craig, X.~Lu, and D.~Sutherland, ``{A New Basis for the Two Higgs
  Doublet Model}.'' To appear.

\bibitem{higglons}
I.~Banta, T.~Cohen, N.~Craig, X.~Lu, and D.~Sutherland, ``{Can New Particles
  Acquire All Their Mass from the Higgs?}.'' To appear.

\bibitem{Finn:2020nvn}
K.~Finn, S.~Karamitsos, and A.~Pilaftsis, ``{Frame Covariant Formalism for
  Fermionic Theories},'' \href{http://arxiv.org/abs/2006.05831}{{\tt
  arXiv:2006.05831 [hep-th]}}.

\bibitem{Coleman:1969sm}
S.~R. Coleman, J.~Wess, and B.~Zumino, ``{Structure of phenomenological
  Lagrangians. 1.},''
\href{http://dx.doi.org/10.1103/PhysRev.177.2239}{{\em Phys. Rev.} {\bf 177}
  (1969)  2239--2247}.

\bibitem{Callan:1969sn}
C.~G. Callan, Jr., S.~R. Coleman, J.~Wess, and B.~Zumino, ``{Structure of
  phenomenological Lagrangians. 2.},''
\href{http://dx.doi.org/10.1103/PhysRev.177.2247}{{\em Phys. Rev.} {\bf 177}
  (1969)  2247--2250}.

\bibitem{krantz2002primer}
S.~Krantz and H.~Parks, {\em A Primer of Real Analytic Functions}.
\newblock Advanced Texts Series. Birkh{\"a}user Boston, 2002.
\newblock \url{https://books.google.it/books?id=i4vw2STJl2QC}.

\bibitem{krantz2012implicit}
S.~G. Krantz and H.~R. Parks, {\em The implicit function theorem: history,
  theory, and applications}.
\newblock Springer Science \& Business Media, 2012.

\bibitem{Henning:2014wua}
B.~Henning, X.~Lu, and H.~Murayama, ``{How to use the Standard Model effective
  field theory},'' \href{http://dx.doi.org/10.1007/JHEP01(2016)023}{{\em JHEP}
  {\bf 01} (2016)  023},
\href{http://arxiv.org/abs/1412.1837}{{\tt arXiv:1412.1837 [hep-ph]}}.

\bibitem{deBlas:2014mba}
J.~de~Blas, M.~Chala, M.~Perez-Victoria, and J.~Santiago, ``{Observable Effects
  of General New Scalar Particles},''
  \href{http://dx.doi.org/10.1007/JHEP04(2015)078}{{\em JHEP} {\bf 04} (2015)
  078}, \href{http://arxiv.org/abs/1412.8480}{{\tt arXiv:1412.8480 [hep-ph]}}.

\bibitem{Gorbahn:2015gxa}
M.~Gorbahn, J.~M. No, and V.~Sanz, ``{Benchmarks for Higgs Effective Theory:
  Extended Higgs Sectors},''
  \href{http://dx.doi.org/10.1007/JHEP10(2015)036}{{\em JHEP} {\bf 10} (2015)
  036}, \href{http://arxiv.org/abs/1502.07352}{{\tt arXiv:1502.07352
  [hep-ph]}}.

\bibitem{Chiang:2015ura}
C.-W. Chiang and R.~Huo, ``{Standard Model Effective Field Theory: Integrating
  out a Generic Scalar},''
  \href{http://dx.doi.org/10.1007/JHEP09(2015)152}{{\em JHEP} {\bf 09} (2015)
  152}, \href{http://arxiv.org/abs/1505.06334}{{\tt arXiv:1505.06334
  [hep-ph]}}.

\bibitem{Brehmer:2015rna}
J.~Brehmer, A.~Freitas, D.~Lopez-Val, and T.~Plehn, ``{Pushing Higgs Effective
  Theory to its Limits},''
  \href{http://dx.doi.org/10.1103/PhysRevD.93.075014}{{\em Phys. Rev. D} {\bf
  93} (2016) no.~7, 075014}, \href{http://arxiv.org/abs/1510.03443}{{\tt
  arXiv:1510.03443 [hep-ph]}}.

\bibitem{Jiang:2016czg}
Y.~Jiang and M.~Trott, ``{On the non-minimal character of the SMEFT},''
  \href{http://dx.doi.org/10.1016/j.physletb.2017.04.053}{{\em Phys. Lett. B}
  {\bf 770} (2017)  108--116}, \href{http://arxiv.org/abs/1612.02040}{{\tt
  arXiv:1612.02040 [hep-ph]}}.

\bibitem{Corbett:2017ieo}
T.~Corbett, A.~Joglekar, H.-L. Li, and J.-H. Yu, ``{Exploring Extended Scalar
  Sectors with Di-Higgs Signals: A Higgs EFT Perspective},''
  \href{http://dx.doi.org/10.1007/JHEP05(2018)061}{{\em JHEP} {\bf 05} (2018)
  061}, \href{http://arxiv.org/abs/1705.02551}{{\tt arXiv:1705.02551
  [hep-ph]}}.

\bibitem{Dawson:2017vgm}
S.~Dawson and C.~W. Murphy, ``{Standard Model EFT and Extended Scalar
  Sectors},'' \href{http://dx.doi.org/10.1103/PhysRevD.96.015041}{{\em Phys.
  Rev.} {\bf D96} (2017) no.~1, 015041},
\href{http://arxiv.org/abs/1704.07851}{{\tt arXiv:1704.07851 [hep-ph]}}.

\bibitem{Ellis:2017jns}
S.~A.~R. Ellis, J.~Quevillon, T.~You, and Z.~Zhang, ``{Extending the Universal
  One-Loop Effective Action: Heavy-Light Coefficients},''
  \href{http://dx.doi.org/10.1007/JHEP08(2017)054}{{\em JHEP} {\bf 08} (2017)
  054},
\href{http://arxiv.org/abs/1706.07765}{{\tt arXiv:1706.07765 [hep-ph]}}.

\bibitem{Jiang:2018pbd}
M.~Jiang, N.~Craig, Y.-Y. Li, and D.~Sutherland, ``{Complete One-Loop Matching
  for a Singlet Scalar in the Standard Model EFT},''
  \href{http://dx.doi.org/10.1007/JHEP02(2019)031}{{\em JHEP} {\bf 02} (2019)
  031},
\href{http://arxiv.org/abs/1811.08878}{{\tt arXiv:1811.08878 [hep-ph]}}.

\bibitem{Haisch:2020ahr}
U.~Haisch, M.~Ruhdorfer, E.~Salvioni, E.~Venturini, and A.~Weiler, ``{Singlet
  night in Feynman-ville: one-loop matching of a real scalar},''
\href{http://arxiv.org/abs/2003.05936}{{\tt arXiv:2003.05936 [hep-ph]}}.

\bibitem{DHoker:1984izu}
E.~D'Hoker and E.~Farhi, ``{Decoupling a Fermion Whose Mass Is Generated by a
  Yukawa Coupling: The General Case},''
\href{http://dx.doi.org/10.1016/0550-3213(84)90586-8}{{\em Nucl. Phys.} {\bf
  B248} (1984)  59--76}.

\bibitem{Huo:2015exa}
R.~Huo, ``{Standard Model Effective Field Theory: Integrating out Vector-Like
  Fermions},'' \href{http://dx.doi.org/10.1007/JHEP09(2015)037}{{\em JHEP} {\bf
  09} (2015)  037}, \href{http://arxiv.org/abs/1506.00840}{{\tt
  arXiv:1506.00840 [hep-ph]}}.

\bibitem{Chen:2017hak}
C.-Y. Chen, S.~Dawson, and E.~Furlan, ``{Vectorlike fermions and Higgs
  effective field theory revisited},''
  \href{http://dx.doi.org/10.1103/PhysRevD.96.015006}{{\em Phys. Rev.} {\bf
  D96} (2017) no.~1, 015006},
\href{http://arxiv.org/abs/1703.06134}{{\tt arXiv:1703.06134 [hep-ph]}}.

\bibitem{Ellis:2020ivx}
S.~A. Ellis, J.~Quevillon, P.~N.~H. Vuong, T.~You, and Z.~Zhang, ``{The
  Fermionic Universal One-Loop Effective Action},''
  \href{http://arxiv.org/abs/2006.16260}{{\tt arXiv:2006.16260 [hep-ph]}}.

\bibitem{Angelescu:2020yzf}
A.~Angelescu and P.~Huang, ``{Integrating Out New Fermions at One Loop},''
  \href{http://arxiv.org/abs/2006.16532}{{\tt arXiv:2006.16532 [hep-ph]}}.

\bibitem{Gunion:2002zf}
J.~F. Gunion and H.~E. Haber, ``{The CP conserving two Higgs doublet model: The
  Approach to the decoupling limit},''
  \href{http://dx.doi.org/10.1103/PhysRevD.67.075019}{{\em Phys. Rev. D} {\bf
  67} (2003)  075019}, \href{http://arxiv.org/abs/hep-ph/0207010}{{\tt
  arXiv:hep-ph/0207010}}.

\bibitem{Egana-Ugrinovic:2015vgy}
D.~Egana-Ugrinovic and S.~Thomas, ``{Effective Theory of Higgs Sector Vacuum
  States},''
\href{http://arxiv.org/abs/1512.00144}{{\tt arXiv:1512.00144 [hep-ph]}}.

\bibitem{Belusca-Maito:2016dqe}
H.~Belusca-Maito, A.~Falkowski, D.~Fontes, J.~C. Romao, and J.~P. Silva,
  ``{Higgs EFT for 2HDM and beyond},''
  \href{http://dx.doi.org/10.1140/epjc/s10052-017-4745-5}{{\em Eur. Phys. J.}
  {\bf C77} (2017) no.~3, 176},
\href{http://arxiv.org/abs/1611.01112}{{\tt arXiv:1611.01112 [hep-ph]}}.

\bibitem{Faro:2020qyp}
F.~Faro, J.~C. Romao, and J.~P. Silva, ``{Nondecoupling in Multi-Higgs doublet
  models},'' \href{http://dx.doi.org/10.1140/epjc/s10052-020-8217-y}{{\em Eur.
  Phys. J. C} {\bf 80} (2020) no.~7, 635},
  \href{http://arxiv.org/abs/2002.10518}{{\tt arXiv:2002.10518 [hep-ph]}}.

\bibitem{Davidson:2005cw}
S.~Davidson and H.~E. Haber, ``{Basis-independent methods for the
  two-Higgs-doublet model},''
  \href{http://dx.doi.org/10.1103/PhysRevD.72.099902}{{\em Phys. Rev. D} {\bf
  72} (2005)  035004}, \href{http://arxiv.org/abs/hep-ph/0504050}{{\tt
  arXiv:hep-ph/0504050}}. [Erratum: Phys.Rev.D 72, 099902 (2005)].

\bibitem{Khandker:2012zu}
Z.~U. Khandker, D.~Li, and W.~Skiba, ``{Electroweak Corrections from Triplet
  Scalars},'' \href{http://dx.doi.org/10.1103/PhysRevD.86.015006}{{\em Phys.
  Rev. D} {\bf 86} (2012)  015006}, \href{http://arxiv.org/abs/1201.4383}{{\tt
  arXiv:1201.4383 [hep-ph]}}.

\bibitem{Ellis:2016enq}
S.~A.~R. Ellis, J.~Quevillon, T.~You, and Z.~Zhang, ``{Mixed heavy--light
  matching in the Universal One-Loop Effective Action},''
  \href{http://dx.doi.org/10.1016/j.physletb.2016.09.016}{{\em Phys. Lett. B}
  {\bf 762} (2016)  166--176}, \href{http://arxiv.org/abs/1604.02445}{{\tt
  arXiv:1604.02445 [hep-ph]}}.

\bibitem{Englert:2014uua}
C.~Englert, A.~Freitas, M.~Mühlleitner, T.~Plehn, M.~Rauch, M.~Spira, and
  K.~Walz, ``{Precision Measurements of Higgs Couplings: Implications for New
  Physics Scales},''
  \href{http://dx.doi.org/10.1088/0954-3899/41/11/113001}{{\em J. Phys. G} {\bf
  41} (2014)  113001}, \href{http://arxiv.org/abs/1403.7191}{{\tt
  arXiv:1403.7191 [hep-ph]}}.

\bibitem{Coleman:1973jx}
S.~R. Coleman and E.~J. Weinberg, ``{Radiative Corrections as the Origin of
  Spontaneous Symmetry Breaking},''
\href{http://dx.doi.org/10.1103/PhysRevD.7.1888}{{\em Phys. Rev.} {\bf D7}
  (1973)  1888--1910}.

\bibitem{Henning:2016lyp}
B.~Henning, X.~Lu, and H.~Murayama, ``{One-loop Matching and Running with
  Covariant Derivative Expansion},''
  \href{http://dx.doi.org/10.1007/JHEP01(2018)123}{{\em JHEP} {\bf 01} (2018)
  123},
\href{http://arxiv.org/abs/1604.01019}{{\tt arXiv:1604.01019 [hep-ph]}}.

\end{thebibliography}\endgroup
\end{spacing}

\end{document}